\documentclass[iop]{emulateapj}
\usepackage{amsthm}
\usepackage{amsmath}

\shorttitle{Radiation hydrodynamics in {\it Athena}}
\shortauthors{M. A. Skinner and E. C. Ostriker}

\begin{document}

\title{A TWO-MOMENT RADIATION HYDRODYNAMICS MODULE IN {\it ATHENA} \\ USING A TIME-EXPLICIT GODUNOV METHOD}

\author{M. Aaron Skinner\altaffilmark{1,}\altaffilmark{2} and Eve C. Ostriker\altaffilmark{3,}\altaffilmark{4}}
\affil{Department of Astronomy, University of Maryland, College Park, MD 20742-2421}
\altaffiltext{1}{Applied Mathematics \& Statistics, and Scientific Computation Program, University of Maryland,
College Park, MD 20742-4015}
\altaffiltext{2}{askinner@astro.umd.edu}
\altaffiltext{3}{Department of Astrophysical Sciences, Princeton University, Princeton, NJ 08544-1001}
\altaffiltext{4}{eco@astro.princeton.edu}

\begin{abstract}
We describe a module for the {\it Athena} code that solves the gray equations of radiation hydrodynamics (RHD), based on the first two moments of the radiative transfer equation.  We use a combination of explicit Godunov methods to advance the gas and radiation variables including the non-stiff source terms, and a local implicit method to integrate the stiff source terms.  We adopt the $M_1$ closure relation and include all leading source terms to $\mathcal{O}(\beta\tau)$.  We employ the reduced speed of light approximation (RSLA) with subcycling of the radiation variables in order to reduce computational costs.  Our code is dimensionally unsplit in one, two, and three space dimensions and is parallelized using MPI.  The streaming and diffusion limits are well-described by the $M_1$ closure model, and our implementation shows excellent behavior for a problem with a concentrated radiation source containing both regimes simultaneously.  Our operator-split method is ideally suited for problems with a slowly varying radiation field and dynamical gas flows, in which the effect of the RSLA is minimal.  We present an analysis of the dispersion relation of RHD linear waves highlighting the conditions of applicability for the RSLA.  To demonstrate the accuracy of our method, we utilize a suite of radiation and RHD tests covering a broad range of regimes, including RHD waves, shocks, and equilibria, which show second-order convergence in most cases.  As an application, we investigate radiation-driven ejection of a dusty, optically thick shell in the interstellar medium (ISM).  Finally, we compare the timing of our method with other well-known iterative schemes for the RHD equations.  Our code implementation, {\it Hyperion}, is suitable for a wide variety of astrophysical applications and will be made freely available on the Web.
\end{abstract}


\section{Introduction} \label{intro}

The importance of radiation to gaseous evolution in many astrophysical systems is well known.  To name but a few, these include star formation in a variety of environments \citep[e.g.,][]{Thompson:2005,Murray:2010}, cosmological structure formation via radiative heating/cooling processes and ionization \citep[e.g.,][]{Barkana:2001}, the dynamics of accretion disks around supermassive black holes \citep{Hirose:2009}, and galaxy evolution with central black hole feedback \citep{Ciotti:2007}.  For example, in star-forming regions of galactic disks with very high surface density $\Sigma$, radiation pressure may contribute significantly to the vertical support of the disk \citep{Thompson:2005,Krumholz:2012}, which would lead to a surface density of star formation $\Sigma_{\rm SFR}$ that is correlated linearly with $\Sigma$, rather than quadratically as expected for disks dominated by supernova feedback \citep{Ostriker:2011}.  For the most massive giant molecular clouds (GMCs), radiation pressure may dominate the disruption process \citep{Murray:2010}, and in the inner regions of accretion disks, radiation pressure may dominate gas pressure by up to a factor of 10 \citep{Hirose:2009}.  To properly gauge the effects of radiation in these and other systems, it is necessary to solve the equations of radiation hydrodynamics (RHD) in fully three-dimensional, time-dependent numerical models.

The equations of RHD consist of the Euler equations of gas dynamics and the time-dependent equation of radiative transfer.  Although the equations and methods for gas dynamics are well-known, there continues to be intensive investigation over the form of the radiative transfer equation to solve and how to incorporate it with gas dynamics.  The solution of the full time-dependent transfer equation, which consists of a six-dimensional integro-differential equation for each frequency of radiation, remains beyond the reach of modern computing.  However, the transfer equation is commonly simplified by truncating a hierarchy of moments at the second order, resulting in a system of time-dependent evolution equations for the radiation energy density and flux.  

To solve the flux equation, the radiation pressure tensor (or Eddington tensor, which is the ratio of pressure to energy density) must be supplied at each physical location, and various methods have been proposed to compute this variable Eddington tensor (VET).  For example, the Eddington tensor can be computed directly from the formal solution of the time-independent transfer equation in the optically thin case from a small, static set of radiation sources \citep{Gnedin:2001}, or along short characteristics taken over a set of preferred directions \citep[][and references therein]{Davis:2012}.  An alternative approach is to make a simplifying geometric assumption regarding the angular dependence of the underlying radiation intensity field itself.  The $M_1$ closure, originally proposed by \cite{Levermore:1984} and recently implemented by \cite{Gonzalez:2007} and \cite{Aubert:2008}, is consistent with the angular dependence of a Lorentz-boosted, isotropic distribution.  For a single source, the $M_1$ closure can describe the limiting cases of optically thin, free-streaming radiation (with $F \to \mathcal{E}/c$) and optically thick, diffusing radiation (with $F \to 0$) exactly, while smoothly connecting these limits in intermediate regimes.  In this work, we shall adopt the $M_1$ closure, while also comparing with isotropic closure relations for some tests.

An alternative to the two-moment formalism is the flux-limited diffusion (FLD) approximation, where the radiative flux is assumed proportional to the gradient of the radiation energy density field (as in Fick's law of diffusion), with special flux limiters put in place to prevent superluminal transport of radiation \citep{Levermore:1981}.  Although this is by far the most popular method currently used in RHD applications \citep{Fryxell:2000,Turner:2001,Krumholz:2007,Gittings:2008,Reynolds:2009,Swesty:2009,Commercon:2011,van-der-Holst:2011,Zhang:2011}, it can potentially lead to serious physical errors in optically thin regions, e.g., due to FLD's inability to create and follow shadows.  Furthermore, because the direction of the radiation flux is always parallel to the gradient in the radiation energy density, radiation forces may accelerate gas in the wrong direction.  Although the two-moment formalism may increase the computational requirements compared to FLD, it can potentially rectify these serious, unphysical effects.  Of course, two-moment methods may themselves have limitations, either from the computational cost of computing the VET when a large number of angles are required to resolve the radiation distribution, or from the inadequacy of adopted closure relations to capture the field arising from complex source geometries.  Therefore, it is important to compare the same problems using different RHD methods to obtain a better understanding of each approach's sensitivity to assumptions and approximations.

In addition to the moment-closure problem, there is debate over the frame in which to integrate the transfer equation.  The absorption and emission coefficients are isotropic in the Lagrangian frame (i.e., the frame comoving with the gas), hence their angular moments are trivial.  However, photons are observed to move along curved trajectories with varying frequencies in this frame, which complicates the solution of the transfer equation.  Moreover, in the Eulerian frame (i.e., the inertial ``laboratory'' frame), the photons move along straight lines with fixed frequencies, but the material property coefficients are no longer isotropic.  \cite{Mihalas:1982} introduce the approach of solving the moment equations in the so-called \emph{mixed-frame}, where material properties are measured in the Lagrangian frame, but intensities, frequencies, lengths, and times are all measured in the Eulerian frame.  We adopt this formulation and include all terms of $\mathcal{O}(\beta\tau)$, where $\beta \equiv v/c$ and $\tau$ is the optical depth.  The importance of including these terms has been extensively described \citep{Mihalas:1982,Krumholz:2007}.

Finally, the dynamics of RHD systems vary substantially in different physical regimes, depending on, among other things, the typical optical depth and sound speed of the gas, and relative contributions of the gas and radiation to the total energy density and momentum of the system.  The equations are well-conditioned to explicit solution methods in some regimes, but often the source terms coupling the gas and radiation subsystems are so stiff that alternate methods must be sought to ensure stability.  In this work, we employ the existing high-order Godunov methods implemented in the {\it Athena} code \citep{Gardiner:2005,Gardiner:2008,Stone:2008} along with an operator splitting between the source terms and transport terms.  We solve the radiation momentum equation semi-explicitly; therefore, we must consider the typically large difference in dynamical time scales between the gas and radiation fields, which can render the solution of the RHD equations computationally infeasible.  \cite{Gnedin:2001} have described a reduced speed of light approximation (RSLA), in which the propagation speed of the radiation field is reduced to some computationally feasible level, while seeking to preserve all relevant dynamical properties of the system.  We adopt the RSLA in this work and also formally evaluate its regime of applicability.  As we shall show, the validity requirements of the RSLA render our method best suited for systems with moderate optical depth.  Many applications involving star formation lie within this regime.

We verify our algorithm and its implementation in our code {\it Hyperion} using a suite of established and novel test problems in RHD spanning a wide range of dynamical regimes.  Among them are tests of angular resolution and shadowing, and convergence of propagating radiation and diffusion waves in problems with a radiation field that is partially coupled to a static gas field.  We also perform a basic timing benchmark to compare the performance of our code to that of a well-known FLD code on a problem involving only the partially coupled radiation subsystem.  We then test the fully coupled RHD system by investigating the radiation force in both optically thin and -thick flows, by examining the role of the $\mathcal{O}(\beta\tau)$ terms in the strong advection of radiation in an optically thick gas, by exploring the propagation of radiation-modified acoustic waves in a wide range of optical depths and energy regimes, and by investigating the structure of sub-critical shocks compared to existing semi-analytic solutions.  Finally, as a preliminary application, we examine the expulsion and subsequent driven expansion of a dusty shell of gas by radiation momentum as described by \cite{Ostriker:2011}.   

The structure of our paper is organized as follows.  In Section~\ref{mixedframe}, we give a detailed derivation of the equations of RHD, examine the various physical regimes spanned by this system, and discuss the $M_1$ closure in the context of an unsplit multidimensional Godunov method.  In Section~\ref{implementation}, we give an outline of our algorithm, review the RSLA and examine its implications for the preservation of relevant dynamical behavior, discuss the hyperbolic transport of radiation in the context of the $M_1$ model, and discuss the treatment of the various source terms according to their mathematical properties and their role in the RHD equations.  Finally, we present our code verification test suite in Section~\ref{tests}, beginning with tests of the uncoupled and partially coupled radiation subsystem in Section~\ref{radsubsystem}, and ending with tests of the fully coupled gas and radiation subsystems in Section~\ref{fullsystem}.  In Section~\ref{conclusion} we give a brief summary of our algorithm and implementation.

\section{The Mixed-Frame Equations of Radiation Hydrodynamics}  \label{mixedframe}

Following \cite{Mihalas:1982}, \cite{Mihalas:1999}, and \cite{Mihalas:2001}, we express the moments of the radiative transfer equation in the \emph{mixed-frame}, where coordinates, differential operators, frequencies, and gas and radiation variables are measured in the inertial lab frame, but material optical properties such as absorption and emission are measured in the frame comoving with the gas.  The advantage of this hybrid approach is that the differential operators remain hyperbolic in the inertial frame, while in the comoving frame the material properties are effectively isotropic.  For simplicity, we assume a gray atmosphere such that the opacities are frequency-independent.  Our method can be extended to multigroup RHD in a straightforward manner \citep{Vaytet:2011}, although this is beyond the scope of our paper.

\vspace{1em}
\subsection{Gas and Radiation Moment Equations}  \label{momenteqns}

The lab-frame equations of RHD consist of the Euler equations of hydrodynamics combined with the frequency-integrated zeroth- and first-order angular moments of the radiative transfer equation given by
\begin{subequations}  \label{mixedframe:fullsystem}
\begin{eqnarray}
  \partial_t \rho + \nabla \cdot (\rho \mathbf{v}) &=& 0, \label{mixedframe:fullsystem:density} \\
  \partial_t (\rho \mathbf{v}) + \nabla \cdot (\rho \mathbf{v} \mathbf{v} + P\mathbb{I}) &=& - \rho\nabla\Phi + \mathbf{G}, \quad \label{mixedframe:fullsystem:momentum} \\
  \partial_t E + \nabla \cdot \left[ (E+P)\mathbf{v} \right] &=& - \rho\mathbf{v} \cdot \nabla \Phi + c G^0, \label{mixedframe:fullsystem:energy} \\
  \frac{1}{\hat{c}} \,\partial_t \mathcal{E} + \nabla \cdot \left(\frac{\mathbf{F}}{c}\right) &=& -G^0,  \label{mixedframe:fullsystem:radenergy} \\
  \frac{1}{\hat{c}} \,\partial_t \left(\frac{\mathbf{F}}{c}\right) + \nabla \cdot \mathbb{P} &=& -\mathbf{G},  \label{mixedframe:fullsystem:radmomentum}
\end{eqnarray}
\end{subequations}
where $\rho$, $\mathbf{v}$, and $P$ are the gas density, velocity, and pressure, $E \equiv e + \case{1}{2} \rho |\mathbf{v}|^2$ is the gas total energy, and $\Phi$ is the gravitational potential.  Here, $e$ is the gas internal energy, which is related to the gas pressure via $e=P/(\gamma-1)$ for an ideal gas ($\gamma \ne 1$).  We assume the material is a perfect gas obeying the law
\begin{equation}
  P = \frac{\rho \, k_{\rm B} T}{\mu},  \label{momenteqns:perfectgas}
\end{equation}
where $\mu$ is the mean particle mass, and $k_{\rm B}$ is the Boltzmann constant.  In Equations~\eqref{mixedframe:fullsystem:radenergy} and~\eqref{mixedframe:fullsystem:radmomentum}, $\mathcal{E}$, $\mathbf{F}$, and $\mathbb{P}$ are the radiation energy density, flux vector, and pressure tensor, respectively, defined as frequency-integrated angular moments of the specific intensity in the inertial frame by
\begin{equation}
  \left[ \begin{array}{c} \mathcal{E} \\ \mathbf{F}/c \\ \mathbb{P} \end{array} \right] \equiv \frac{1}{c} \int_0^\infty \oint \left[ \begin{array}{c} 1 \\ \hat{\mathbf{k}} \\ \hat{\mathbf{k}} \hat{\mathbf{k}} \end{array} \right] I(\hat{\mathbf{k}},\nu) \,d\Omega \,d\nu, \label{mixedframe:radmoments}
\end{equation}
where $I(\hat{\mathbf{k}},\nu)$ is the specific intensity of the radiation field in the direction of unit vector $\hat{\mathbf{k}}$ at frequency $\nu$.  The specific radiation four-force density in Equations~\eqref{mixedframe:fullsystem:momentum}-\eqref{mixedframe:fullsystem:radmomentum} is given by
\begin{eqnarray}
  \left[ \begin{array}{c} G^0 \\ \mathbf{G} \end{array} \right] &\equiv& \frac{\rho}{c} \int_0^\infty \oint \left[ \begin{array}{c} 1 \\ \hat{\mathbf{k}} \end{array} \right] \nonumber \\ 
  && \times \left( \chi(\hat{\mathbf{k}},\nu) I(\hat{\mathbf{k}},\nu) - \eta(\hat{\mathbf{k}},\nu) \right) \,d\Omega\,d\nu, \label{mixedframe:fourforce}
\end{eqnarray}
where $\eta(\hat{\mathbf{k}},\nu)$ and $\chi(\hat{\mathbf{k}},\nu)$ are the specific emission and absorption coefficients, respectively, as measured in the inertial frame.  Note that we use $c$ to denote the ratio of photon energy to photon momentum, but in anticipation of adopting a reduced propagation speed for the radiation fluid (see Section~\ref{rsla}), we introduce $\hat{c}$ in the time-dependent terms in Equations~\eqref{mixedframe:fullsystem:radenergy} and~\eqref{mixedframe:fullsystem:radmomentum}.

Equations~\eqref{mixedframe:fullsystem:radenergy} and~\eqref{mixedframe:fullsystem:radmomentum} are often called the \emph{radiation energy} and \emph{radiation momentum} equations, since they describe the dynamic evolution of $\mathcal{E}$ and $\mathbf{F}/c$, respectively.  Note that by adding Equation~\eqref{mixedframe:fullsystem:energy} and $c$ times Equation~\eqref{mixedframe:fullsystem:radenergy}, and by neglecting external work, the source terms on the right-hand sides cancel and we obtain a strong conservation law for a combined energy density, $E + (c/\hat{c})\mathcal{E}$.  Similarly, by adding Equations~\eqref{mixedframe:fullsystem:momentum} and~\eqref{mixedframe:fullsystem:radmomentum}, and by neglecting external forces, the source terms on the right-hand sides again cancel and we obtain a strong conservation law for a combined momentum density, $\rho\mathbf{v} + (1/\hat{c})\mathbf{F}/c$.  When $\hat{c}=c$, the combined terms are the total energy density and total momentum density of the gas plus radiation, respectively.  Although they are not the focus of this work, note also that magnetic terms can be added in conservation law form to Equations~\eqref{mixedframe:fullsystem}.  The {\it Athena} code includes an unsplit evolution of magnetic fields via constrained transport \citep{Gardiner:2005,Gardiner:2008}.

For simplicity, we neglect scattering, and we assume that in the comoving frame (denoted by ``0'' subscripts) the material property coefficients are isotropic and characterized by a local temperature $T$.\footnote{This local temperature is assumed to be that of the gas.  However, in certain cases (e.g., the low gas temperature regime) the emission is set by the dust temperature rather than the gas temperature.}  Hence, $\chi(\hat{\mathbf{k}},\nu) = \kappa(\nu_0)$, where $\nu_0$ is the comoving-frame frequency, and by Kirchhoff's Law, $\eta(\hat{\mathbf{k}}_0,\nu_0) = \kappa(\nu_0) B(\nu_0,T)$, where $B(\nu_0,T) = (2h\nu_0^3/c^2)/(e^{h\nu_0/k_{\rm B}T} - 1)$ is the Planck function, $T$ is the material temperature, and $k_{\rm B}$ is the Boltzmann constant.  These assumptions would be valid, e.g., for thermal radiation in a sufficiently dense region of the interstellar medium (ISM).  

Following \cite{Krumholz:2007}, we expand the specific radiation four-force density for a direction-independent flux spectrum \citep[see][equations~54b and~54d]{Mihalas:2001} to $\mathcal{O}(v/c)^2$.  The result is
\begin{subequations}  \label{mixedframe:expandedfourforce} 
	\begin{eqnarray}
		G^0 &=& \rho \left( \kappa_{0\mathcal{E}}\mathcal{E} - \kappa_{\rm 0P}a_{\rm R} T^4 \right) + \rho \left( \kappa_{\rm 0F} - 2\kappa_{0\mathcal{E}} \right) \frac{\mathbf{v}}{c} \cdot \frac{\mathbf{F}}{c} \nonumber \\
		&& +\: \frac{1}{2} \rho \left[ 2 \left( \kappa_{0\mathcal{E}} - \kappa_{\rm 0F} \right) \mathcal{E} + \left( \kappa_{0\mathcal{E}}\mathcal{E} - \kappa_{\rm 0P} a_{\rm R} T^4 \right) \right] \left( \frac{v}{c} \right)^2 \nonumber \\
		&& +\: \rho \left( \kappa_{0\mathcal{E}} - \kappa_{\rm 0F} \right) \frac{\mathbf{v}}{c} \frac{\mathbf{v}}{c} : \mathbb{P}, \label{mixedframe:expandedfourforce:energy} \\
		\mathbf{G} &=& \rho\kappa_{\rm 0F} \frac{\mathbf{F}}{c} + \rho \left( \kappa_{0\mathcal{E}}\mathcal{E} - \kappa_{\rm 0P} a_{\rm R} T^4 \right) \frac{\mathbf{v}}{c} \nonumber \\
	  && -\: \rho\kappa_{\rm 0F} \frac{\mathbf{v}}{c} \cdot \left( \mathcal{E} \mathbb{I} + \mathbb{P} \right) + \frac{1}{2} \rho \kappa_{\rm 0F} \frac{\mathbf{F}}{c} \left( \frac{v}{c} \right)^2  \nonumber \\
		&& +\: 2 \rho \left( \kappa_{\rm 0F} - \kappa_{0\mathcal{E}} \right) \left( \frac{\mathbf{v}}{c} \cdot \frac{\mathbf{F}}{c} \right) \frac{\mathbf{v}}{c}, \label{mixedframe:expandedfourforce:momentum} 
	\end{eqnarray}
\end{subequations}
where
\begin{subequations}  \label{mixedframe:kappas}
\begin{eqnarray}
  \kappa_{\rm 0P} &\equiv& \frac{\int_0^\infty \kappa(\nu_0) B(\nu_0,T) \,d\nu_0}{B}, \\
  \kappa_{0\mathcal{E}} &\equiv& \frac{\int_0^\infty \kappa(\nu_0) \mathcal{E}(\nu_0) \,d\nu_0}{\mathcal{E}_0}, \\
  \kappa_{\rm 0F} &\equiv& \frac{\int_0^\infty \kappa(\nu_0) F(\nu_0) \,d\nu_0}{F_0},
\end{eqnarray}
\end{subequations}
are the frequency-integrated specific opacities weighted by the Planck function, energy density, and flux in the comoving frame, respectively.  The frequency-integrated Planck function is related to the temperature by
\begin{equation}
  B = \int_0^\infty B(\nu_0,T) \,d\nu_0 = \frac{ca_{\rm R} T^4}{4\pi},
\end{equation}
where $a_{\rm R} = 4\sigma_{\rm SB}/c$ is the radiation constant and $\sigma_{\rm SB}$ is the Stefan-Boltzmann constant.

For simplicity, we will henceforth take $\kappa_{\rm 0P}=\kappa_{\rm 0F}=\kappa_{\rm 0\mathcal{E}}\equiv \kappa_0$ and retain only leading-order terms to obtain the system
\begin{subequations} \label{mixedframe:graysystem}
\begin{eqnarray}
  \partial_t \rho + \nabla \cdot (\rho \mathbf{v}) &=& 0, \label{mixedframe:graysystem:density} \\
  \partial_t (\rho \mathbf{v}) + \nabla \cdot (\rho \mathbf{v} \mathbf{v} + P\mathbb{I}) &=& -\rho\nabla\Phi + \rho\kappa_0\frac{\mathbf{F}}{c} \nonumber \\
  && -\: \rho\kappa_0 \frac{\mathbf{v}}{c} \cdot (\mathcal{E}\mathbb{I} + \mathbb{P}),  \label{mixedframe:graysystem:momentum} \\
  \partial_t E + \nabla \cdot \left[ (E+P)\mathbf{v} \right] &=& - \rho\mathbf{v} \cdot \nabla \Phi \nonumber \\
  && -\: c\rho\kappa_0(a_{\rm R} T^4 - \mathcal{E}) \nonumber \\
  && -\: c\rho\kappa_0\frac{\mathbf{v}}{c}\cdot\frac{\mathbf{F}}{c}, \label{mixedframe:graysystem:energy} \\
  \frac{1}{\hat{c}} \,\partial_t \mathcal{E} + \nabla \cdot \left(\frac{\mathbf{F}}{c}\right) &=& \rho\kappa_0(a_{\rm R} T^4 - \mathcal{E}) \nonumber \\
  && +\: \rho\kappa_0\frac{\mathbf{v}}{c}\cdot\frac{\mathbf{F}}{c},  \label{mixedframe:graysystem:radenergy} \\
  \frac{1}{\hat{c}} \,\partial_t \left(\frac{\mathbf{F}}{c}\right) + \nabla \cdot \mathbb{P} &=& -\rho\kappa_0\frac{\mathbf{F}}{c} \nonumber \\
  && +\: \rho\kappa_0 \frac{\mathbf{v}}{c} \cdot (\mathcal{E}\mathbb{I} + \mathbb{P}).  \label{mixedframe:graysystem:radmomentum}
\end{eqnarray}
\end{subequations}
Note that in going from Equation~\eqref{mixedframe:expandedfourforce:momentum} to Equations~\eqref{mixedframe:graysystem:momentum} and~\eqref{mixedframe:graysystem:radmomentum}, we take $a_{\rm R} T^4 \to \mathcal{E}$ for the $\mathcal{O}(v/c)$ source terms, as the latter does not require an additional solution to obtain the gas temperature in the radiation subcycle (see Section~\ref{m1}).

With $\hat{c} \ne c$, our scheme does not conserve either the total energy or total momentum of the matter-plus-radiation. Instead, the method is designed to be able to recover the same quasi-steady radiation field as would be found when the terms $(1/c) \,\partial_t \mathcal{E}$ and $(1/c) \,\partial_t (\mathbf{F}/c)$ are small compared to other terms in the radiation energy and momentum equations.  Provided that the radiation propagation speed $\hat{c}$ is sufficiently large compared to other signal speeds, the radiation field is able to approach this quasi-steady equilibrium configuration rapidly with respect to the characteristic gas time scales.  Note that for the commonly adopted diffusion limit, $(1/c) \,\partial_t (\mathbf{F}/c)$ is set to zero.  In cases where thermal time scales are short compared to dynamical time scales, the thermal state of the gas does not depend on the energy exchange rate but primarily on other properties such as the radiation temperature.  In particular, the approximations we adopt are suitable for modeling radiation reprocessed by dust.

\vspace{1em}
\subsection{Physical Regimes for Source Terms} \label{regimes}

Following \cite{Mihalas:1982}, \cite{Mihalas:1999}, \cite{Mihalas:2001}, and \cite{Krumholz:2007}, we refer to three limiting regimes based upon the relative sizes of two dimensionless parameters:  the optical depth, $\tau \equiv L/\ell$, where $L$ is a characteristic flow scale and $\ell \equiv 1/(\rho\kappa_0)$ is the photon mean free path, and $\beta \equiv v/c$, a measure of how relativistic the gas bulk flow is.

Where $\tau \ll 1$, the gas and radiation are weakly coupled, and the radiation streams freely through the medium.  In this case, the specific intensity in the comoving frame $I_{0}$, is strongly concentrated about some direction of propagation $\hat{\mathbf{n}}_0$, hence $\mathbf{F}_0 \to c\mathcal{E}_0 \hat{\mathbf{n}}_0$ and $\mathbb{P}_0 \to \mathcal{E}_0 \hat{\mathbf{n}}_0 \hat{\mathbf{n}}_0$.  We refer to this as the \emph{streaming} limit.  

Conversely, where $\tau \gg 1$, the gas and radiation are strongly coupled, and the radiation diffuses through the medium.  In this case, $I_{0}$ is nearly isotropically distributed in the comoving frame, i.e., where the gas is locally at rest, hence $\mathbf{F}_0 \sim (c\mathcal{E}_0/\tau)\hat{\mathbf{n}}_0 \to \mathbf{0}$ and $\mathbb{P}_0\to \case{1}{3} \mathcal{E}_0 \mathbb{I}$.  Therefore, in a steady state, it follows from Equation~\eqref{mixedframe:graysystem:radenergy} that $(a_{\rm R} T^4 - \mathcal{E}_0) \sim \mathcal{E}_0/\tau^2$ in this frame, i.e., the mean intensity approaches that of a blackbody at high optical depth.

In the classical Newtonian limit, $\beta \ll 1$, hence, unless $\tau$ is very large, terms of $\mathcal{O}(\beta\tau)$ in Equations~\eqref{mixedframe:graysystem} can be neglected.  In this case, the radiation is primarily transported by diffusing through the gas as if through a completely static medium.  However, if $\tau$ is sufficiently large, the $\mathcal{O}(\beta\tau)$ terms may contribute significantly to the dynamical behavior of the system, in which case the radiation is so strongly coupled to the gas that it is primarily transported by gas advection.  We refer to the case $\beta\tau \ll 1$ as the \emph{static diffusion} limit and the case $\beta\tau \gg 1$ as the \emph{dynamic diffusion} limit.  In terms of the characteristic flow-crossing time scale $t_{\rm flow} \sim L/v$ and the characteristic radiation-diffusion time scale $t_{\rm diff} \sim L^2/(c\ell) = \tau L/c$, $\beta\tau \sim t_{\rm diff}/t_{\rm flow}$ so that $t_{\rm diff} \ll t_{\rm flow}$ in the static diffusion limit and $t_{\rm diff} \gg t_{\rm flow}$ in the dynamic diffusion limit.

To clarify the distinction between these limits, we Lorentz-transform the comoving-frame radiation energy, flux, and pressure, expressing them in the lab frame to $\mathcal{O}(\beta^2)$ to obtain for a one-dimensional flow
\begin{subequations}  \label{regimes:lorentz1}
	\begin{eqnarray}
		\mathcal{E} &=& \mathcal{E}_0 + 2\beta F_0/c + \beta^2 ( \mathcal{E}_0 + P_0 ), \label{regimes:lorentz1:energy} \\
		F/c &=& F_0/c + \beta ( \mathcal{E}_0 + P_0 ) + 2\beta^2 F_0/c , \label{regimes:lorentz1:flux} \\
		P &=& P_0 + 2\beta F_0/c + \beta^2 ( \mathcal{E}_0 + P_0 ).
	\end{eqnarray}
\end{subequations}
Recall that in the diffusion regime, $F_0 \sim c\mathcal{E}_0/\tau$, $P_0 \sim \case{1}{3} \mathcal{E}_0$, and $a_{\rm R} T^4 - \mathcal{E}_0 \sim \mathcal{E}_0/\tau^2$ in the comoving frame.  Thus, Equation~\eqref{regimes:lorentz1:energy} implies that
\begin{equation}
  a_{\rm R} T^4 - \mathcal{E} \sim \mathcal{O}\left(\frac{\mathcal{E}_0}{\tau^2}\right) + \mathcal{O}\left(\frac{\beta\mathcal{E}_0}{\tau}\right) - \frac{4}{3} \beta^2 \mathcal{E}_0, \label{regimes:diffusionenergy}
\end{equation}
to $\mathcal{O}(\beta^2)$ in this regime.  In the static diffusion limit, $\beta\tau \ll 1$ implies that $a_{\rm R} T^4 - \mathcal{E} \sim \mathcal{O}(\mathcal{E}_0 \tau^{-2})$, and in the dynamic diffusion limit, $\beta\tau \gg 1$ implies that $a_{\rm R} T^4 - \mathcal{E} \sim \mathcal{O}(\beta^2 \mathcal{E}_0)$.  Using these scaling arguments, in the static diffusion limit it follows that the terms $\rho\kappa_0(a_{\rm R} T^4 - \mathcal{E})$ in Equation~\eqref{mixedframe:graysystem:radenergy} and $\rho\kappa_0\mathbf{F}/c$ in Equation~\eqref{mixedframe:graysystem:radmomentum} are dominant over the remaining terms, which are all higher-order in their respective equations.  Furthermore, in the dynamic diffusion limit, it follows that each term in Equation~\eqref{mixedframe:graysystem:radenergy} is $\mathcal{O}(\beta^2\tau)$, and each term in Equation~\eqref{mixedframe:graysystem:radmomentum} is $\mathcal{O}(\beta\tau)$, when compared to $\mathcal{E}/L$.  Therefore, in general, all of the higher-order terms in these equations (and the corresponding terms in Equations~\eqref{mixedframe:graysystem:energy} and~\eqref{mixedframe:graysystem:momentum}) must be retained when $\beta\tau \gg 1$ or even $\beta\tau \gtrsim 1$.

\vspace{1em}
\subsection{The $M_1$ Closure Relation}  \label{m1}

The two-moment hierarchy of Equations~\eqref{mixedframe:graysystem:radenergy} and~\eqref{mixedframe:graysystem:radmomentum} can not readily be solved, since it contains moments of three orders.  To proceed, we specify a closure relation of the form
\begin{equation}
  \mathbb{P}=\mathcal{E}\,\mathbb{T}(\mathcal{E},\mathbf{F}),
\end{equation}
where the Eddington tensor $\mathbb{T}$ describes the angular dependence of the radiation pressure, and by assumption depends only on the lower-order moments $\mathcal{E}$ and $\mathbf{F}$.

The simplest choice is the $P_1$ closure relation, which is derived from an assumption that the specific intensity is isotropic in the laboratory frame, i.e., $\mathbb{T} \propto \mathbb{I}$.  This completely symmetric model is appropriate to describe the diffusion limit, but fails in the streaming limit, for example, by allowing directed radiation to leak around an obstruction instead of casting a shadow.  A better choice is the $M_1$ closure relation \citep{Levermore:1984}, which is derived by assuming the specific intensity is rotationally invariant about some preferred direction $\hat{\mathbf{n}}$, which is taken to be the direction of the radiative flux.  This implies that $\mathbb{T}$ is a linear combination of the isotropic unit tensor $\mathbb{I}$, and the directional tensor $\hat{\mathbf{n}}\hat{\mathbf{n}}$, describing a radiation field that is Dirac-distributed in the direction of $\hat{\mathbf{n}}$.

It follows from the moment definitions in Equations~\eqref{mixedframe:radmoments} that $\mathcal{E}$ and $\mathbf{F}$ must always satisfy the relation
\begin{equation}
	\| \mathbf{F} \| \le \|\hat{\mathbf{k}}\| \left| \oint I \,d\Omega \right| = c\mathcal{E}. \label{m1:fluxlimit}
\end{equation}
\cite{Levermore:1984} showed that two sufficient conditions ensuring the \emph{flux-limiting} condition of Equation~\eqref{m1:fluxlimit} is satisfied are given by
\begin{subequations} \label{m1:suffconds}
\begin{eqnarray}
	{\rm tr} \mathbb{T} &=& 1, \label{m1:cond1} \\
	\hat{\mathbf{x}}\cdot(\mathbb{T}-\mathbf{f}\mathbf{f}) \cdot \hat{\mathbf{x}} &\ge& 0, \quad \forall \, \hat{\mathbf{x}}, \label{m1:cond2}
\end{eqnarray}
\end{subequations}
where $\mathbf{f} \equiv \mathbf{F}/(c\mathcal{E})$ denotes the reduced flux.  Under the assumptions of the $M_1$ model, Equations~\eqref{m1:suffconds} imply that $\mathbb{T}$ must have the form
\begin{equation}
	\mathbb{T} = \frac{1-\chi}{2}\mathbb{I} + \frac{3\chi-1}{2}\hat{\mathbf{n}}\hat{\mathbf{n}}, \label{m1:m1tns}
\end{equation}
where
\begin{equation}
  \hat{\mathbf{n}} = \frac{\mathbf{F}}{\|\mathbf{F}\|}  \label{m1:nhat}
\end{equation}
is a unit vector in the direction of the flux and 
\begin{equation}
	\chi = \frac{1}{c\mathcal{E}} \oint [\hat{\mathbf{k}}\cdot\hat{\mathbf{n}}]^2 I \,d\Omega
\end{equation}
is the \emph{Eddington factor}.  Levermore further showed that if $I$ is isotropic in some inertial frame, i.e., that the radiation field can be described as a Lorentz-boosted, isotropic distribution in the laboratory frame, then $\chi$ is related to the norm of the \emph{reduced flux}, $f=\|\mathbf{F}\|/(c\mathcal{E})$, by the function
\begin{equation}
	\chi(f) = \frac{3+4f^2}{5+2\sqrt{4-3f^2}}.  \label{m1:chi}
\end{equation}
It can easily be verified that Equations~\eqref{m1:m1tns} and~\eqref{m1:chi} satisfy Equations~\eqref{m1:cond1} and~\eqref{m1:cond2}, hence the $M_1$ closure scheme is flux-limited.

In the diffusion limit, $\|\mathbf{F}\| \ll c\mathcal{E}$, hence $f \to 0$ and $\chi \to \case{1}{3}$.  From Equation~\eqref{m1:m1tns}, it follows that $\mathbb{T} \to \case{1}{3}\mathbb{I}$, hence this regime is described exactly.  Furthermore, in the streaming limit, $\|\mathbf{F}\| \to c\mathcal{E}$, hence $f \to 1$ and $\chi \to 1$.  From Equation~\eqref{m1:m1tns}, it follows that $\mathbb{T} \to \hat{\mathbf{n}}\hat{\mathbf{n}}$, hence this regime is also described exactly.  It follows from Equation~\eqref{m1:chi} that $f \in [0,1]$ implies $\chi \in [\case{1}{3},1]$.  It has been remarked by \cite{Sincell:1999} that certain distributions of radiation may have Eddington factors that fall outside this range, such as in the case of very high Mach number radiative shocks.  However, these distributions are not isotropic in any inertial frame, hence the $M_1$ model is only approximate in these situations anyway.  

It is important to note that the closure relation described by Equations~\eqref{m1:m1tns}, \eqref{m1:nhat}, and~\eqref{m1:chi} under the $M_1$ closure is based entirely on local data, in contrast to other schemes such as OTVET \citep{Gnedin:2001} or the solver of \cite{Davis:2012} that use non-local data to obtain an approximate local Eddington tensor.  While a local closure relation is computationally advantageous, it is also inherently limited and may not be able to accurately describe complex radiation fields.  The simplifying assumptions of the $M_1$ closure allow it to capture the behavior of radiation well in simple diffusing and streaming limits, but complex radiation field geometries may be better described using other non-local schemes.  It is known, for example, that the $M_1$ closure is subject to the two-beam instability \citep{frank:2012}, and more generally it cannot be expected to produce an accurate solution in situations where radiation from distributed sources interacts in an optically thin region, as we have verified.  Nonetheless, the $M_1$ scheme is relatively simple, is immediately parallelizable using MPI, has well-demonstrated performance \citep{Gonzalez:2007,Aubert:2008}, and has a comparatively low computational cost (see Section~\ref{benchmark}).  These features motivate the application of $M_1$ to identify the range of radiation regimes and problems where it is most advantageous.  We note that although we have adopted the $M_1$ scheme for this paper and the corresponding implementation in {\it Athena}, it is straightforward to substitute alternate approaches for obtaining an estimate for $\mathbb{P}$ (including non-local methods) to extend the range of our semi-explicit update method to applications for which $M_1$ is insufficiently accurate.  Also, the $M_1$ scheme has been adopted in methods that use fully implicit rather than semi-explicit update of the radiation moment equations \citep{Gonzalez:2007}.

Finally, note that we can simplify the application of the $\mathcal{O}(\beta\tau)$ source term $\rho\kappa_0 (\mathbf{v}/c) \cdot (\mathcal{E}\mathbb{I} + \mathbb{P})$ in Equations~\eqref{mixedframe:graysystem:momentum} and~\eqref{mixedframe:graysystem:radmomentum} by examining its behavior in the diffusion regime, i.e., in the only regime where it is non-negligible.  For static diffusion, $f \sim \tau^{-1}$ implies that $\chi = \case{1}{3} + \mathcal{O}(\tau^{-2})$.  Similarly, for dynamic diffusion, $f \sim \beta$ implies that $\chi = \case{1}{3} + \mathcal{O}(\beta^2)$.  From Equation~\eqref{m1:m1tns}, it follows that $\mathbb{T} \sim \case{1}{3}\mathbb{I}$ with off-diagonal terms of either $\mathcal{O}(\tau^{-2})$ or $\mathcal{O}(\beta^2)$, respectively, in these regimes.  When compared to $\mathcal{E}/L$ in Equations~\eqref{mixedframe:graysystem:momentum} and~\eqref{mixedframe:graysystem:radmomentum}, these off-diagonal terms are of order $\mathcal{O}(\beta\tau^{-1})$ and $\mathcal{O}(\beta^3\tau)$, respectively.  These terms can be neglected since they are not of leading-order in either regime, hence the $\mathcal{O}(\beta\tau)$ source term can be simplified as
\begin{equation}
	 \rho\kappa_0 \frac{\mathbf{v}}{c} \cdot \left( \mathcal{E}\mathbb{I} + \mathbb{P} \right) \to \frac{4}{3} \mathcal{E} \rho \kappa_0 \frac{\mathbf{v}}{c}.  \label{m1:simplifiedsource}
\end{equation}
The source term given in Equation~\eqref{m1:simplifiedsource} is much more efficient, since it does not require computing the radiation pressure tensor $\mathbb{P}$ explicitly.  Also, the source term in Equation~\eqref{m1:simplifiedsource} is related to the ``relativistic work term'' described in \cite{Krumholz:2007},\footnote{Note that in \cite{Krumholz:2007}, the analogous term appears in their radiation energy diffusion equation as a work term, whereas it appears here as a force term in our radiation momentum equation.} which is shown to be important in non-equilibrium, non-uniform dynamic diffusion systems with $\beta\tau \sim 1$.  They cite as a motivating example the structure of a radiation-dominated shock, the solution of which will contain errors within the shock itself (but neither upstream nor downstream where conditions become uniform and approach equilibrium) if this term is omitted.

{\section{Numerical Implementation}  \label{implementation}}

\vspace{1em}
\subsection{Algorithm Overview} \label{algorithm}

The system described in Equations~\eqref{mixedframe:graysystem} has the form of a nonlinear, hyperbolic conservation law plus source terms, which can be expressed compactly as
\begin{equation}
  \partial_t \mathbf{U} + \nabla \cdot \mathbb{F} = \mathbf{S},  \label{algorithm:compactform}
\end{equation}
where
\begin{equation}
  \mathbf{U} \equiv \left[ \begin{array}{c}
    \rho \\
    \rho\mathbf{v} \\
    E \\
    \mathcal{E} \\
    \mathbf{F}
  \end{array} \right], 
\end{equation}
\begin{equation}
    \mathbb{F} \equiv \left[ \begin{array}{c}
    \rho\mathbf{v} \\
    \rho\mathbf{v}\mathbf{v} + P\mathbb{I} \\
    (E+P)\mathbf{v} \\
    \hat{c}\mathbf{F}/c \\
  \hat{c}c\mathbb{P} \end{array} \right],
\end{equation}
\begin{equation}
  \mathbf{S} \equiv \left[ \begin{array}{c}
    0 \\
    -\rho\nabla\Phi + \rho\kappa_0 \mathbf{F}/c - \case{4}{3} \rho\kappa_0 (\mathbf{v}/c) \mathcal{E} \\
    -\mathbf{v} \cdot \nabla\Phi - c\rho\kappa_0(a_{\rm R} T^4 - \mathcal{E}) - c\rho\kappa_0 \mathbf{v}/c \cdot \mathbf{F}/c \\
    \hat{c}\rho\kappa_0(a_{\rm R} T^4 - \mathcal{E}) + \hat{c}\rho\kappa_0 \mathbf{v}/c \cdot \mathbf{F}/c \\
    -\hat{c}c\rho\kappa_0 \mathbf{F}/c + \hat{c}c\case{4}{3}\mathcal{E} \rho\kappa_0 (\mathbf{v}/c)
  \end{array} \right].  \label{algorithm:variables}
\end{equation}
Note that in Equation~\eqref{algorithm:variables} we use the simplified $\mathcal{O}(\beta\tau)$ source term given in Equation~\eqref{m1:simplifiedsource}.

The various source terms may cause the differential system to become stiff in certain regimes.  In this case, the numerical solution of Equations~\eqref{algorithm:compactform} may become sensitive to perturbations, hence prone to ringing \citep{Leveque:2002}.  Furthermore, the criteria for stability in explicit integration schemes may place too severe a restriction on the time step.  For these reasons, many algorithms adopt implicit integration schemes, which offer stability and larger time steps at the price of lower accuracy and higher computational cost per time step. One common approach is to use a fractional-step or operator-split method in which one alternately solves the two subproblems
\begin{subequations} \label{algorithm:splitsystem}
\begin{eqnarray}
  \partial_t \mathbf{U} + \nabla \cdot\mathbb{F} &=& \mathbf{S}_{\rm e},  \label{algorithm:explicit} \\
  \partial_t \mathbf{U} &=& \mathbf{S}_{\rm i},  \label{algorithm:implicit}
\end{eqnarray}
\end{subequations}
where $\mathbf{S}_{\rm e} \equiv \mathbf{S}_{\rm e,gas} + \mathbf{S}_{\rm e,rad}$, and
\begin{equation}
	\mathbf{S}_{\rm e,gas} \equiv \left[ \begin{array}{c}
    0 \\
    -\rho\nabla\Phi + \rho\kappa_0 \mathbf{F}/c - \case{4}{3} \rho\kappa_0 (\mathbf{v}/c) \mathcal{E} \\
    -\mathbf{v} \cdot \nabla\Phi - c\rho\kappa_0 \mathbf{v}/c \cdot \mathbf{F}/c \\
    0 \\
    0
  \end{array} \right], \label{algorithm:source:explicit:gas}
\end{equation}
\begin{equation}  
	\mathbf{S}_{\rm e,rad} \equiv \left[ \begin{array}{c}
    0 \\
    0 \\
    0 \\
    \hat{c}\rho\kappa_0 \mathbf{v}/c \cdot \mathbf{F}/c \\
    -\hat{c}c\rho\kappa_0 \mathbf{F}/c + \hat{c}c\case{4}{3}\mathcal{E}\rho\kappa_0 (\mathbf{v}/c)
  \end{array} \right], \label{algorithm:source:explicit:rad}
\end{equation}
\begin{equation}  
	\mathbf{S}_{\rm i} \equiv	\left[ \begin{array}{c}
    0 \\
    0 \\
    - c\rho\kappa_0(a_{\rm R} T^4 - \mathcal{E}) \\
    \hat{c}\rho\kappa_0(a_{\rm R} T^4 - \mathcal{E}) \\
    0
  \end{array} \right].  \label{algorithm:source:implicit}
\end{equation}
The separation of source terms into explicit (equations~\ref{algorithm:source:explicit:gas} and~\ref{algorithm:source:explicit:rad}) and implicit (equation~\ref{algorithm:source:implicit}) terms is explained in Section~\ref{source}.  Note that Equation~\eqref{algorithm:explicit} is a non-stiff subsystem of hyperbolic partial differential equations (PDE) and Equation~\eqref{algorithm:implicit} is a stiff subsystem of nonlinear ordinary differential equations (ODE).  Solution methods are discussed in Section~\ref{source}.

The splitting error of this method is formally first-order in time, regardless of the order of the method used to solve each subproblem.  Specifically, the error is proportional to the commutator bracket of the split differential operators \citep{Leveque:2002}.  For example, we demonstrate in Section~\ref{radwave} that for the simple case of the advection of a free-streaming radiation wave in a purely absorbing, homogeneous background medium, Equations~\eqref{algorithm:splitsystem} reduce to a system of constant-coefficient, linear ODE.  In this case, since neither the amount of radiation energy nor momentum absorbed by the medium depends on the location of the wave (i.e., since $\rho$ and $\kappa_0$ are held constant), we get the same result whether the wave is first advected before being absorbed or vice-versa; hence, the differential operators commute exactly, and there is no splitting error.  It is more difficult to measure the splitting error in the general case.  However, for the other test problems we have explored, the first-order splitting error seems to have such a small coefficient that the total error is dominated by that of the individual numerical methods used for each subproblem.  For this reason, we have not found it particularly advantageous to pursue higher-order fractional-step methods such as Strang splitting.

Explicit Godunov methods offer a high-order accurate, conservative, and relatively inexpensive method for solving the hyperbolic transport subproblem in Equation~\eqref{algorithm:explicit}.  However, since the gas and radiation fluids may be transported on very different time scales, it is useful to apply an additional operator splitting to the subsystems describing the gas and radiation dynamics.  In this manner, we alternately evolve the subsystem
\begin{equation}
  \partial_t \mathbf{U}_{\rm gas} + \nabla \cdot\mathbb{F}_{\rm gas} = \mathbf{S}_{\rm e,gas},  \label{algorithm:gastransport}
\end{equation}
for the hydrodynamic variables $\rho$, $\rho\mathbf{v}$, and $E$ over a time step $\Delta t_{\rm gas} \sim \Delta x/v_{\rm max}$, where $v_{\rm max}$ is the maximum signal speed for the gas variables, and the subsystem
\begin{equation}
  \partial_t \mathbf{U}_{\rm rad} + \nabla \cdot\mathbb{F}_{\rm rad} = \mathbf{S}_{\rm e,rad},  \label{algorithm:radtransport}
\end{equation}
for the radiation variables $\mathcal{E}$ and $\mathbf{F}$ over a series of time steps $\Delta t_{\rm rad} \sim \Delta x/\hat{c}$, where $\hat{c}$ is the propagation speed of the radiation variables, until both subsystems have been formally advanced to the same time.  This allows the use of a stable explicit method to advance the radiation subsystem without having to advance the hydrodynamic subsystem over unnecessarily small a time step.  Furthermore, the existing code framework of {\it Athena} is designed for a hydrodynamic subsystem such as Equation~\eqref{algorithm:gastransport}, is second-order accurate, and can handle the radiation subsystem in Equation~\eqref{algorithm:radtransport} with only slight modification.  Finally, note that a hyperbolic solution of the two-moment radiation subsystem has the desirable property that wave solutions naturally propagate at finite speeds.  However, splitting the gas and radiation subsystems means that conservation of combined energy and momentum can not be strictly maintained, since the source terms are not integrated on the same time scales.

For the stiff subproblem in Equation~\eqref{algorithm:implicit}, we must use an implicit method such as Backward Euler to ensure stability on a reasonable time scale.  If the system is nonlinear, an iterative method such as Newton-Raphson must be used.  Note that although the $\mathcal{O}(\beta\tau)$ source terms in Equations~\eqref{mixedframe:graysystem:radenergy} and~\eqref{mixedframe:graysystem:radmomentum} may become dynamically important in certain regimes, as we have demonstrated in Section~\ref{regimes}, they are typically small compared to the dominant source terms.  Therefore, we can treat these terms explicitly without adversely affecting the overall stability of the method.  We discuss in Section~\ref{source} which source terms are stiff and must be updated implicitly, and which can be updated explicitly.

An alternative approach is to drop the temporal derivative in Equation~\eqref{mixedframe:graysystem:radmomentum}, yielding $\mathbf{F} = -[c/(\rho\kappa_0)] \,\nabla \cdot \mathbb{P}$ for the $\beta\tau \ll 1$ case, which when inserted in Equation~\eqref{mixedframe:graysystem:radenergy} results in a parabolic diffusion equation for the radiation energy density.  To ensure finite-speed propagation in this approach, some form of flux-limiting must be employed.  Furthermore, any approach that introduces a spatial differential operator to the right-hand side source terms results in a numerical method containing non-local information.  To treat stiff terms implicitly, this requires an additional iterative solver such as GMRES \citep{Saad:1986} to invert a sparse matrix as well as corresponding boundary conditions.  The resulting FLD approach is currently the most common method used for RHD in astrophysics \citep{Fryxell:2000,Turner:2001,Krumholz:2007,Gittings:2008,Reynolds:2009,Swesty:2009,Commercon:2011,van-der-Holst:2011,Zhang:2011}.

The hydrodynamic time step $\Delta t_{\rm gas}$, determined using the standard Courant-Friedrichs-Lewy (CFL) condition based on the fastest signal speed, must be modified to account for the effect of radiation pressure on the propagation of acoustic waves.  \cite{Krumholz:2007} give an approximate expression for the effective sound speed,
\begin{equation}
  c_{\rm eff} \equiv \sqrt{\frac{\gamma P + \case{4}{9}\mathcal{E}(1-e^{-\rho\kappa_0\,\Delta x})}{\rho}},  \label{algorithm:ceff}
\end{equation}
which interpolates between the limit for optically thick cells, where radiation pressure contributes to the total pressure and increases the effective speed of acoustic waves, and optically thin cells where the radiation pressure does not contribute.  The hydrodynamic time step is then set to
\begin{equation}
  \Delta t_{\rm gas} = K_0 \,\frac{\Delta x}{v_{\rm max}},  \label{algorithm:gastimestep}
\end{equation}
where $K_0$ is the Courant number, usually $0.4$ for a Van Leer (VL) integration scheme, and $v_{\rm max} \equiv \max\{|\mathbf{v}| + c_{\rm eff}\}$ is the maximum effective signal speed over all grid cells.  When a reduced speed of light is used for the hyperbolic radiation subsystem, we usually set $\hat{c}$ so that
\begin{equation}
  R \equiv \frac{\hat{c}}{v_{\rm max}} \gg 1,  \label{algorithm:speedratio}
\end{equation}
with typical radiation-to-gas signal propagation speed ratio $R\sim 10$ for optically thin cases.  In the diffusion regime, there may be additional constraints on $\hat{c}$ and $R$ (see Section~\ref{rsla}).  Alternatively, in situations where $c$ is not too large compared to $v_{\rm max}$, we instead take $\hat{c}\to c$ and $R \equiv c/v_{\rm max}$.  In our code, the radiation time step is set to
\begin{equation}
  \Delta t_{\rm rad} = K_0 \,\frac{\Delta x}{\hat{c}} = \frac{\Delta t_{\rm gas}}{R},  \label{algorithm:radtimestep}
\end{equation}
hence for every gas integration cycle of time step $\Delta t_{\rm gas}$, roughly $R$ radiation integration subcycles of time step $\Delta t_{\rm rad}$ must be performed.  Note that $R$ is not fixed, since the gas time step is set by the (variable) maximum acoustic signal speed, $v_{\rm max}$, but the radiation time step is set by the (constant) reduced speed of light, $\hat{c}$. 

Our algorithm can be summarized as follows:
\begin{enumerate}
	\item Calculate the gas time step, $\Delta t_{\rm gas}$, at time $t^n$ using the radiation-modified CFL condition as described in Equation~\eqref{algorithm:gastimestep}.  Then calculate the radiation time step, $\Delta t_{\rm rad}$, using $\Delta t_{\rm gas}$ and $\hat{c}$ as described in Equation~\eqref{algorithm:radtimestep}. \label{algorithm:timesteps}
	\item Integrate the source term $\mathbf{S}_{\rm i}$ in Equation~\eqref{algorithm:implicit} over the time step $\Delta t_{\rm gas}$ using an implicit solver.\label{algorithm:radimplicit}
	\item Integrate the gas subsystem in Equation~\eqref{algorithm:gastransport} over the time step $\Delta t_{\rm gas}$ using an explicit, hyperbolic Godunov solver, adding in the source term $\mathbf{S}_{\rm e,gas}$ at first-order using the radiation variables at time $t^n$.
	\item Integrate the radiation subsystem in Equation~\eqref{algorithm:radtransport} over the time step $\Delta t_{\rm rad}$ using an explicit, hyperbolic Godunov solver, adding in the source term $\mathbf{S}_{\rm e,rad}$ at second-order.\label{algorithm:radexplicit}
	\item Repeat Step~\ref{algorithm:radexplicit} ($\approx R$ times) until the gas and radiation variables have been formally advanced to the same time, $t^{n+1}=t^n+\Delta t_{\rm gas}$.
	\item Correct the source term $\mathbf{S}_{\rm e,gas}$ to second-order in the gas subsystem in Equation~\eqref{algorithm:gastransport} using the radiation variables at time $t^{n+1}$.\label{algorithm:secondordercorrection}
	\item Repeat Steps~\ref{algorithm:timesteps} through~\ref{algorithm:secondordercorrection} until time $t_{\rm final}$ is reached.
\end{enumerate}

\vspace{1em}
\subsection{The Reduced Speed of Light Approximation} \label{rsla}

In many astrophysical settings, the ratio of the radiation propagation speed, $c$, to the maximum acoustic signal speed of the gas, $v_{\rm max} \equiv \max\{|\mathbf{v}|+c_{\rm eff}\}$, can be quite large.  Consequently, the ratio of the corresponding CFL time steps for explicit integration of the gas and radiation transport subsystems, $\Delta t_{\rm gas}/\Delta t_{\rm rad} \sim c/v_{\rm max}$, may be many orders of magnitude greater than 1.  An explicit scheme for the radiation subsystem, such as the one described in Section~\ref{algorithm}, can be rendered impractical by such a large ratio.  Fortunately, in many situations we can reduce the signal propagation speed of the radiation fluid to some value $\hat{c} \ll c$, which in turn reduces the gas-to-radiation explicit time step ratio to a computationally tractable level, while preserving the essential dynamical behavior of the RHD system.  This is the essence of the RSLA, originally described by \cite{Gnedin:2001} and recently implemented by \cite{Gonzalez:2007}, \cite{Aubert:2008}, and \cite{Petkova:2011}.

Stated more precisely, {\it local dynamics are insensitive to the RSLA as long as the relevant ordering of time scales in a given dynamical regime is preserved}.  First, consider the Newtonian (i.e., non-relativistic) limit, in which the speed of light is taken to be effectively infinite (i.e., $c \gg v_{\rm max}$).  Under the RSLA, the RHD system will remain within a first-order approximation of the Newtonian limit provided 
\begin{equation}
  \hat{c} \gg v_{\rm max}.  \label{rsla:newtoncrit}
\end{equation}
Second, consider the static diffusion limit in which the gas dynamical time scale $t_{\rm dyn} \equiv L/v_{\rm max}$, is large compared to the radiation-diffusion time scale $t_{\rm diff} \equiv L\tau/c$.  In this regime, reducing the speed of light to $\hat{c}$ corresponds to increasing the characteristic radiation-diffusion time scale to $\hat{t}_{\rm diff} \equiv L\tau/\hat{c} = (c/\hat{c}) t_{\rm diff}$.  To ensure that the original ordering of time scales is not altered under the RSLA, we must impose an effective lower limit on $\hat{c}$ so that $\hat{t}_{\rm diff} \ll t_{\rm dyn}$ whenever $t_{\rm diff} \ll t_{\rm dyn}$.  This is satisfied provided
\begin{equation}
  \hat{c} \gg v_{\rm max}\tau_{\rm max},  \label{rsla:diffcrit}
\end{equation}
where $\tau_{\rm max}$ is the maximum optical depth in a given problem.  Equations~\eqref{rsla:newtoncrit} and~\eqref{rsla:diffcrit} can be combined to form the \emph{RSLA static diffusion criterion} given by
\begin{equation}
  \hat{c} \gg v_{\rm max} \max\{1,\tau_{\rm max}\}.  \label{rsla:staticdiffreq}
\end{equation} 
It is clear that $\hat{c}$ satisfying Equation~\eqref{rsla:staticdiffreq} will be much larger than all other signal propagation speeds, and that the gas dynamical time scale will remain large compared to the radiation-diffusion time scale when $\tau_{\rm max} \gg 1$.   

In the regime which is of most practical interest for the application of our code (i.e., star-formation/ISM), static diffusion applies and we set $\hat{c}$ according to Equation~\eqref{rsla:staticdiffreq}, hence the gas-to-radiation time step ratio $R$ is given by
\begin{equation}
  R \equiv \frac{\hat{c}}{v_{\rm max}} \gg \max\{1,\tau_{\rm max}\}.  \label{rsla:speedratio}
\end{equation}
For problems in the optically thin regime, Equation~\eqref{rsla:staticdiffreq} is satisfied provided $R \gg 1$; we typically choose $R \sim 10$, corresponding to roughly 10 radiation subcycles per gas cycle.  For problems in the diffusion regime with optical depths up to $\tau_{\rm max} \sim 10$, Equation~\eqref{rsla:staticdiffreq} is satisfied for $R$ in the range $\sim$10-100.  Recall that $\hat{c}$ only enters as a factor in the time-dependent terms of the radiation Equations~\eqref{mixedframe:graysystem:radenergy} and~\eqref{mixedframe:graysystem:radmomentum}; the true speed of light $c$ is used in all source terms and in the ratio of radiation flux to energy.  One important consequence of this is that \emph{the spatial structure of quasi-steady radiation solutions is insensitive to the RSLA}.

Finally, note that since $\hat{c}$ is held constant throughout the computation, in certain situations it may be difficult to know a priori exactly what $\tau_{\rm max}$ will be.  Therefore, we can first make a conservative choice of $\hat{c}$ by assuming that $\tau_{\rm max}$ is a few times $\bar{\rho} \kappa_{0,{\rm max}} L$, where $\bar{\rho}$ is the mean density, $\kappa_{0,{\rm max}}$ is an upper-bound on what $\kappa_0$ may become, and $L$ is the size of the computational domain or other relevant spatial scale in a given problem.  Second, we can analyze the structure of the output to assess the actual value of $\tau_{\rm max}$; the value of $\hat{c}$ can then be adjusted up or down accordingly.  The first run can be done at lower resolution and the second at higher resolution to save computational costs.  In particular, in studying star formation, sink particles can be used to represent collapsed cores \citep[e.g.,][]{Gong:2013}, providing a maximum cutoff density to facilitate the selection of $\hat{c}$.

\vspace{1em}
\subsection{Hyperbolic Transport of Radiation}  \label{hyperbolic}

To evolve the transport Equation~\eqref{algorithm:radtransport}, we use the VL integrator implemented in {\it Athena} \citep{Stone:2009}, a high-order Godunov finite-volume method based on a variation of the MUSCL-Hancock scheme described by \cite{Falle:1991}.  To advance the radiation field, we use the second-order, piecewise-linear spatial reconstruction implemented in {\it Athena} along with a Harten-Lax-van Leer (HLL) Riemann solver such as the one described by \cite{Gonzalez:2007}.  

To compute the HLL flux, e.g., in the $x$-direction, we first compute the fluxes $\boldsymbol{\mathcal{F}}_{\rm R/L} = \mathbf{F}_{\rm R/L} - S_{\rm R/L} \mathbf{U}_{\rm R/L}$ along characteristics, where $\mathbf{F}_{\rm R/L} = \hat{\mathbf{x}} \cdot \mathbb{F}_{\rm R/L}$ is the flux in the $x$-direction, $\mathbf{U}_{\rm L/R}$ is the volume-averaged state vector, and $S_{\rm L/R}$ is the fastest left/right-going signal propagation speed on either side of the cell interface.  The intermediate-state flux is then given by
\begin{equation}
	\boldsymbol{\mathcal{F}}^{*} = \frac{1}{2} \left( \boldsymbol{\mathcal{F}}_{\rm L} + \boldsymbol{\mathcal{F}}_{\rm R} \right) + \frac{1}{2} \left( \mathbf{\boldsymbol{F}}_{\rm L} - \boldsymbol{\mathcal{F}}_{\rm R} \right) \left( \frac{S_{\rm R} + S_{\rm L}}{S_{\rm R} - S_{\rm L}} \right). \label{hyperbolic:hllflux1}
\end{equation}
For numerical stability, we must upwind the HLL flux whenever $S_{\rm L}$ and $S_{\rm R}$ have the same sign.  Thus, the proper HLL flux is then given by
\begin{equation}
	\mathbf{F}_{x;\,i-1/2}^{\rm HLL} = \left\{ \begin{array}{ll} \boldsymbol{\mathcal{F}}_{\rm L}, & S_{\rm L} > 0 \\ \boldsymbol{\mathcal{F}}^*, & S_{\rm L} \le 0 \le S_{\rm R} \\ \boldsymbol{\mathcal{F}}_{\rm R}, & S_{\rm R} < 0 \end{array} \right. . \label{hyperbolic:hllflux1:upwinded}
\end{equation}

Alternatively, the HLL flux in Equation~\eqref{hyperbolic:hllflux1:upwinded} can be written as
\begin{equation}
	\boldsymbol{\mathcal{F}}_{x;\,i-1/2}^{\rm HLL} = \frac{S_{\rm R}^+\mathbf{F}_{\rm L} - S_{\rm L}^-\mathbf{F}_{\rm R} + S_{\rm R}^+ S_{\rm L}^- (\mathbf{U}_{\rm R} - \mathbf{U}_{\rm L})}{S_{\rm R}^+ - S_{\rm L}^-}, \label{hyperbolic:hllflux2}
\end{equation}
where $S_{\rm R} \equiv \max\left\{\lambda^{\rm max}(\mathbf{U}_{\rm L}),\lambda^{\rm max}(\mathbf{U}_{\rm R})\right\}$ and $S_{\rm L} \equiv \min\left\{\lambda^{\rm min}(\mathbf{U}_{\rm L}),\lambda^{\rm min}(\mathbf{U}_{\rm R})\right\}$ are estimates of the fastest right- and left-moving wave speeds, respectively, of the linearized, hyperbolic radiation subsystem projected in the $x$-direction, and $S_{\rm R}^+ \equiv \max\{S_{\rm R},0\}$ and $S_{\rm L}^- \equiv \min\{S_{\rm L},0\}$ are their properly upwinded values.  In Equation~\eqref{hyperbolic:hllflux2}, the indices $({\rm L},{\rm R})$ correspond to $(i-1,i)$ for the first-order fluxes and to $(i-\case{1}{2},i-\case{1}{2})$ for the second-order fluxes.  Note that the HLL scheme uses a single intermediate state, thus it can not resolve isolated contact discontinuities.  This makes it more dissipative than schemes with additional intermediate states, i.e., schemes that track additional waves.  Nonetheless, the HLL scheme is fairly simple, and it is robust and positivity-preserving for one-dimensional problems, making it an attractive choice of Riemann solver for our method.  

The radiation transport subsystem for $\mathbf{U}_{\rm rad}$ can be written compactly as 
\begin{equation}
	\partial_t \mathbf{U}_{\rm rad} + A\, \partial_x \mathbf{U}_{\rm rad} = 0, \label{hyperbolic:linear}
\end{equation}
where $A(\mathbf{U}_{\rm rad}) = \partial\mathbf{F}/\partial\mathbf{U}_{\rm rad}$ is the $4 \times 4$ Jacobian matrix for the fluxes in the $x$-direction.  By taking $A$ constant about some state $\mathbf{U}_{\rm rad}^n$, Equation~\eqref{hyperbolic:linear} becomes a linear system.  The wave speeds $\lambda$ are the eigenvalues of $A$, which are real for a hyperbolic system.  Furthermore, by the axisymmetry assumption of the $M_1$ model, described in Section~\ref{m1}, these eigenvalues can only depend on $\hat{c}$, on the norm of the reduced flux, $f$, and on the angle $\theta$ that $\hat{\mathbf{n}} = \mathbf{f}/f$ makes with the interface normal $\hat{\mathbf{x}}$, but not on $\hat{\mathbf{n}}$ itself.  Without loss of generality, we can rotate our local coordinate system about $\hat{\mathbf{x}}$, transforming from coordinates $(x,y,z)$ to $(x,y',z')$, so that $\hat{\mathbf{z}}' \cdot \hat{\mathbf{n}} = 0$ in the new coordinate system.  Since $A$ depends only on $f_x = \mathbf{f}\cdot\hat{\mathbf{x}}$ and $f_{y'} = \mathbf{f}\cdot\hat{\mathbf{y}}'$, i.e., on $\mathcal{E}$, $F_x$, and $F_{y'}$ only, there can be at most three linearly independent eigenvectors (i.e., two of the four eigenvectors are always linearly dependent).  The three corresponding eigenvalues can be shown to be
\begin{subequations} \label{hyperbolic:evals}
\begin{eqnarray}
	\frac{\lambda_{1,3}}{\hat{c}} &=& \Bigg\{\mu f \pm \Bigg[\frac{2}{3}\left(4-3f^2-\sqrt{4-3f^2}\right)  \nonumber \\
	&& + 2\mu^2\left(2-f^2-\sqrt{4-3f^2}\right)\Bigg]^{1/2}\Bigg\} \nonumber \\
	&& \div \sqrt{4-3f^2}, \label{hyperbolic:evals13} \\
	\frac{\lambda_{2}}{\hat{c}} &=& \frac{\mu}{f}\left(2-\sqrt{4-3f^2}\right), \label{hyperbolic:evals2} 
\end{eqnarray}
\end{subequations}
where $\mu \equiv \cos\theta = \hat{\mathbf{x}}\cdot\hat{\mathbf{n}}$, and where $\lambda_1$ and $\lambda_3$ correspond to the $(-)$ and $(+)$ roots, respectively.  It can be shown that the three eigenvalues given in Equations~\eqref{hyperbolic:evals} are always ordered $\lambda_1 \le \lambda_2 \le \lambda_3$.  These eigenvalues are the closed-form, multidimensional analogs of the wave speeds given explicitly by \citet[][equations~35a,b]{Audit:2002} for a one-dimensional flow ($\mu = 1$).

\begin{figure}
  \centering
  \epsscale{1}
  \plotone{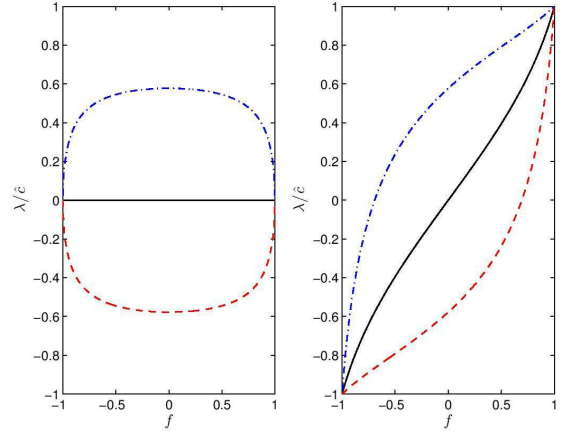}
  \caption{Eigenvalues of the hyperbolic radiation wave matrix, scaled by $\hat{c}$, as a function of the norm of the reduced flux, $f \equiv |\mathbf{F}|/(c\mathcal{E})$, for the $M_1$ closure relation in the cases of perpendicular transport ($\mu=0$, left) and parallel transport ($\mu=1$, right).  The dashed and dash-dotted lines show the wave speeds $\lambda_1$ and $\lambda_3$ (Equation~\eqref{hyperbolic:evals13}), respectively, and the solid lines show the wave speed $\lambda_2$ (Equation~\eqref{hyperbolic:evals2}), which is unused in our implementation.  \label{hyperbolic:evalsfig}}
\end{figure}

Equation~\eqref{hyperbolic:linear} is hyperbolic, but not \emph{strictly} so, hence its eigenvalues are not necessarily distinct.  In the streaming limit, it follows from Equations~\eqref{hyperbolic:evals} with $f \to 1$ that $\lambda_{1,2,3} \to \hat{c}\mu$, so that when $\hat{\mathbf{n}}$ and $\hat{\mathbf{x}}$ are parallel, the fastest signal speed is given by the reduced speed of light, $\hat{c}$, and when $\hat{\mathbf{n}}$ and $\hat{\mathbf{x}}$ are perpendicular, there is zero transport in the $x$-direction.  In the diffusion limit, it follows from Equations~\eqref{hyperbolic:evals} with $f \to 0$ that $\lambda_{2} \to 0$ and $\lambda_{1,3} \to \mp \hat{c}/\sqrt{3}$, so that we recover the fastest signal speeds given by diffusion theory.  

Figure~\ref{hyperbolic:evalsfig} shows the dependence of the eigenvalues $\lambda_{1,2,3}$ on the norm of the reduced flux, $f$, for the cases of parallel ($\mu=1$) and perpendicular ($\mu=0$) transport in a given direction.  As emphasized by \cite{Gonzalez:2007}, the proper dependence of the eigenvalues on $\mu$ in the streaming limit is necessary for capturing shadowing.  Note that the eigenvalue $\lambda_{2}$ represents the intermediate wave speed of an entropy mode while the eigenvalues $\lambda_{1,3}$ represent the speeds of the fastest left- and right-moving radiation waves.  All waves become degenerate as $f \to 1$, which physically represents the fact that all photons propagate in the same direction in the streaming limit.  Since only the fastest left- and right-moving wave speeds are needed to compute the HLL flux in Equation~\eqref{hyperbolic:hllflux2}, we only need to compute $\lambda_{1,3}$.

\vspace{1em}
\subsection{Treatment of Source Terms}  \label{source}

As mentioned in Section~\ref{algorithm}, some of the radiation source terms in Equations~\eqref{algorithm:variables} must be handled carefully in certain physical regimes where they may become stiff.  In this case, stability requirements may become too restrictive on the time step for explicit methods to remain feasible and one must resort to lower-order implicit methods.  Yet in other regimes, the stability requirements can often be relaxed or even neglected.  

In our treatment of the $\mathcal{O}(\beta\tau)$ source terms, we assume they are never stiff, i.e., that we are confined to the static diffusion regime with $\beta\tau \ll 1$ as described in Section~\ref{regimes}.  Thus, for the update of the radiation subsystem from Equation~\eqref{algorithm:variables}, we first consider only the (potentially) stiff source terms
\begin{subequations} \label{source:stiff}
	\begin{eqnarray}
	  \partial_t \mathcal{E} &=& \hat{c}\rho\kappa_0(a_{\rm R} T^4 - \mathcal{E}),  \label{source:stiff:radenergy} \\
	  \partial_t \mathbf{F} &=& -\hat{c}\rho\kappa_0\mathbf{F}. \label{source:stiff:radmomentum}
	\end{eqnarray}
\end{subequations}

Equation~\eqref{source:stiff:radmomentum} represents the process of radiative momentum absorption by the gas, which does not directly affect the gas density, $\rho$.  Thus, by taking $\rho$ constant over the radiation time step, $\Delta t_{\rm rad}$, Equation~\eqref{source:stiff:radmomentum} can be solved using a standard $\theta$-scheme update given by
\begin{equation}
	\mathbf{F}^{n+1} =  \mathbf{F}^{n} \left[ \frac{1 - (1-\theta)\hat{c}\rho\kappa_0\Delta t_{\rm rad}}{1 + \theta\hat{c}\rho\kappa_0\Delta t_{\rm rad}} \right].  \label{source:stiff:radmomentum:soln}
\end{equation}
Equation~\eqref{source:stiff:radmomentum:soln} represents the unconditionally stable, first-order Backward Euler Method for $\theta = 1$, and the marginally stable, second-order Trapezoidal Method for $\theta = \case{1}{2}$.  In most cases, we can set $\theta = 0.51$ to achieve nearly second-order accuracy while avoiding the ringing associated with the completely time-centered Trapezoidal Method.  In cases where Equation~\eqref{source:stiff:radmomentum} may become stiff, stability of the update demands that we use $\theta = 1$; however, the solution is always direct rather than iterative since the equation is linear in $\mathbf{F}$.  With this caveat regarding the choice of $\theta$, we categorize Equation~\eqref{source:stiff:radmomentum} as ``non-stiff'' and include the corresponding source term in Equation~\eqref{algorithm:source:explicit:rad}.

Furthermore, Equation~\eqref{source:stiff:radenergy} represents the exchange of the radiation and gas energies via absorption and emission of radiation.  Since $\rho$ is also unaffected by gas-radiation energy exchange, Equation~\eqref{source:stiff:radenergy} represents a nonlinear ODE in the two scalar variables $\mathcal{E}$ and $T$, which can be solved using standard iterative methods.

We can further reduce Equation~\eqref{source:stiff:radenergy} to a single-variable, nonlinear ODE as follows.  First, we use Equation~\eqref{momenteqns:perfectgas} to relate $T$ to the gas internal energy, $e$.  We do this for both the radiation energy update in Equation~\eqref{source:stiff:radenergy} and the corresponding gas internal energy update in Equation~\eqref{algorithm:implicit} (from equation~\ref{mixedframe:graysystem:energy}) to obtain the system
\begin{subequations} \label{source:energyexchange}
\begin{eqnarray} 
  \partial_t e &=& -c\rho\kappa_0 (\alpha e^4 - \mathcal{E}), \label{source:energyexchange:gas} \\
  \partial_t \mathcal{E} &=& \hat{c}\rho\kappa_0 (\alpha e^4 - \mathcal{E}), \label{source:energyexchange:rad}
\end{eqnarray}
\end{subequations}
where $\alpha \equiv a_{\rm R}[(\gamma-1)\mu/(\rho \,k_{\rm B})]^4$ is constant over the energy exchange update.  Note that we write Equation~\eqref{source:energyexchange:gas} as an update to the gas internal energy only; the gas kinetic energy is not directly affected by the processes of absorption and emission of radiation.  Second, by adding Equations~\eqref{source:energyexchange:gas} and $c/\hat{c}$ times Equation~\eqref{source:energyexchange:rad}, it follows that the quantity
\begin{equation}
	\bar{E} = e + \frac{c}{\hat{c}}\mathcal{E},  \label{source:combinedenergy}
\end{equation}
is constant over the energy exchange update.  We can then eliminate $e$ in Equation~\eqref{source:energyexchange:rad} using Equation~\eqref{source:combinedenergy} to obtain
\begin{equation}
	\partial_t \mathcal{E} = \hat{c}\rho\kappa_0 \left[ \alpha \left( \bar{E} - \frac{c}{\hat{c}}\mathcal{E} \right)^4 - \mathcal{E} \right], \label{source:energyexchange:combined}
\end{equation}
a nonlinear ODE in the single variable $\mathcal{E}$.

In certain physical regimes where Equation~\eqref{source:energyexchange:combined} may become stiff, we must resort to implicit solution methods to provide stable solutions on the larger time scale of the gas.  We use a standard $\theta$-scheme update given by
\begin{eqnarray}
  \frac{\mathcal{E}^{n+1} - \mathcal{E}^n}{\Delta t_{\rm rad}} = \hat{c}\rho\kappa_0 \Bigg\{ \theta \left[ \alpha \left( \bar{E} - \frac{c}{\hat{c}}\mathcal{E}^{n+1} \right)^4 - \mathcal{E}^{n+1} \right] \nonumber \\
  + (1-\theta) \left[ \alpha \left( \bar{E} - \frac{c}{\hat{c}}\mathcal{E}^n \right)^4 - \mathcal{E}^n \right] \Bigg\},  \label{source:energyupdate}
\end{eqnarray}
which reduces to the Backward Euler Method for $\theta = 1$, and to the Trapezoidal Method for $\theta = 1/2$.

It follows from Equation~\eqref{source:combinedenergy} that $\mathcal{E}^{n+1}$ is related to $e^{n+1}$ via
\begin{equation}
	\mathcal{E}^{n+1} = \frac{\hat{c}}{c}(\bar{E} - e^{n+1}). \label{source:eradupdate}
\end{equation}
Equation~\eqref{source:eradupdate} can be substituted for $\mathcal{E}^{n+1}$ (along with an analogous expression for $\mathcal{E}^n$) on the right-hand side of Equation~\eqref{source:energyupdate}, to obtain the new left-hand side $-(\hat{c}/c)[e^{n+1}-e^n]/\Delta t_{\rm rad}$.  Thus, Equation~\eqref{source:energyupdate} reduces to a fourth-order polynomial equation in the single variable $x \equiv e^{n+1}$, the solution of which can be found using standard root-finding methods \citep{Turner:2001}.  It can be shown that this polynomial equation has the form $f_\theta(x) = 0$, where
\begin{subequations}  \label{source:polynomial} 
\begin{eqnarray}
  f_\theta(x) &\equiv& c_4 x^4 + c_1 x + c_0,  \\
  c_4 &\equiv& \alpha\eta\theta, \\
	c_1 &\equiv& 1 + \frac{\hat{c}}{c}\eta\theta, \\
	c_0 &\equiv& (1-\theta)\eta \left[ \alpha (e^n)^4 - \mathcal{E}^n \right] \nonumber \\
	&&- \left( e^n + \frac{\hat{c}}{c}\eta\theta\bar{E} \right), \\
	\eta &\equiv& c \rho \kappa_0 \Delta t_{\rm rad}.
\end{eqnarray}
\end{subequations}
Since $x > 0$, $c_4 > 0$, and $c_1 > 0$, it follows immediately that $f_\theta$ is strictly increasing and convex, and it can be further shown that $f_\theta$ is bracketed on some feasible domain between $0$ and $\min\{|c_0/c_1|,|c_0/c_4|^{1/4}\}$, provided $c_0 \le 0$.  This is guaranteed for $\theta = 1$ or for a system initially in radiative equilibrium, i.e., one for which $\alpha(e^n)^4 = a_{\rm R} (T^n)^4 = \mathcal{E}^n$.  For $\theta < 1$, $f_\theta$ may fail to be bracketed on a feasible domain if Equation~\eqref{source:energyexchange:combined} is stiff, in which case there may exist no solution to Equation~\eqref{source:polynomial}.  By default, we use the unconditionally stable value $\theta=1$, although in most of our code tests we are able to use the value $\theta = 0.51$ to achieve higher-order accuracy without introducing instability (see Section~\ref{tests}).

When a solution to Equation~\eqref{source:polynomial} does exist, Newton-Raphson iteration can be used to solve for the root, typically with rapid convergence.  If that fails, we can resort to the Bisection Method, which is slower but guaranteed to converge.  Once either method has converged to the root $x=e^{n+1}$, within a relative error tolerance of $\epsilon$, the update for $\mathcal{E}^{n+1}$ is completed by applying Equation~\eqref{source:eradupdate}.  By default, we use the value $\epsilon = 10^{-10}$.  It can be shown that the relative error of the solution for $\mathcal{E}^{n+1}$ is approximately $K\epsilon$, where
\begin{equation}
	K \equiv \frac{e^{n+1}}{(c/\hat{c})\mathcal{E}^{n+1}} \label{source:kappacond}
\end{equation}
is the condition number of the update for $\mathcal{E}^{n+1}$ via Equation~\eqref{source:eradupdate}.  On the one hand, if $K \gg 1$ for a relatively weak but non-negligible radiation field, then there may be a significant loss of numerical precision of the solution for $\mathcal{E}^{n+1}$ upon application of Equation~\eqref{source:eradupdate}, even if the relative error of the solution for $e^{n+1}$ is small.  In this case, it may be preferable to estimate $K \approx e^n/[(c/\hat{c})\mathcal{E}^n]$ a priori, and preemptively reduce $\epsilon$, the relative error tolerance for the solution of $e^{n+1}$, so that $\epsilon$ and $K\epsilon$ yield acceptable levels of relative error of the solutions for $e^{n+1}$ and $\mathcal{E}^{n+1}$, respectively.  Note that this affects the precision of the implicit energy exchange update but has no effect on $\bar{E}$, which by construction is conserved to the level of machine precision.  On the other hand, if $K \gg 1$ for a \emph{negligible} radiation field, the update for $\mathcal{E}^{n+1}$ may be ill-conditioned, but the relative error of the solution for $e^{n+1}$ will be at the level of $\epsilon \ll 1$.  Our algorithm is designed to track a radiation field that is at least weakly coupled to the gas; in the purely uncoupled limit, including the purely hydrodynamic limit, one can not reasonably expect to resolve precisely the dynamics of such an extremely weak radiation field independently of the gas.  

The above describes the most general approach to updating the terms in Equations~\eqref{source:energyexchange}, which we categorize as ``stiff'' for the purposes of Equation~\eqref{algorithm:implicit}.  For certain cases, we instead adopt a different approach.  In the case of a purely absorbing medium with no (effective) emission, e.g., absorption of UV or optical radiation by dust (which would be re-emitted in the infrared), we neglect source terms for the gas energy equation and Equation~\eqref{source:stiff:radenergy} reduces to
\begin{equation}
	\partial_t \mathcal{E} = -\hat{c}\rho\kappa_0 \mathcal{E}.  \label{source:noemission}
\end{equation}
As before with Equation~\eqref{source:stiff:radmomentum}, Equation~\eqref{source:noemission} can be solved via the $\theta$-scheme given by
\begin{equation}
	\mathcal{E}^{n+1} =  \mathcal{E}^{n} \left[ \frac{1 - (1-\theta)\hat{c}\rho\kappa_0\Delta t_{\rm rad}}{1 + \theta\hat{c}\rho\kappa_0\Delta t_{\rm rad}} \right],  \label{source:stiff:radenergy:soln}
\end{equation}
which we use in lieu of the implicit solution of Equation~\eqref{source:energyexchange:combined} described above.  Note that the implicit energy exchange update (equations~\ref{source:energyexchange}) is always computed on the gas time step, $\Delta t_{\rm gas}$, whereas the alternative absorption-only update (equation~\ref{source:noemission}) is computed on the radiation time step, $\Delta t_{\rm rad}$.  A second special case important for applications involving the interaction of IR with the dusty ISM is the condition of \emph{radiative equilibrium}.  In this case, Equation~\eqref{source:stiff:radenergy} is omitted altogether, and the corresponding energy exchange term for the gas is also omitted (i.e, the right-hand sides of both equations~\ref{source:energyexchange:gas} and~\ref{source:energyexchange:rad} are zero).

Next, we consider the $\mathcal{O}(\beta\tau)$ non-stiff source terms from Equation~\eqref{algorithm:source:explicit:rad} in the update of the radiation subsystem in Equation~\eqref{algorithm:radtransport}.  We add in these contributions explicitly without regard to stability since they are only dominant in the dynamic diffusion regime.  Recall that we use the VL unsplit integrator to advance the radiation state $\mathbf{U}_{\rm rad}^n$ from time $t^n$ through $R$ subcycles to time $t^{n+1} = t^n + \Delta t_{\rm gas}$, while holding the gas state $\mathbf{U}_{\rm gas}^n$ fixed.  For each subcycle, we advance the radiation state $\mathbf{U}_{\rm rad}^m$ from time $t^m$ to time $t^{m+1} = t^m + \Delta t_{\rm rad}$, where the index $m=0,\ldots,R$ runs over the $R$ radiation subcycles so that $m=0$ corresponds to time $t^n$ and $m=R$ corresponds to time $t^{n+1}$.  In each radiation subcycle, all of the non-stiff radiation source terms in Equation~\eqref{algorithm:source:explicit:rad} are computed twice as described in \cite{Stone:2009}:  at first-order during the prediction step using the radiation state $\mathbf{U}_{\rm rad}^{m}$, and then again at second-order during the correction step using the radiation state $\mathbf{U}_{\rm rad}^{m+1/2}$ advanced to the half-time step $t^{m+1/2} = t^m + \case{1}{2} \Delta t_{\rm rad}$.  Note that since the gas state $\mathbf{U}_{\rm gas}^n$ remains constant throughout the radiation subcycles, there is a first-order splitting error of $\mathcal{O}(\Delta t_{\rm gas})$.  Nevertheless, we include the source term contributions on the smaller time step $\Delta t_{\rm rad}$ to improve the code's ability to approach a quasi-steady radiation state when the $\mathcal{O}(\beta\tau)$ terms become significant.
       
Finally, we consider the non-stiff source term update of the gas subsystem in Equation~\eqref{algorithm:gastransport}.  As for the radiation subsystem, the gas subsystem is advanced using an unsplit integrator:  either the VL integrator described in \cite{Stone:2009} or the corner transport upwind (CTU) integrator described in \cite{Gardiner:2008}.  During the prediction step, the non-stiff source terms are added explicitly at first-order using the gas state $\mathbf{U}_{\rm gas}^n$ and the radiation state $\mathbf{U}_{\rm rad}^n$ at time $t^n$.  However, during the correction step, the source terms are added explicitly at first-order again, this time using the gas state $\mathbf{U}_{\rm gas}^{n+1/2}$ advanced to the half-time step $t^{n+1/2} = t^n + \case{1}{2} \Delta t_{\rm gas}$ and the unadvanced radiation state $\mathbf{U}_{\rm rad}^n$, which is still at time $t^n$.  At this point, the first-order gas state $\mathbf{U}_{\rm gas}^{n+1}$, which is now held fixed, is used during the radiation subcycles to advance the radiation state from time $t^n$ to time $t^{n+1}$ in an operator-split manner as described above.  Finally, the gas state $\mathbf{U}_{\rm gas}^{n+1}$ is corrected to second-order using the advanced radiation state $\mathbf{U}_{\rm rad}^{n+1}$ via the update
\begin{equation}
	\frac{\Delta t_{\rm gas}}{2} \left[ \mathbf{S}_{\rm e,gas}(\mathbf{U}_{\rm gas}^{n+1/2},\mathbf{U}_{\rm rad}^{n+1}) - \mathbf{S}_{\rm e,gas}(\mathbf{U}_{\rm gas}^{n+1/2},\mathbf{U}_{\rm rad}^n) \right].
\end{equation}
The net result is that the non-stiff source terms in the gas subsystem are time-centered in all variables.

To summarize, except for the gas-radiation energy exchange term that is updated implicitly (or explicitly if emission is neglected, or not at all if radiative equilibrium is assumed), all other source terms are applied via direct (i.e., non-iterative) updates.  Because certain source terms are important only in particular regimes, we have implemented code switches so that source terms (e.g., the $\mathcal{O}(\beta\tau)$ terms) can be turned on or off.

{\section{Code Verification Tests}  \label{tests}}

We now present a suite of tests designed to verify the methods used in our code.  These tests are presented in increasing order of the amount of physics and code features they exercise.  Where tests are borrowed from other authors, we try to preserve their overall character as much as possible, adhering to published parameter sets, to initial and boundary conditions, and to grid resolutions, within the confines of our particular algorithm, in order to provide a standard basis of comparison with other published methods.

Note that in the non-dimensionalization process, we frequently make use of the parameter $\mathcal{P}_0 \equiv a_{\rm R} T_0^4/(\rho_0 a_0^2)$ \citep{Lowrie:2001}, where $T_0$, $\rho_0$, and $a_0$ are characteristic values of the gas temperature, density, and sound speed.  We refer to $\mathcal{P}_0$ here as the dimensionless, radiation-to-gas \emph{pressure ratio}, although more accurately, $\mathcal{P}_0$ is proportional to this ratio or to the ratio of radiation-to-gas energies.  Unless otherwise specified, all tests use the value $\theta=0.51$ in the energy exchange update of Equation~\eqref{source:energyupdate}.  Except as noted for specific tests, we omit the $\mathcal{O}(\beta\tau)$ radiation terms.

\vspace{1em}
{\subsection{Tests of the Radiation Subsystem}  \label{radsubsystem}}

In these first tests, the gas is held motionless and only the radiation subsystem is evolved.  The evolution of the radiation subsystem will include an exchange of energies via the solution of Equations~\eqref{source:stiff} if emission terms are included, but there will be no change to gas density or momentum.  Since the overall time step is set only by the CFL condition for radiation, which does not depend on gas velocity or sound speed, we can take $\hat{c} \to c$ in these tests.

\vspace{1em}
{\subsubsection{Shadowing by a Dense Cloud}  \label{radshadow}}

As argued by \cite{Hayes:2003}, the ability to reproduce and preserve strong angular variations in the radiation field is an important feature of an RHD method.  To that end, we begin by reproducing their test of shadowing by a dense cloud.  This two-dimensional test consists of a domain of length $L_x = 1.0 \text{ cm}$ and height $L_y = 0.12 \text{ cm}$ filled with gas at an ambient density of $\rho_0 = 10^{-3} \text{ g cm}^{-3}$ in which is placed an ellipsoidal cloud of density $\rho_1 = 1.0 \text{ g cm}^{-3}$ with density structure given by
\begin{equation}
	\rho_{\rm cloud} (x,y) = \rho_0 + \frac{\rho_1 - \rho_0}{1 + e^\Delta},  \label{radshadow:rho}
\end{equation}
where
\begin{equation}
	\Delta \equiv 10 \left[ \left(\frac{x-x_c}{x_0}\right)^2 + \left(\frac{y-y_c}{y_0}\right)^2 - 1 \right].  \label{radshadow:delta}
\end{equation}
Equations~\eqref{radshadow:rho} and~\eqref{radshadow:delta} describe a cloud with a thin, ``fuzzy'' surface instead of one whose density transitions instantaneously from $\rho_1$ to $\rho_0$.  The cloud is centered at $(x_c,y_c) = (0.5,0)$ with major and minor axes given by $x_0 = 0.10$ and $y_0 = 0.06$, respectively.

The system is initially in radiative equilibrium with temperature $T_{\rm gas} = T_{\rm rad} = T_0 = 290 \text{ K}$, where $T_{\rm gas}$ and $T_{\rm rad}$ are the gas and radiation temperatures, respectively.  At time $t=0$, a uniform source with temperature $T_1 = 6T_0 = 1740 \text{ K}$ directed toward the right illuminates the left boundary.  We use the specific absorption opacity
\begin{equation}
	\kappa(T_{\rm gas},\rho) = \kappa_0 \left(\frac{T_{\rm gas}}{T_0}\right)^{-3.5} \left(\frac{\rho}{\rho_0}\right),
\end{equation}
with $\kappa_0 = 100 \text{ cm}^2 \text{ g}^{-1}$, which gives a nearly transparent ambient medium and a highly opaque cloud.  We use a Dirichlet boundary condition on the left, an outflow boundary condition on the right and top, and reflecting boundary conditions on the bottom.  We use a grid resolution of $N_x \times N_y = 280 \times 80$ and evolve the system for $10$ horizontal light-crossing times.  Finally, we measure the temperature at the right boundary.

\begin{figure}
  \centering
  \epsscale{1}
  \plotone{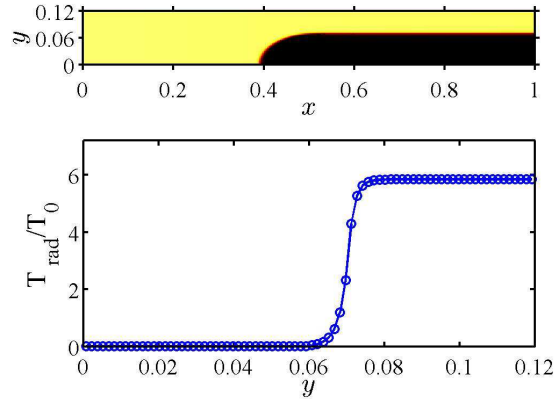}
  \caption{Top:  pseudocolor image indicating radiation temperature $T_{\rm rad}$ in units of $T_0$ after 10 light-crossing times, with red high and blue low, for the shadow test.  Bottom:  radiation temperature profile measured at the far right boundary.  Emission is neglected, yielding a very sharply defined shadow behind the cloud.  The width of the transition corresponds to the transition from optically thick to thin conditions at the surface of the cloud.  \label{radshadow:noemission}}
\end{figure}

In our first version of the test, we neglect the energy emission terms, solving Equation~\eqref{source:noemission} for the radiation source term update.  As seen in Figure~\ref{radshadow:noemission}, this yields a well-defined shadow with a very sharp radiation temperature profile behind the cloud, demonstrating the code's ability to maintain sharp angular features a good distance behind the target.  The characteristic width of the radiation temperature gradient is consistent with the width of the transition from optically thick to -thin conditions at the surface of the cloud.

\begin{figure}
  \centering
  \epsscale{1}
  \plotone{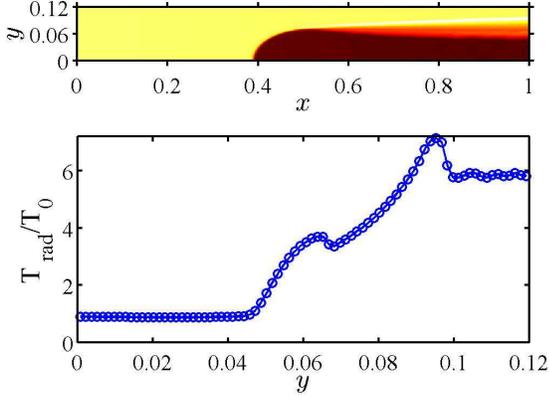}
  \caption{Same as Figure~\ref{radshadow:noemission}, but with thermal emission terms included.  The angular resolution of the shadow is not as sharp due to increased numerical diffusion caused by the operator-split implicit solver.  \label{radshadow:yesemission}}
\end{figure}

In our second version of the test, we add in the thermal emission terms (now solving Equations~\eqref{source:stiff} for the source term update), obtaining the somewhat less sharp radiation temperature profile shown in Figure~\ref{radshadow:yesemission}.  Although the angular resolution is not as sharp now due to increased numerical diffusion caused by the operator-split implicit solver, the shadow is still fairly well-preserved a distance behind the target.  

As noted by \cite{Hayes:2003} and \cite{Gonzalez:2007}, methods that are not angularly well-resolved will fail to preserve a shadow in this test.  For example, FLD fails immediately since the radiation pressure tensor in the diffusion approximation, $\mathbb{P} = \case{1}{3}\mathcal{E}\mathbb{I}$, is inherently isotropic, allowing radiation to ``leak'' around the back of the cloud.  Our solution is comparable to that obtained by \cite{Gonzalez:2007} using the $M_1$ closure relation.

\vspace{1em}
{\subsubsection{Radiation Wave Propagation}  \label{radwave}}

As a simple test of hyperbolic transport of the radiation subsystem, we investigate the propagation of small-amplitude, free-streaming radiation waves in a purely absorbing, homogeneous medium with low optical depth.  This test is similar to the two-dimensional hydrodynamic linear wave propagation test described in \cite{Gardiner:2005} and its three-dimensional analog described in \cite{Gardiner:2008}.  Ignoring the hydrodynamic equations and emission terms, the radiation subsystem reduces to
\begin{subequations}  \label{radwave:reducedsystem}
\begin{eqnarray}
  \frac{1}{\hat{c}} \partial_t \mathcal{E} + \nabla \cdot \left(\frac{\mathbf{F}}{c}\right) &=& -\rho\kappa_0\mathcal{E}, \\
  \frac{1}{\hat{c}} \partial_t \left(\frac{\mathbf{F}}{c}\right) + \nabla \cdot \mathbb{P} &=& -\rho\kappa_0 \frac{\mathbf{F}}{c}.
\end{eqnarray}
\end{subequations}
with $\mathbb{P} = \mathcal{E} \hat{\mathbf{n}}\hat{\mathbf{n}}$ in the streaming limit.  We consider temporally damped, plane-wave solutions of the form $e^{i(\mathbf{k}\cdot\mathbf{x} - \omega t)}$ with $\mathbf{k} \in \mathbb{R}^3$ and $\omega \in \mathbb{C}$, which leads to the dispersion relation
\begin{equation}
  \omega = \pm \hat{c}k - i\hat{c}\rho\kappa_0.
\end{equation}
Thus, the solutions to Equations~\eqref{radwave:reducedsystem} consist of weakly damped, linear radiation waves propagating with a phase speed equal to $\hat{c}$ and a damping rate equal to $\hat{c}\rho\kappa_0$.

It is convenient to describe the initial wave state vector in the rotated coordinates $(x',y',z')$, which are chosen such that the wave propagates in the $x'$-direction.  These coordinates are related to the grid coordinates $(x,y,z)$ by the transformation
\begin{subequations}  \label{radwave:transformation}
 \begin{eqnarray}
   x' &=& x \,\cos\alpha\,\cos\beta + y \,\cos\alpha\,\sin\beta + z \,\sin\alpha, \\
   y' &=& -x \,\sin\beta + y \,\cos\beta, \\
   z' &=& -x \,\sin\alpha\,\cos\beta - y \,\sin\alpha\,\sin\beta + z \,\cos\alpha,
 \end{eqnarray}
\end{subequations}
where the angle $\beta$ measures the inclination of the wave vector in the $xy$-plane with respect to the $x$-axis, and the angle $\alpha$ measures the inclination of the wave vector above the $xy$-plane.  We set the initial state vector to
\begin{equation}
  \mathbf{U} = \bar{\mathbf{U}}_{\rm rad} + \varepsilon \sin\left(\frac{2\pi x'}{\lambda}\right), \label{radwave:initialstate}
\end{equation}
where $\bar{\mathbf{U}}_{\rm rad}$ is the mean background state, $\varepsilon \ll 1$ is the wave amplitude, and $\lambda$ is the wavelength.  We allow the wave to propagate a distance of one wavelength in a time equal to one wave period, $t_\lambda \equiv \lambda/\hat{c}$, and then we compare the result to the analytic solution $\mathbf{U}^*(x',t_\lambda) = \bar{\mathbf{U}}_{\rm rad} + \varepsilon e^{-\rho\kappa_0 \lambda}\sin(2\pi x'/\lambda)$.

For the one-dimensional version of this test, we use a domain of size $L$ with a grid of resolution $N$.  The wave propagates along the $x$-axis ($\alpha=\beta=0$), and we set $L=\lambda$ so that there is one complete wave period in the $x$-direction.  For the two-dimensional version, we use a domain of size $2L \times L$ with a grid of resolution $2N \times N$.  The wave is inclined at an angle $\beta=\tan^{-1}(2) \approx 63\fdg4$ with respect to the $x$-axis and lies in the $xy$-plane ($\alpha=0$).  We set $L=(\sqrt{5}/2)\lambda$ so that there is one complete wave period in each of the $x$- and $y$-directions.  For the three-dimensional version, we use a domain of size $2L \times L \times L$ with a grid resolution of $2N \times N \times N$.  The wave is inclined at an angle $\beta=\tan^{-1}(2)\approx 63\fdg4$ with respect to the $x$-axis and is inclined at an angle of $\alpha=\tan^{-1}(2/\sqrt{5}) \approx 41\fdg8$ with respect to the $xy$-plane.  We set $L=\case{3}{2}\lambda$ so that there is one complete wave period in each coordinate direction.  For each test, we use a wave amplitude of $\varepsilon = 10^{-6}$, an optical depth per wavelength of $\tau_\lambda \equiv \rho_0 \kappa_0 \lambda = 0.1$, and periodic boundary conditions everywhere.

\begin{figure}
  \centering
  \epsscale{1}
  \plotone{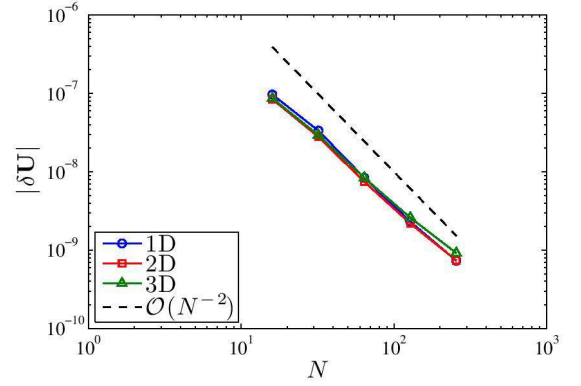}
  \caption{Convergence of $|\delta\mathbf{U}|$ for various levels of discretization of the radiation wave propagation test in one, two, and three dimensions.  For reference, we plot a line of slope $-2$ (dashed) to show that the convergence is second-order in $1/N$. \label{radwave:convergence}}
\end{figure}

The $L_1$-error vector for the $d$-dimensional solution at time $t=t_\lambda$ is defined as
\begin{equation}
  \delta \mathbf{U} \equiv \frac{1}{2N^d} \sum_i |\mathbf{U}_i - \mathbf{U}_i^*|,
\end{equation}
where $\mathbf{U}_i$ and $\mathbf{U}_i^*$ are the computed and analytic solutions, respectively, and the summation runs over all zones.  Figure~\ref{radwave:convergence} shows a plot of $|\delta\mathbf{U}|$ for various values of $N$ and for the one-, two-, and three-dimensional tests.  Since there is no emission, the source term calculation is exact, hence, we observe a nearly second-order convergence rate as expected from the second-order integration method (Section~\ref{algorithm}).

\vspace{1em}
{\subsubsection{Non-equilibrium Marshak Wave}  \label{radmarshak}}

In this problem, we investigate non-equilibrium diffusion of radiation in a cold, homogeneous, absorbing medium occupying the right half-plane, $0 \le x < \infty$.  This time-dependent diffusion problem is originally described by \cite{Marshak:1958} and a semi-analytic solution is given by \cite{Su:1996}.  As for the previous two tests, the gas density is fixed and the gas velocity is zero, thus we take $\hat{c} \to c$.  The gas temperature and radiation energy density are also zero initially.

At time $t=0$, a constant flux, $F_{\rm inc}\hat{x}$, impinges upon the boundary at $x=0$ and a radiation wave diffuses into the medium.  Exchange between radiative and thermal energies (or equivalently, temperatures) is given by the equations
\begin{subequations} \label{marshak:energyexchange}
\begin{eqnarray}
  c_v \partial_t T &=& -c\rho\kappa_0(a_R T^4 - \mathcal{E}), \label{marshak:energyexchange:T} \\
  \partial_t \mathcal{E} + \partial_x F &=& c\rho\kappa_0(a_R T^4 - \mathcal{E}), \label{marshak:energyexchange:Er}
\end{eqnarray}
\end{subequations}
where $c_v \equiv \partial e/\partial T$ is the constant-volume heat capacity of the gas, $e$ is the gas internal energy, and $T$ is the gas temperature.  Equation~\eqref{marshak:energyexchange:T} replaces the material energy equation of Equation~\eqref{mixedframe:graysystem:energy} in this problem.  Two simplifications to Marshak's original description due to \cite{Pomraning:1979} are to assume that the specific absorption coefficient $\kappa_0$, is independent of $T$ and that $c_v = \alpha T^3$ for some constant $\alpha$ so that the thermal emission depends linearly on the internal energy.

Additionally, Marshak and subsequently Su \& Olson make the diffusion and Eddington approximations, which lead to a parabolic ODE describing a diffusion process.  Since our code is hyperbolic in nature, we must independently solve the radiation momentum Equation~\eqref{mixedframe:graysystem:radmomentum}, 
\begin{equation}
  \frac{1}{c} \,\partial_t \left(\frac{F}{c}\right) + \partial_x P_{xx} = -\rho\kappa_0 \frac{F}{c}, \label{marshak:F}
\end{equation}
where the radiation pressure component $P_{xx}$ is derived from $\mathcal{E}$ and $F$ via the $M_1$ closure relation.

The so-called Marshak boundary condition, which imposes the constraint of constant radiative flux on the surface $x=0$, is given by
\begin{equation}
  c\mathcal{E}(0,t) + 2 F(0,t) = 4 F_{\rm inc}. \label{marshak:bc_ix1}
\end{equation}
This, together with the boundary condition
\begin{equation}
  \mathcal{E}(x,t) \to 0 \qquad \text{as} \qquad x \to \infty, \label{marshak:bc_ox1}
\end{equation}
the initial condition
\begin{equation}
  \mathcal{E}(x,0) = T(x,0) = 0,
\end{equation}
and Equations~\eqref{marshak:energyexchange} and~\eqref{marshak:F} define the radiation subsystem that we solve numerically.

The semi-analytic solution of Su \& Olson is given in terms of the dimensionless independent variables $\chi \equiv \sqrt{3}\rho\kappa_0 x$ and $\tau \equiv \epsilon c\rho\kappa_0 t$, and the dependent variables $u(\chi,\tau) \equiv c\mathcal{E}(x,t)/(4F_{\rm inc})$ and $v(\chi,\tau) \equiv c a_R T^4(x,t)/(4F_{\rm inc})$, where $\epsilon \equiv 4 a_R/\alpha$ is a retardation parameter.  By choosing a system of units in which $a_R=c=1$ and $F_{\rm inc}=\case{1}{4}$, we can identify $(u,v)$ with $(\mathcal{E},T^4)$, respectively.  With these definitions, the equations to be integrated become
\begin{subequations} \label{marshak:system}
	\begin{eqnarray}
		\partial_\tau v &=& -(v-u), \label{marshak:system:v} \\
		\epsilon \, \partial_\tau u + \sqrt{3} \,\partial_\chi \left( \frac{F}{4 F_{\rm inc}} \right) &=& (v-u), \label{marshak:system:u} \\
		\epsilon \, \partial_\tau \left( \frac{F}{4 F_{\rm inc}} \right) + \sqrt{3} \,\partial_\chi \left( \frac{c P_{xx}}{4 F_{\rm inc}} \right) &=& -\left( \frac{F}{4 F_{\rm inc}} \right).
	\end{eqnarray}
\end{subequations}

We impose the Marshak boundary condition in Equation~\eqref{marshak:bc_ix1} indirectly via the semi-analytic solution for $u$ of Equations~\eqref{marshak:system}, given by
\begin{eqnarray}
  &&u_{\rm soln}(\chi,\tau) = \frac{c \mathcal{E}(x,t)}{4 F_{\rm inc}} \nonumber \\
  &&= 1 - \frac{2\sqrt{3}}{\pi} \int_0^1 d\eta\, e^{-\tau \eta^2} \left\{ \frac{\sin[\chi\Gamma_1(\eta) + \Theta_1(\eta)]}{\eta \sqrt{3+4\Gamma_1^2(\eta)}} \right\} \nonumber \\
  &&- \frac{\sqrt{3}}{\pi} e^{-\tau} \int_0^1 d\eta\, e^{-\tau/\epsilon \eta} \left\{ \frac{\sin[\chi\Gamma_2(\eta) + \Theta_2(\eta)]}{\eta (1+\epsilon\eta) \sqrt{3+4\Gamma_2^2(\eta)}} \right\}, \label{marshak:suolson}
\end{eqnarray} 
where
\begin{subequations}
\begin{eqnarray}
  \Gamma_1(\eta) &=& \eta \sqrt{\epsilon + \frac{1}{1-\eta^2}}, \\
  \Gamma_2(\eta) &=& \sqrt{(1-\eta)\left(\epsilon + \frac{1}{\eta}\right)}, \\
  \Theta_n(\eta) &=& \cos^{-1} \sqrt{\frac{3}{3+4\Gamma_n^2(\eta)}}, \qquad n=1,2,
\end{eqnarray}
\end{subequations}
evaluated at $\chi=0$ \citep[see][equation~36]{Su:1996}.  Once $\mathcal{E}(0,t) = u_{\rm soln}(0,\tau)$ has been so obtained, we compute $F(0,t)$ via Equation~\eqref{marshak:bc_ix1}.  Note that we need not compute $v(0,\tau)$ at the left boundary since it is neither required to compute $u(0,\tau)$ nor to compute $\kappa_0$, which is constant.

Since the solution in Equation~\eqref{marshak:suolson} represents a parabolic approximation to the hyperbolic behavior of our radiation subsystem at $\chi=0$, at least to the extent that the radiation is actually in the streaming regime at low optical depth, we evaluate the integrals using simple midpoint quadrature in lieu of some more elaborate scheme.  On the right side, we use the Dirichlet boundary condition $u=v=0$.  The domain is chosen sufficiently large that the asymptotic boundary condition in Equation~\eqref{marshak:bc_ox1} is reasonably approximated.

To evolve the internal energy of the gas, we use a $\theta$-scheme update of the gas energy Equation~\eqref{marshak:energyexchange:T} given by
\begin{subequations} \label{marshak:update}
\begin{eqnarray}
  v^{n+1} &=& v^n + \Delta v, \\
  u^{n+1} &=& u^n + \Delta u, \\
  \Delta v &=& \frac{-(v^n-u^n)\,\Delta \tau}{1+\theta(1+1/\epsilon)\Delta \tau}, \\
  \Delta u &=& -\Delta v/\epsilon,  \label{marshak:conservation}
\end{eqnarray}
\end{subequations}
where we have used conservation of energy to write Equation~\eqref{marshak:conservation}.  This is done for the source terms in Equations~\eqref{marshak:system:v} and~\eqref{marshak:system:u} in lieu of the usual energy balance source term step as described in Section~\ref{source}.

\begin{figure}
  \centering
  \plotone{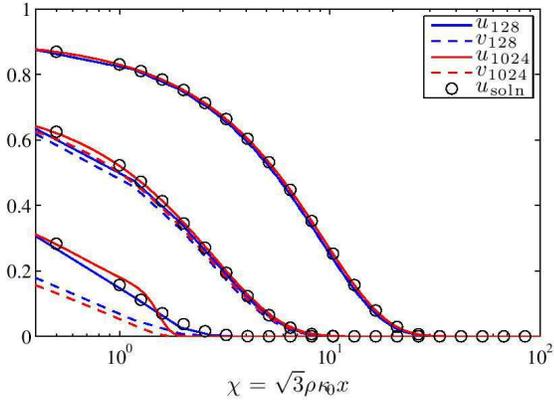}
  \caption{Computed solution of $u \equiv \mathcal{E}$ and $v \equiv T^4$ for the non-equilibrium Marshak wave problem with $N = 128$, as well as a reference solution using $N = 1024$, at times $\tau=\{1,10,100\}$ (left to right) on a log-linear scale.  The semi-analytic (equation~\ref{marshak:suolson}) solution $u_{\rm soln}$ of the diffusion equation is shown for comparison.  \label{radmarshak:plot:soln:linear}}
\end{figure}

\begin{figure}
  \centering
  \plotone{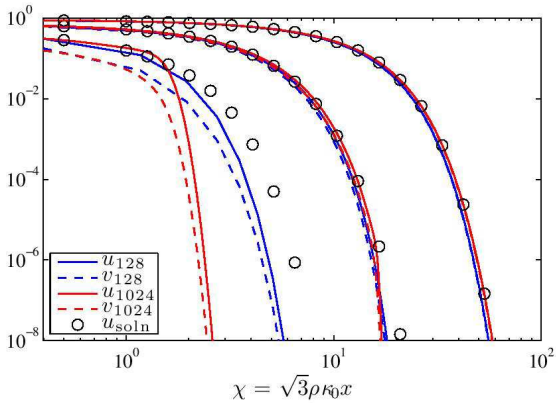}
  \caption{Same as Figure~\ref{radmarshak:plot:soln:linear} on a log-log scale. \label{radmarshak:plot:soln:log}}
\end{figure}

We use a one-dimensional grid of resolution $N=128$ on the domain $\chi \in [0,100]$, with background density $\rho_0 = 1$, specific absorption opacity $\kappa_0 = 1$, and retardation parameter $\epsilon=1$.  The computed results for $u$ and $v$ at times $\tau=\{1,10,100\}$ are shown on log-linear and log-log scales in Figures~\ref{radmarshak:plot:soln:linear} and~\ref{radmarshak:plot:soln:log}, respectively.  For comparison, we have also plotted reference solutions for a grid of resolution $N = 1024$ as well as the semi-analytic solution $u_{\rm soln}$ in Equation~\eqref{marshak:suolson}.  The plots show good agreement between the solutions, which improves at later time, i.e., at larger optical depth, as the system approaches the equilibrium diffusion regime.  However, at earlier time, i.e., at small optical depth, the system is still in the streaming regime.  Thus, our computed solution is expected to differ from the semi-analytic solution of Su \& Olson, which is based on the diffusion approximation.  Also, since we solve a hyperbolic system of PDE, our wave solution propagates at finite speed.  On the contrary, Su \& Olson solve a \emph{parabolic} system of PDE, hence their solution propagates instantaneously \citep[see][equations~9 and~10]{Su:1996}.  This is especially evident in the higher-resolution reference solution $u_{1024}$ at $\tau=1$ (i.e., $t=1$), which contains less numerical diffusion than the lower-resolution solution, hence the wave front at $\chi=\sqrt{3}$ (i.e., $x=ct=1$) is more sharply defined there.

\vspace{1em}
{\subsection{Performance Benchmark Test}  \label{benchmark}}

Next, we perform a timing benchmark, comparing results from our code to results obtained with the FLD module of the well-known  code {\it Enzo} \citep{Reynolds:2009}.  Our aim is to compare the performance of our algorithm for solving the radiation moment equations, which combines explicit Godunov transport with implicit source term treatment, against the fully implicit iterative methods typically used in FLD codes.  We choose for our benchmark the non-equilibrium Marshak wave problem, in which only the radiation energy and momentum, and the gas energy are integrated; the gas density and momentum are held constant.  

\begin{figure}
  \centering
  \epsscale{1}
  \plotone{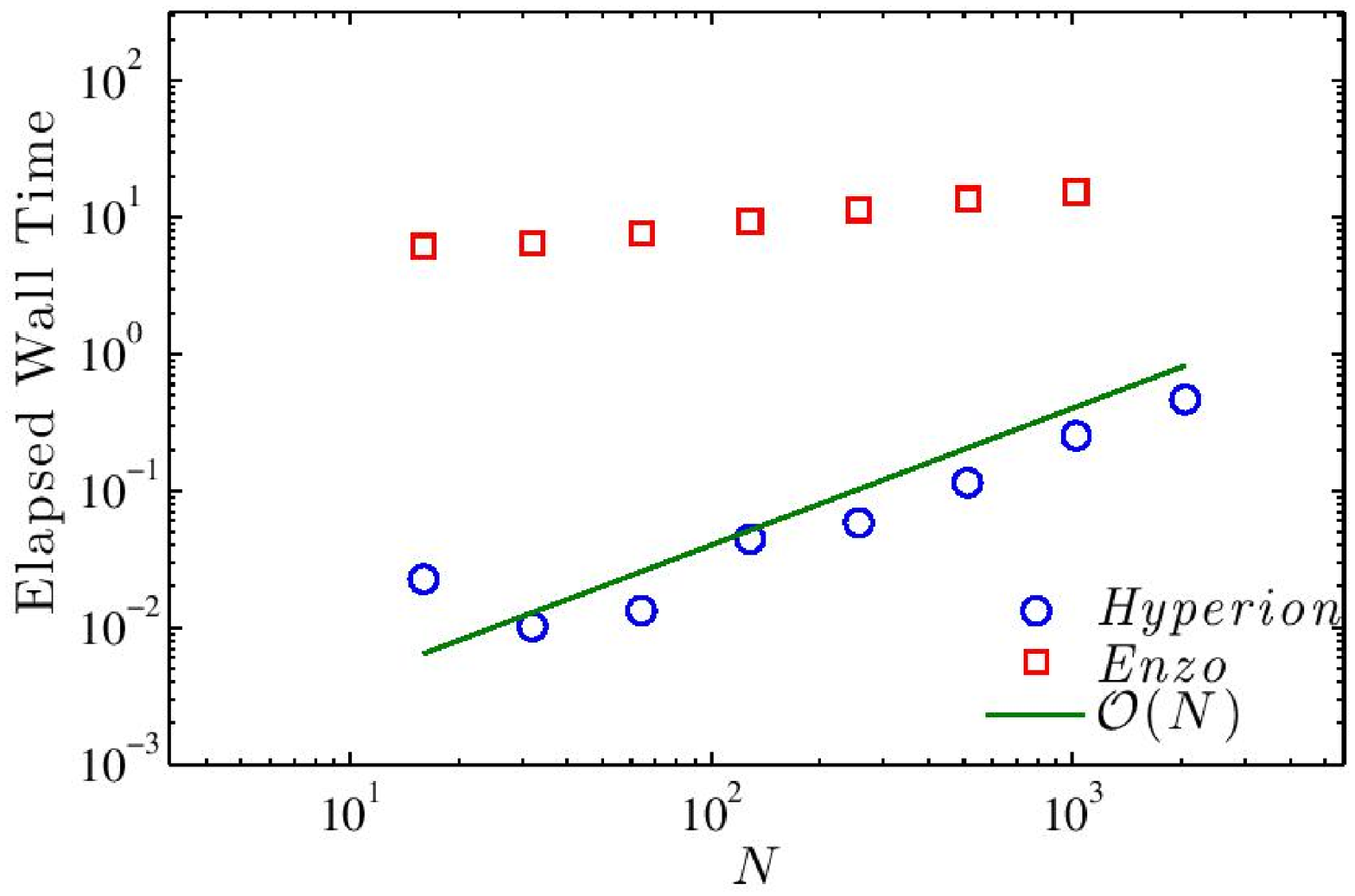}
  \plotone{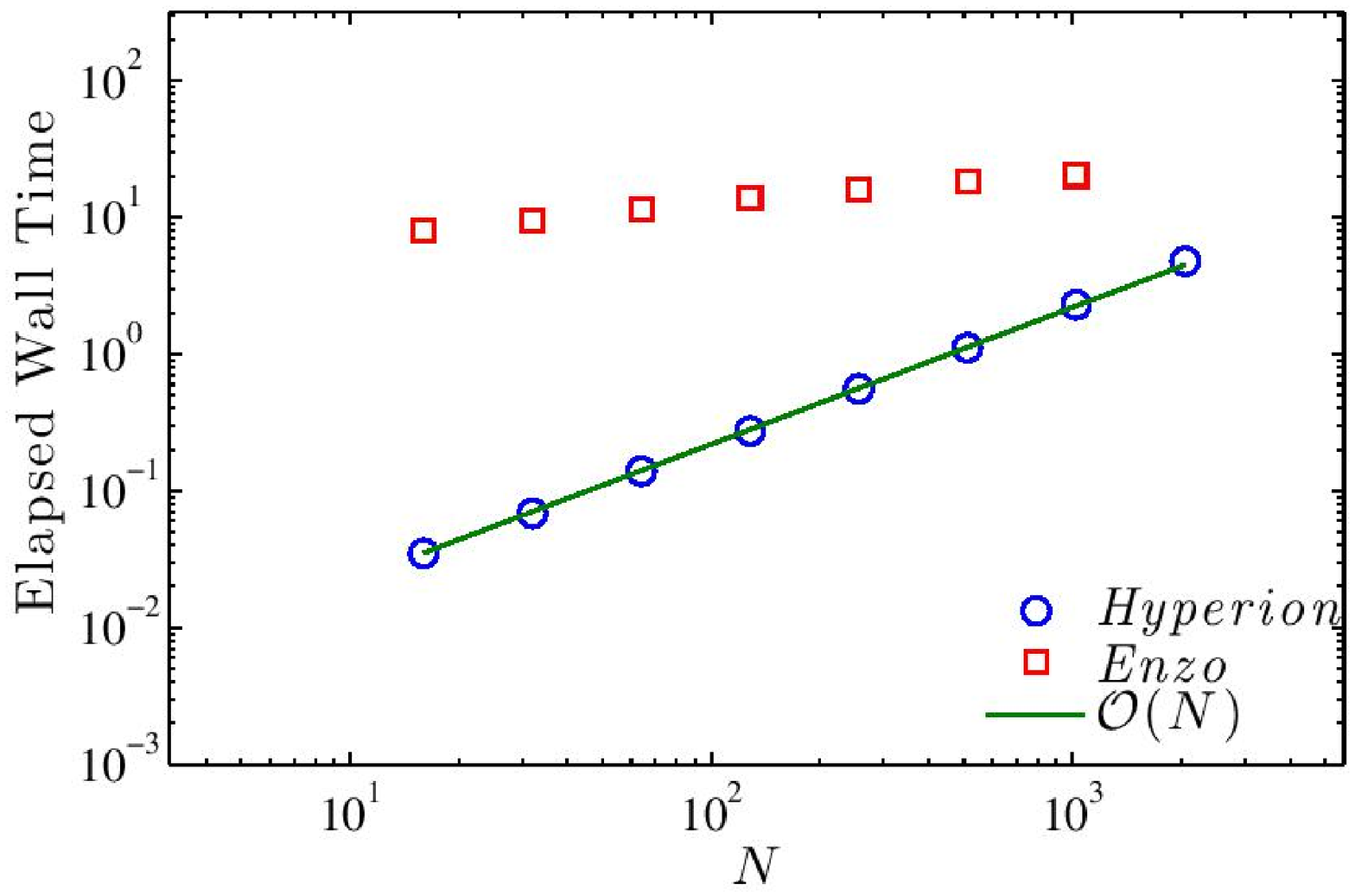}
  \plotone{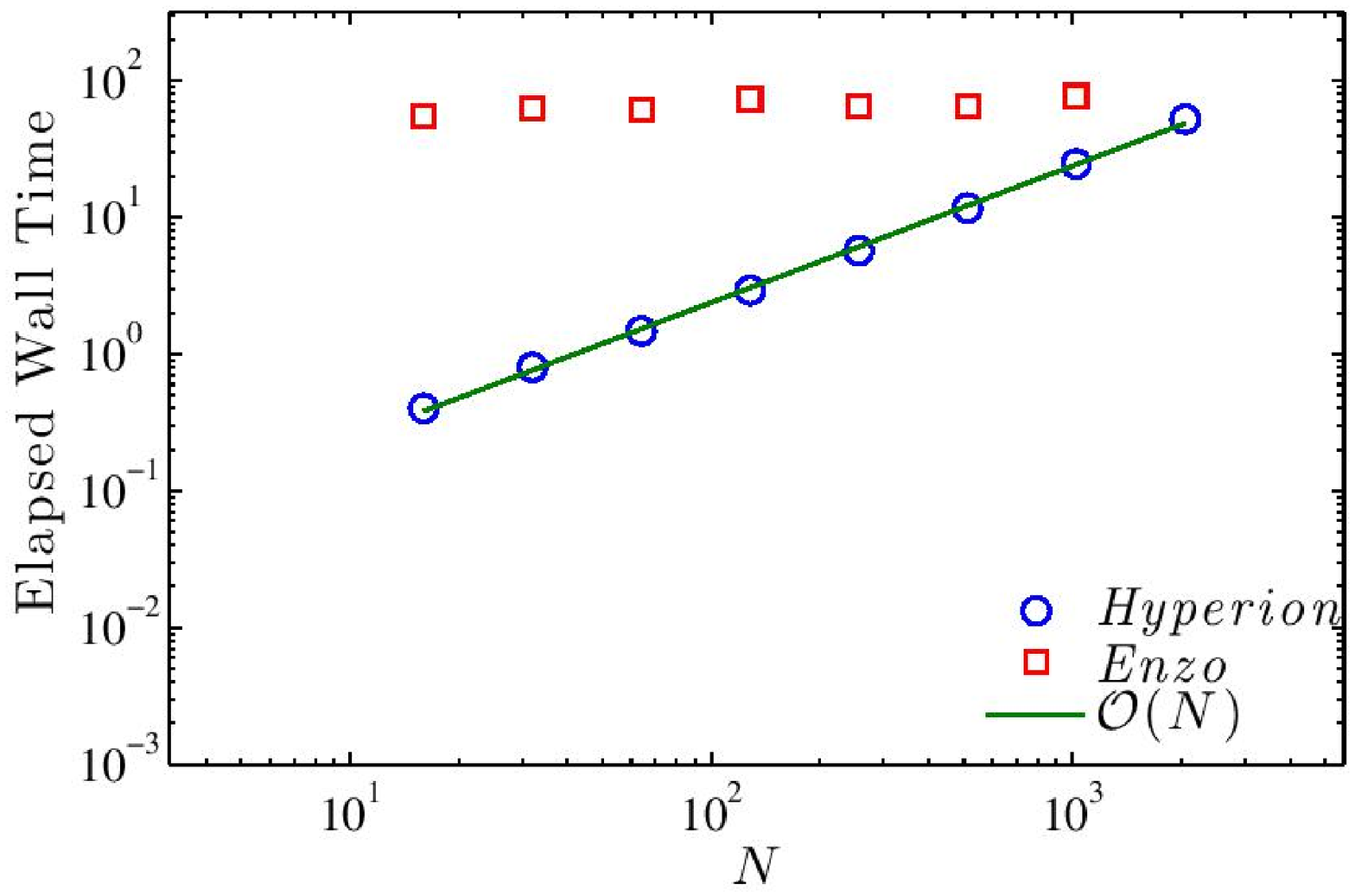}
  \caption{Timing benchmark comparison of our code, {\it Hyperion}, with the FLD module of {\it Enzo} on the one-dimensional non-equilibrium Marshak wave problem evolved to $\tau=\{1,10,100\}$ (top, middle, bottom).  The number of zones varies from $N = 32$ to $N = 2048$.  All tests were run on a single 2.8 GHz Intel Core i7 processor.  \label{benchmark:timings}}
\end{figure}

We use the same parameter set and boundary conditions as described in Section~\ref{radmarshak}, and run the problem to dimensionless time $\tau = \{1,10,100\}$ on one-dimensional grids of resolution varying from $N = 16$ to $N = 2048$.  We perform the same test using both our code, which we have named {\it Hyperion}, and the {\it Enzo} code on the same 2.8 GHz Intel Core i7 processor, and record the total wall-clock time elapsed during each run.  The adaptive time step for the {\it Enzo} code is primarily controlled by prescribed accuracy requirements; in this case, an accuracy tolerance of $\varepsilon = 10^{-7}$ is used.  We use the same tolerance for the iterative solution of the energy balance update given in Equation~\eqref{source:stiff:radenergy}.  

Figure~\ref{benchmark:timings} shows the timings for both {\it Hyperion} (circles) and {\it Enzo} (squares) versus the grid resolution $N$, along with a reference curve of $\mathcal{O}(N)$ (solid line) for the Marshak wave evolved to $\tau=\{1,10,100\}$.  Since the Marshak wave problem is inherently one-dimensional, all runs were performed using the one-dimensional integrator in each respective code.\footnote{We were not able to perform the last run at resolution $N=2048$ using {\it Enzo}.}  The data clearly show that the timing of the {\it Hyperion} code scales linearly with grid resolution, as one would expect for an algorithm whose execution time is dominated by the explicit Godunov method.  In contrast, the timing of the {\it Enzo} code is approximately constant, which \cite{Reynolds:2009} suggest may be attributed to the fact that the Inexact Newton's Method used to iteratively solve the nonlinear radiation subsystem has been shown to be independent of spatial resolution for diffusive problems, such as the non-equilibrium Marshak wave.  This suggests that there is some resolution beyond which the {\it Enzo} code will outperform the {\it Hyperion} code; however, this threshold seems to be at a higher resolution than most practical applications would require. 

\vspace{1em}
{\subsection{Fully Coupled Radiation Hydrodynamics Tests}  \label{fullsystem}}

In the next tests, the gas and radiation subsystems are fully coupled.  These tests are designed to verify the interplay between the gas and radiation dynamics in the context of the RSLA.  In each test, the reduced speed of light $\hat{c}$ must first be determined in order to preserve the relevant ordering of characteristic time scales while allowing for computationally feasible explicit time subcycling.

\vspace{1em}
{\subsubsection{Radiatively Inhibited Accretion and Radiatively Driven Wind}  \label{radbondi}}

To test the accuracy of the radiation force in the optically thin regime, we present a one-dimensional, planar version of the radiatively inhibited Bondi accretion problem of \cite{Krumholz:2007}.  We consider the steady flow of an isothermal gas under the assumption that it is neither heated nor cooled by the radiation.  We set $\kappa_0$ to a sufficiently small value such that the gas is optically thin throughout the computational domain.  We consider a constant radiation field in the streaming limit with $\mathbf{F} = F_0\hat{z}$ and $\mathcal{E} = F_0/c$.  The radiation applies a specific force of
\begin{equation}
  \mathbf{f}_{\rm rad} = \kappa_0 \frac{F_0}{c}\hat{z},
\end{equation}
to the gas.  We also consider a linear gravitational potential of the form $\Phi_{\rm grav} = g_0 z$, for some constant $g_0 > 0$, which applies a specific force of
\begin{equation}
  \mathbf{f}_{\rm grav} = -g_0\hat{z},
\end{equation}
to the gas.  Thus, the total specific force on the gas is given by
\begin{equation}
  \mathbf{f}_{\rm total} = \mathbf{f}_{\rm rad} + \mathbf{f}_{\rm grav} = -(1-\eta_{\rm Edd})g_0\hat{z},  \label{radbondi:totalforce}
\end{equation}
where $\eta_{\rm Edd}$ is the fraction of the Eddington-limit flux defined by
\begin{equation}
  \eta_{\rm Edd} \equiv \frac{\kappa_0 F_0}{cg_0}.
\end{equation}
For $\eta_{\rm Edd} = 1$, the radiation and gravitational forces balance and the system is in hydrostatic equilibrium; for $0 \le \eta_{\rm Edd} < 1$, the gravitational force dominates and the gas is steadily accreted inward; for $\eta_{\rm Edd} > 1$, the radiation force dominates and the gas is steadily driven outward in a wind.

To set the initial conditions, we use the constants of motion given by
\begin{eqnarray}
  a_0 &=& \sqrt{\frac{P}{\rho}},  \label{radbondi:soundspeed} \\
  \dot{M} &=& \rho v.  \label{radbondi:accretionrate}
\end{eqnarray}
The flow must satisfy the Bernoulli equation, $\mathcal{B} = \text{constant}$, along streamlines for
\begin{equation}
  \mathcal{B} = \frac{1}{2}v^2 + h + \Phi_{\rm total},  \label{radbondi:bernoulli}
\end{equation}
where
\begin{equation}
  h \equiv \int \frac{dP}{\rho} = a_0^2 \,\ln{\rho}
\end{equation}
is the specific enthalpy of an isothermal gas derived from Equation~\eqref{radbondi:soundspeed}, and $\Phi_{\rm total} = (1-\eta_{\rm Edd})g_0z$ is the potential of the total force given in Equation~\eqref{radbondi:totalforce}.  Note that for a hydrostatic, isothermal atmosphere with no radiation, Equation~\eqref{radbondi:bernoulli} implies that $\rho = \rho_0 e^{-z/H}$, where
\begin{equation}
  H \equiv \frac{a_0^2}{g_0}
\end{equation}
is the isothermal scale height.

For our problem, we scale the gas density to $\rho_0$, its value at $z=0$, the gas velocity to the background sound speed, $a_0$, and the $z$-coordinate to the isothermal scale height, $H$.  In terms of the dimensionless density $\alpha \equiv \rho/\rho_0$, Mach number $\mathcal{M} \equiv v/a_0$, height $\chi \equiv z/H$, and mass-accretion rate $\lambda \equiv \dot{M}/\rho_0 a_0$, it follows from Equation~\eqref{radbondi:accretionrate} that
\begin{equation}
  \lambda = \alpha\mathcal{M},  \label{radbondi:continuity}
\end{equation}
and from Equation~\eqref{radbondi:bernoulli} that
\begin{equation}
  \tilde{\mathcal{B}} = \frac{1}{2}\mathcal{M}^2 + \ln\alpha + (1-\eta_{\rm Edd})\chi.  \label{radbondi:dimensionlessbernoulli}
\end{equation}
Once a value for the Mach number $\mathcal{M}_0$ at $\chi=0$ is chosen, we have $\tilde{\mathcal{B}} = \case{1}{2} \mathcal{M}_0^2$ and $\lambda=\mathcal{M}_0$.  The initial conditions are obtained by solving
\begin{equation}
	\frac{\mathcal{M}^2 - \mathcal{M}_0^2}{2} + \ln \left( \frac{\mathcal{M}_0}{\mathcal{M}} \right) + (1 - \eta_{\rm Edd})\chi = 0 \label{bondi:initialconds}
\end{equation}
for $\mathcal{M}$ as a function of $\chi$ via Newton--Raphson iteration and using $\alpha=\mathcal{M}_0/\mathcal{M}$.

\begin{figure}
  \centering
  \epsscale{1}
  \plotone{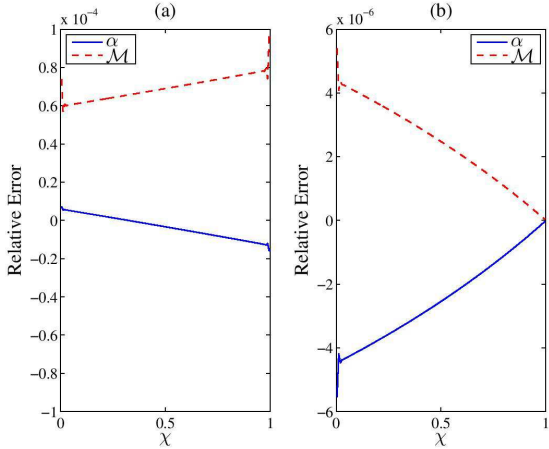}
  \caption{Relative errors of the computed solutions of the dimensionless density $\alpha$ and velocity $\mathcal{M}$ for the radiatively inhibited accretion problem with $\eta_{\rm Edd} = 0.5$.  The relative errors are plotted for both a subsonic flow (left) with $\mathcal{M}_0 = 0.1$ and a supersonic flow (right) with $\mathcal{M}_0 = 2.5$.  The maximum relative error is approximately $0.0098\%$ for all solutions.  \label{radbondi:soln:accretion}}
\end{figure}

\begin{figure}
  \centering
  \epsscale{1}
  \plotone{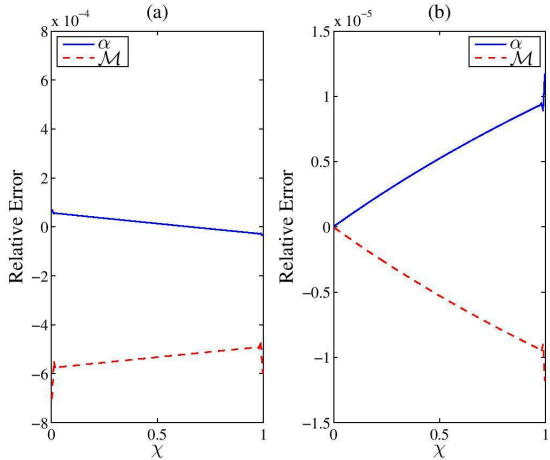}
  \caption{Same as Figure~\ref{radbondi:soln:accretion} for a wind solution with $\eta_{\rm Edd} = 1.5$.  The maximum relative error is approximately $0.070\%$ for all solutions.  \label{radbondi:soln:wind}}
\end{figure}

We use Dirichlet boundary conditions based on the initial conditions obtained from Equation~\eqref{bondi:initialconds} for both the gas and radiation on the domain $\chi \in [0,1]$ with resolution $N=128$.  Starting from the semi-analytic solution, we evolve for $10$ grid sound-crossing times.  To obtain a total optical depth over the simulation domain similar to that of \cite{Krumholz:2007}, we set $\kappa_0 = 10^{-6}$.  By computing $d\chi/d\mathcal{M}$ and $d^2\chi/d\mathcal{M}^2$ at the sonic point, i.e., where $\mathcal{M}=1$, for fixed values of $\lambda$ and $\eta_{\rm Edd}$, it can be shown that there are no trans-sonic solutions; only entirely subsonic or supersonic solutions exist.  We compute solutions for both a radiatively inhibited accretion flow with $\eta_{\rm Edd}=0.5$ and a radiatively driven wind flow with $\eta_{\rm Edd}=1.5$, for both a subsonic case with $\mathcal{M}_0=0.1$ and a supersonic case with $\mathcal{M}_0=2.5$.  Since this problem lies squarely within the optically thin regime, we set $\hat{c} = 10 v_{\rm max}$, where $v_{\rm max} = \mathcal{M}_0 + 1$ in problem units, so that $R \sim 10$ radiation subcycles are performed per gas cycle.  Figure~\ref{radbondi:soln:accretion} shows the relative error of the computed solutions for the dimensionless density, $\alpha$, and velocity, $\mathcal{M}$, compared to the semi-analytic solution obtained from Equation~\eqref{bondi:initialconds} for a radiatively inhibited accretion flow in both the subsonic (left) and supersonic (right) cases.  The maximum relative error is $\approx 0.0098\%$ for all solutions.  Figure~\ref{radbondi:soln:wind} shows the same plots as Figure~\ref{radbondi:soln:accretion} for the radiatively driven wind flow with a maximum relative error of $\approx 0.070\%$ for all solutions.  The error in the computed solution for each of these steady flows is dominated by operator splitting error, which causes a slight force imbalance leading to a nearby solution of Equation~\eqref{bondi:initialconds}.  This solution differs slightly from the initial conditions, which are held fixed at the boundaries, and as expected, the discontinuity causes an increase in the error there.  These tests provide good evidence of the code's ability to accurately compute the radiation force in optically thin regimes.

\vspace{1em}
{\subsubsection{Advection of a Radiation Pulse}  \label{radpulse}}

As a test of the $\mathcal{O}(\beta\tau)$ terms in the radiation energy and flux equations in the dynamic diffusion regime, we simulate the strong advection of a diffusing radiation pulse by the gas.  A similar test is described by \cite{Krumholz:2007}.  

We advect a pulse of radiation energy in an optically thick gas with a uniform background flow velocity.  Initially, the system is in both pressure and radiative equilibrium everywhere, implying that $\nabla (P + \case{1}{3}\mathcal{E}) = 0$ and $\mathcal{E} = a_{\rm R} T^4$.  It follows that the initial density and gas temperature are related by
\begin{equation}
	\frac{\rho}{\rho_0} = \frac{T_0}{T} + \frac{\gamma \mathcal{P}_0}{3} \left[ \frac{T_0}{T} - \left( \frac{T}{T_0} \right)^3 \right],  \label{radpulse:rho}
\end{equation}
where $\rho_0$, $T_0$, and $\mathcal{P}_0$ are the background values of density, gas temperature, and dimensionless pressure ratio, respectively, away from the pulse.  The gas temperature is initialized to a constant-plus-Gaussian profile of width $w$, centered at the origin, with peak temperature twice the background value $T_0$, given by
\begin{equation}
	\frac{T}{T_0} = 1 + \exp \left( -\frac{x^2}{2w^2} \right).  \label{radpulse:T}
\end{equation}
From Equation~\eqref{radpulse:rho}, it is clear that the increase in both gas and radiation pressure due to an increase in gas temperature above $T_0$ must be offset by a corresponding decrease in density below $\rho_0$.  As excess radiation diffuses outward from the pulse, pressure equilibrium is lost and gas moves inward.  

For the parameter set of \cite{Krumholz:2007}, in the background state the dimensionless pressure ratio is $\mathcal{P}_0 \approx 0.18$, the characteristic optical depth over a distance $w$ is $\tau_0 = \rho_0 \kappa_0 w = 2900$, the flow Mach number is $\mathcal{M}_0 \equiv v/a_0 \approx 0.053$, and $\beta \equiv v/c \approx 3.3 \times 10^{-5}$.  Thus, $\beta\tau_0 \approx 0.096$ and $\mathcal{P}_0\tau_0 \approx 520 \gg 1$.  Since $\mathcal{P}_0 < 1$ and $\mathcal{M}_0 \ll 1$, the dynamics of this problem are dominated by the gas pressure force; the characteristic dynamical time is $t_{\rm dyn} \sim w/a_0 = w/\sqrt{\gamma k_{\rm B} T_0/\mu}$, and the characteristic diffusion time is $t_{\rm diff} \sim w\tau_0/c$.  The ratio of these time scales is $t_{\rm dyn}/t_{\rm diff} = c/(a_0 \tau_0) \approx 0.55$; for our test, we must require $\hat{c}/(a_0 \tau_0) \approx 0.55$ in order to obtain similar behavior to the \cite{Krumholz:2007} solution.  This is not feasible for our code with an optical depth of $\tau_0 = 2900$, so we choose instead a smaller value of $\tau_0$ such that $\hat{c}/a_0 \approx 100$.  Also, we choose a background flow velocity $v$ so that the flow remains subsonic with $\beta\tau_0 \sim 0.1$, in order to preserve the relative sizes of the $\mathcal{O}(\beta\tau)$ source terms.  Furthermore, since the splitting of the source term integration between the gas and radiation subsystems in our code introduces a non-conservation of momentum of $\mathcal{O}(\mathcal{P}_0\tau_0)$, such a large value of $\mathcal{P}_0$ would lead to significant splitting error.  Instead, we choose parameters so that $\mathcal{P}_0\tau_0 \lesssim 1$.

With these considerations in mind, we use the background density $\rho_0 = 25 \text{ g cm}^{-3}$, temperature $T_0 = 1.1 \times 10^7 \text{ K}$, $w = 20 \text{ cm}$, mean particle mass $\mu = m_{\rm H} = 1.67 \times 10^{-24} \text{ g}$, specific absorption opacity $\kappa_0 = 0.4 \text{ cm}^2 \text{ g}^{-1}$, and background flow velocity $v = 3 \times 10^6 \text{ cm s}^{-1}$.  It follows that $\tau_0 \approx 200$, $\mathcal{P}_0 \approx 0.005$, and $v \ll a_0 \approx 3 \times 10^7 \text{ cm s}^{-1}$, hence the flow remains subsonic.  We also choose $\hat{c} = 100 a_0$, so that with our other parameters $t_{\rm dyn}/\hat{t}_{\rm diff} \sim \hat{c}/(a_0 \tau_0) \approx 0.5$ and $\beta\tau_0 \approx 0.02$, both of which are comparable to the parameter set of \cite{Krumholz:2007}.  Furthermore, we have for our problem $\mathcal{P}_0\tau_0 \approx 1$.  

We use periodic boundary conditions on the one-dimensional domain with $x/w \in [-21.25, 21.25]$, using a grid resolution of $N=512$ zones, and we run the simulation for a time $t = 2w/v$ so that the pulse is advected over twice its width.  Since there is no simple analytic solution for this test, we compare the results to those of an unadvected run with zero background flow velocity.  So that the two runs both end up centered about the origin, we shift the initial profile of the advected run by a distance $vt=2w$ to the left.  Finally, we run the test with and without the $\mathcal{O}(\beta\tau)$ source terms included.  Since $\beta\tau \approx 0.02$, the error in the solution without the $\mathcal{O}(\beta\tau)$ terms should be slightly larger than with the terms included.  

\begin{figure}
  \centering
  \plotone{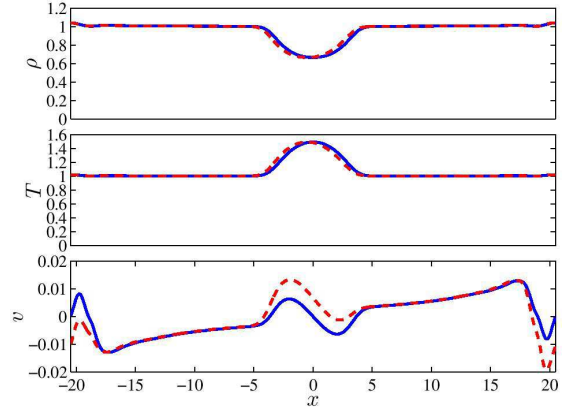}
  \caption{Solution of the advected radiation pulse (solid line) along with the unadvected reference solution (dashed line) for $\rho$, $T$, and $v$ with the $\mathcal{O}(\beta\tau)$ source terms included. \label{radpulse:plot:soln}}
\end{figure}

\begin{figure}
  \centering
  \plotone{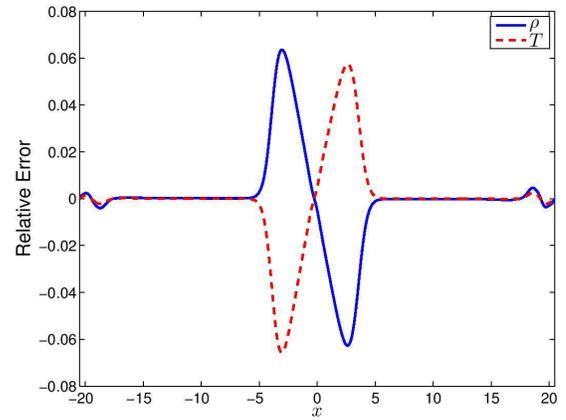}
  \caption{Relative error in the solutions of the density and temperature of the advected radiation pulse versus the unadvected reference solution with the $\mathcal{O}(\beta\tau)$ source terms included.  The relative errors are less than $6.6\%$ across the domain.  \label{radpulse:plot:relerr}}
\end{figure}

\begin{figure}
  \centering
  \plotone{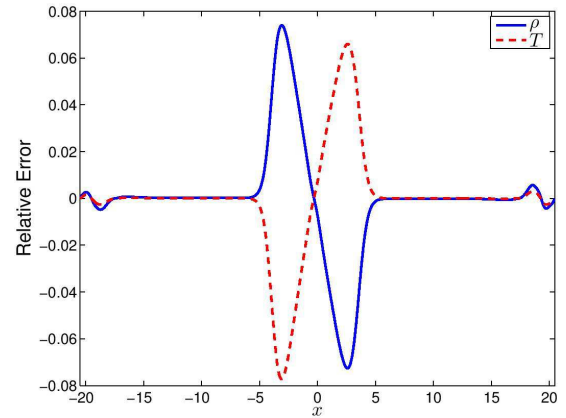}
  \caption{Same as Figure~\ref{radpulse:plot:relerr}, but \emph{without} the $\mathcal{O}(\beta\tau)$ source terms included.  The relative errors are slightly larger in this case, but are less than $7.7\%$ across the domain.  \label{radpulse:plot:noradterms:relerr}}
\end{figure}

Figure~\ref{radpulse:plot:soln} shows the solutions of the density, temperature, and velocity (subtracting out the background) for both the advected radiation pulse and the unadvected reference solution at the same time with the $\mathcal{O}(\beta\tau)$ source terms included.  The high optical depth of the gas in this problem keeps the system so near to radiative equilibrium that we do not distinguish between the gas and radiation temperatures, which are equivalent at the $10^{-3}$ level.  Figure~\ref{radpulse:plot:relerr} shows the relative error in the density and temperature for this run, but we do not compute the relative error in the velocity since the reference value is close to $0$ in places.  The agreement between the advected and unadvected solutions is good, and the relative errors in density and temperature are less than $6.6\%$ across the domain.  Figure~\ref{radpulse:plot:noradterms:relerr} shows the relative error \emph{without} the $\mathcal{O}(\beta\tau)$ terms included.  In this run, the maximum relative errors in the density and temperature are slightly larger, but they are less than $7.7\%$ across the domain, so the agreement is still fair.  This test provides good evidence of the code's ability to reproduce the effect of strong advection of a radiation field by optically thick gas in the static diffusion regime.  This test also indicates the importance of including the $\mathcal{O}(\beta\tau)$ source terms when $\beta\tau \sim 1$.

\vspace{1em}
{\subsubsection{Radiation Pressure Tube}  \label{radtube}}

\cite{Krumholz:2007} describe a simulation to test the accuracy of the radiation pressure force in a one-dimensional tube filled with gas and radiation in static equilibrium.  The gas is optically thick, hence the Eddington approximation holds.  Also, the gas and radiation are in equilibrium, hence their temperatures are equal and we can define $T \equiv T_{\rm gas} = T_{\rm rad}$.  Force balance between the gas and radiation pressures implies that the total pressure is constant throughout the domain, from which it follows that
\begin{equation}
  \left[ \frac{\rho}{\rho_0} + \frac{4}{3} \left( \frac{T}{T_0} \right)^3 \right] \frac{T'}{T_0} + \frac{T}{T_0} \frac{\rho'}{\rho_0} = 0,  \label{radtube:ode1}
\end{equation}
where primes denote spatial derivatives.  By choosing $T_0$ such that $a_{\rm R} T_0^4 = \rho_0 k_{\rm B} T_0 / \mu$, coefficients are absorbed into the problem units.  Furthermore, slab symmetry implies that the radiative flux is constant, from which it follows that
\begin{equation}
  -\frac{\rho_0}{\rho} \frac{\rho'}{\rho_0} \frac{T'}{T_0} + 3 \frac{T_0}{T} \left( \frac{T'}{T_0} \right)^2 + \frac{T''}{T_0} = 0.  \label{radtube:ode2}
\end{equation}

To obtain a semi-analytic solution, we note that this system can be written as a nonlinear, first-order ODE in the variables $\rho$, $T$, and $T'$.  Given the values $\rho_0$, $T_0$, and $T_0'$ at the left boundary, the ODE can be integrated to the right boundary with arbitrary precision using conventional methods.  

\begin{figure}
  \centering
  \epsscale{1}
  \plotone{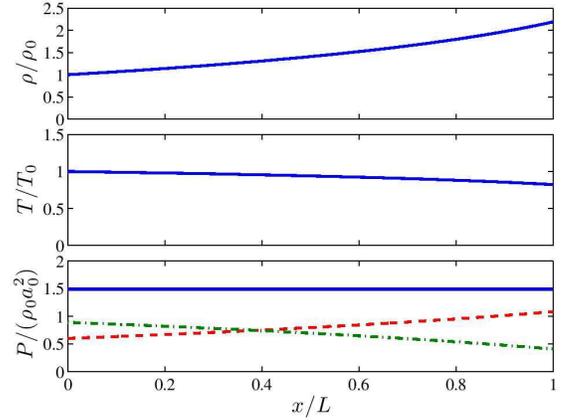}
  \caption{Semi-analytic solution of the density (top), temperature (middle), and pressure (bottom) versus position for the radiation tube problem.  The bottom plot shows the total pressure (solid), gas internal pressure (dashed), and radiation pressure (dot-dashed).  \label{radtube:odesoln}}
\end{figure}

We use the parameter set of \cite{Krumholz:2007}, where $\rho_0 = 1 \text{ g cm}^{-3}$ and $\rho_0' = 5 \times 10^{-3} \text{ g cm}^{-4}$.  The gas is characterized by a mean particle mass of $\mu = 3.9 \times 10^{-24} \text{ g}$ so that $T_0 = 2.75 \times 10^7 \text{ K}$, and by the specific absorption opacity of $\kappa_0 = 100 \text{ cm}^2 \text{ g}^{-1}$.  This yields a system with roughly comparable pressures that is dominated by radiation pressure on the left and by gas pressure on the right.  We use a one-dimensional domain of length $L = 128 \text{ cm}$ with grid resolution $N=128$, and impose Dirichlet boundary conditions on the gas and radiation\footnote{\cite{Krumholz:2007} impose symmetry (reflection) boundary conditions on the gas in order to preserve the total mass, but this is not helpful in our code since the radiation force is not applied symmetrically.  In our test, the relative error in the total mass at the end of the run compared to the initial value is $4.7 \times 10^{-6}$, suggesting that mass loss is not the dominant source of error.}.  We include the energy balance source term in Equation~\eqref{source:stiff:radenergy} so that radiative equilibrium must be maintained numerically rather than enforced, presenting a more rigorous test of the code.  Figure~\ref{radtube:odesoln} shows the semi-analytic solutions for the density and temperature, as well as the resulting gas pressure, radiation pressure, and total pressure.  We set these solutions as the initial condition for our problem and evolve the system for 10 sound crossing times $t_{\rm sound} \equiv L/a_0 = \sqrt{\gamma k_{\rm B} T_0/\mu}$.  Since the flux absorption source term in Equation~\eqref{source:stiff:radmomentum} is very stiff, we use the fully implicit Backward Euler method with $\theta=1$.  

With such a large characteristic optical depth of $\tau \sim \rho_0 \kappa_0 L = 1.28 \times 10^3$ across the simulation domain, one might expect that the RSLA might not be feasible in this problem.  However, the relevant time scales of interest are the characteristic sound-crossing time, $t_{\rm sound}$, and the characteristic diffusion time $t_{\rm diff} \sim \rho_0 \kappa_0 L^2/c$, whose ratio $t_{\rm diff}/t_{\rm sound} \sim \tau \sqrt{k_{\rm B} T_0/\mu}/c \approx 13 \gg 1$ varies by at most a factor of order unity across the domain.  The requirement to preserve the time scale ordering $t_{\rm diff} \gg t_{\rm sound}$ in this case means that for the RSLA we need $\hat{t}_{\rm diff} \gg t_{\rm sound}$, which is satisfied using $\hat{c} =10 a_0$, since $\hat{t}_{\rm diff} = (c/\hat{c}) t_{\rm diff} \gg t_{\rm diff} \gg t_{\rm sound}$.

\begin{figure}
  \centering
  \epsscale{1}
  \plotone{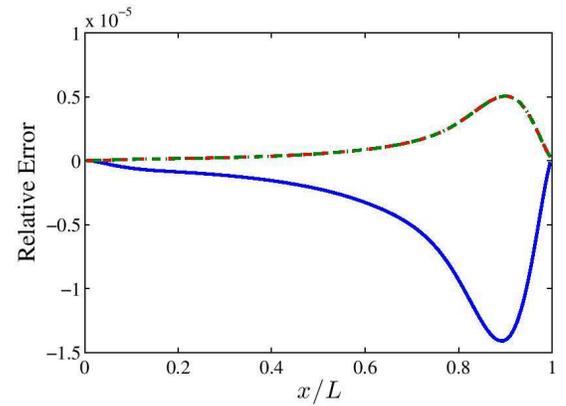}
  \caption{Relative error of the computed solutions for $\rho$ (solid), $T_{\rm gas}$ (dashed), and $T_{\rm rad}$ (dot-dashed), compared to the semi-analytic solution for the radiation tube problem.  The maximum absolute value of the relative error over the grid is $5.1 \times 10^{-6}$ for both the gas and radiation temperature solutions, and is $1.4 \times 10^{-5}$ for the density solution.  \label{radtube:relerr}}
\end{figure}

Figure~\ref{radtube:relerr} shows the relative error in the computed solutions for the gas density and for the gas and radiation temperatures compared to their semi-analytic solutions.  The maximum absolute values of the relative error over the grid is $5.1 \times 10^{-6}$ for both the gas and radiation temperatures, and is $1.4 \times 10^{-5}$ for the density.  This test demonstrates the code's ability to maintain static radiative equilibrium as well as to accurately calculate equilibrium forces in both the gas- and radiation-pressure-dominated regimes in the optically thick limit.

\vspace{1em}
{\subsubsection{RHD Linear Waves}  \label{radlinearwave}}

A rigorous test of the fully coupled, non-equilibrium system is the propagation of linear acoustic waves in a radiating medium.  For a plane-wave disturbance of the form $e^{i(k x-\omega t)}$, with wave number $k$ and frequency $\omega$, the problem can be described either for the case of the spatial damping of a driven wave ($k \in \mathbb{C}$, $\omega \in \mathbb{R}$), often called the boundary value problem (BVP), or for the case of the temporal damping of an initial disturbance ($k \in \mathbb{R}$, $\omega \in \mathbb{C}$), often called the initial value problem (IVP).  The dispersion relation for the BVP, which reduces to a complex quadratic in $k^2$, and the resulting dynamics have been extensively described by \cite{Mihalas:1999}.  Similar analysis of the dispersion relation of the IVP, which reduces to a fifth-order complex polynomial in $\omega$, is described in \cite{Lowrie:1999,Johnson:2010}.  Here, we augment the analyses of both the BVP and IVP to account for the effects of the RSLA.

We begin with the linearized equations of RHD for a medium initially at rest and in radiative equilibrium with background density $\rho_0$, gas temperature $T_0$, and sound speed $a_0$, and adopt the Eddington approximation \citep[see][Section~101]{Mihalas:1999}.  To nondimensionalize the equations, we set the density unit to the background value $\rho_0$, the length unit equal to the wavelength $\lambda$, and the time unit equal to the sound-crossing time $t_{\rm sound} \equiv \lambda/a_0$ of one wavelength.  

For the BVP, this leads to the dispersion relation
\begin{equation}
	c_4 z^4 + c_2 z^2 + c_0 = 0, \label{radlinearwave:bvp:dispersion}
\end{equation}
where $z \equiv a_0 k/\omega$, and
\begin{subequations} \label{radlinearwave:bvp:dispersion:coeffs}
	\begin{eqnarray}
		c_4 &=& 1 + i \frac{2 (\gamma-1) \mathcal{P}_0 \mathcal{C} \tau_\lambda}{\pi},  \label{radlinearwave:bvp:dispersion:coeffs:c4} \\
		c_2 &=& -\frac{3}{\widehat{\mathcal{C}}^2} \left( 1 + i \frac{\widehat{\mathcal{C}}\tau_\lambda}{2\pi} \right)^2 - 1 - i \frac{2 \gamma (\gamma-1) \mathcal{P}_0 \mathcal{C} \tau_\lambda}{\pi} \nonumber \\
		&& +\: \frac{(\gamma-1) \mathcal{P}_0 \tau_\lambda^2}{\pi^2} \left[ 2 + 3\frac{\mathcal{C}}{\widehat{\mathcal{C}}} + \frac{4}{3} \gamma \mathcal{P}_0 \right],  \label{radlinearwave:bvp:dispersion:coeffs:c2} \\
		c_0 &=& \frac{3}{\widehat{\mathcal{C}}^2} \bigg[ \left(1 + i \frac{\widehat{\mathcal{C}}\tau_\lambda}{2\pi} \right)^2 \nonumber \\
		&& +\: i \frac{2 \gamma (\gamma-1) \mathcal{P}_0 \mathcal{C} \tau_\lambda}{\pi} \left( 1 + i \frac{\widehat{\mathcal{C}}\tau_\lambda}{2\pi} \right)\bigg].  \label{radlinearwave:bvp:dispersion:coeffs:c0}
\end{eqnarray}
\end{subequations}
In Equations~\eqref{radlinearwave:bvp:dispersion:coeffs}, $\mathcal{P}_0 \equiv a_{\rm R} T_0^4 / (\rho_0 a_0^2)$ is the dimensionless pressure ratio in the equilibrium background state; $\mathcal{C} \equiv c/a_0$ and $\widehat{\mathcal{C}} \equiv \hat{c}/a_0$ are the original and reduced speeds of light, respectively, in units of the adiabatic sound speed $a_0=\sqrt{\gamma k_{\rm B} T_0/\mu}$; and $\tau_\lambda \equiv \rho\kappa_0\lambda$ is the optical depth across one wavelength of a linear disturbance propagating at the speed $a_0$.  The solutions of Equation~\eqref{radlinearwave:bvp:dispersion} are of the form $z = \pm(z_{\rm R} - i z_{\rm I})$, representing wave modes propagating in the $\pm x$-direction with phase speed $v_{\rm p}/a_0 \equiv 1/z_{\rm R}$ and spatial damping length $L_{\rm damp}/\lambda \equiv z_{\rm R}/(2\pi z_{\rm I})$.

As remarked by \cite{Jiang:2012}, the quantities $\mathcal{P}_0 \tau_\lambda$ and $\mathcal{P}_0 \mathcal{C} \tau_\lambda$, which appear as coefficients of the coupling terms in the gas momentum and energy equations when written in nondimensional form, measure the importance of momentum- and energy-exchange between the gas and radiation fields, respectively.  As either $\mathcal{P}_0 \to 0$ or $\tau_\lambda \to 0$, the gas and radiation subsystems become decoupled and wave damping by radiative processes is weak.  For $\mathcal{P}_0 \mathcal{C} \tau_\lambda \gtrsim 1$ but $\mathcal{P}_0 \tau_\lambda \ll 1$, the gas and radiation energies are strongly coupled, but the momentum carried by the radiation waves is relatively unimportant.  However, for $\mathcal{P}_0 \tau_\lambda \gtrsim 1$, momentum transport by the radiation waves becomes dominant.

The quadratic form of the dispersion relation for the BVP is generally much simpler to analyze than the fifth-order form for the IVP, and reveals some very useful information about the effects of reducing the speed of light on the phase speeds and damping rates of the various wave families in the limiting regimes.  In our code, we are primarily concerned with the behavior of damped acoustic waves propagating at or near the adiabatic sound speed, since radiation and diffusion waves propagating at or near the speed of light typically must be resolved on very small time scales on the order of the light-crossing time of a grid zone.  As we shall demonstrate, the phase speeds and damping rates of acoustic waves are not sensitive to the actual speed of propagation of the radiation in most cases, provided we preserve the temporal ordering of certain physical processes such as that of static radiative diffusion.  

First, we consider the gas-energy-dominated case $\mathcal{P}_0 \ll 1$.  Following the analysis by \cite{Mihalas:1983}, it can be shown that for the optically thin regime with $\tau_\lambda \ll 1/\max\{1,\mathcal{P}_0 \mathcal{C}\}$, Equation~\eqref{radlinearwave:bvp:dispersion} yields a weakly damped acoustic wave with
\begin{equation}
	z \approx 1 + i \frac{(\gamma-1)^2 \mathcal{P}_0 \mathcal{C} \tau_\lambda}{\pi},  \label{radlinearwave:acoustic:thin}
\end{equation}
which implies a phase speed of $v_{\rm p} \approx a_0$ and a damping length of $L_{\rm damp} \approx \lambda/[2 (\gamma-1)^2 \mathcal{P}_0 \mathcal{C} \tau_\lambda]$.  Note that Equation~\eqref{radlinearwave:bvp:dispersion} also yields a radiation mode propagating with a phase speed of $v_{\rm p} \approx \hat{c}/\sqrt{3}$ (instead of $\hat{c}$ due to the Eddington approximation), but the acoustic mode is unaffected by the RSLA in this regime.  Furthermore, for the optically thick regime with $\tau_\lambda \gg \max\{1,\mathcal{P}_0 \mathcal{C}\}$, provided $\widehat{\mathcal{C}} \gg \max\{1,\mathcal{P}_0 \mathcal{C}\}$ is also satisfied, Equation~\eqref{radlinearwave:bvp:dispersion} yields a weakly damped acoustic wave with
\begin{equation}
	z \approx 1 + i \frac{4\pi \gamma (\gamma-1) \mathcal{P}_0 \mathcal{C}}{3 \tau_\lambda},  \label{radlinearwave:acoustic:thick}
\end{equation}
which implies a phase speed of $v_{\rm p} \approx a_0$ and a damping length of $L_{\rm damp} \approx 3\lambda\tau_\lambda/[8\pi^2 \gamma (\gamma-1) \mathcal{P}_0 \mathcal{C}]$.  

Second, we consider the radiation-energy-dominated case $\mathcal{P}_0 \gg 1$.  For the optically thin regime with $\tau_\lambda \ll 1/(\mathcal{P}_0 \mathcal{C})$, Equation~\eqref{radlinearwave:bvp:dispersion} once again yields a weakly damped acoustic wave given by the solution in Equation~\eqref{radlinearwave:acoustic:thin}.  Furthermore, in the optically thick limit with $\tau_\lambda \gg \max\{1,\mathcal{P}_0 \mathcal{C}\}$, provided $\widehat{\mathcal{C}} \gg \mathcal{C}/\mathcal{P}_0$ is also satisfied, Equation~\eqref{radlinearwave:bvp:dispersion} yields a radiation-modified acoustic wave with
\begin{equation}
	z \approx \frac{3}{2} \left( \frac{\mathcal{C}}{\mathcal{P}_0 \widehat{\mathcal{C}}} \right)^{1/2} \left[ 1 + i \frac{3 \pi \mathcal{C}}{4 \mathcal{P}_0 \tau_\lambda} \right],  \label{radlinearwave:bigp:radiationacousticmode}
\end{equation}
with a phase speed of $v_{\rm p} \approx \twothirds(\mathcal{P}_0 \widehat{\mathcal{C}}/\mathcal{C})^{1/2} a_0$ and a damping length of $L_{\rm damp} \approx 2 \lambda \mathcal{P}_0 \tau_\lambda / (3 \pi^2 \mathcal{C})$.  From Equation~\eqref{radlinearwave:bigp:radiationacousticmode}, it is evident that the radiation-modified acoustic mode will \emph{always} be affected by the RSLA.  Thus, only as $\widehat{\mathcal{C}} \to \mathcal{C}$ do we recover the correct acoustic mode phase speed in this regime, which is approximately equal to the true radiation-modified acoustic speed given by
\begin{eqnarray}
	a_0^*  &\equiv& \frac{1}{3} \left[\frac{9 + 60 \mathcal{P}_0 (\gamma-1) + 16 \mathcal{P}_0^2 \gamma (\gamma-1)}{1 + 4 \mathcal{P}_0 \gamma (\gamma-1)} \right]^{1/2} a_0 \nonumber \\
	&\approx& \frac{2}{3} \mathcal{P}_0^{1/2} a_0,  \label{radlinearwave:radiationacousticspeed}
\end{eqnarray}
\citep[see][equation~45]{Lowrie:1999}.  
In this case, our algorithm is only feasible for $\widehat{\mathcal{C}} \approx \mathcal{C} \lesssim 10$ or so.

Note that for the case $\mathcal{P}_0 \ll 1$ with $\tau_\lambda \gg \max\{1,\mathcal{P}_0 \mathcal{C}\}$,\footnote{This case corresponds to ``region a'' in Figure 1 of \cite{Johnson:2010}.} the provision $\widehat{\mathcal{C}} \gg \max\{1,\mathcal{P}_0 \mathcal{C}\}$ needed to obtain Equation~\eqref{radlinearwave:acoustic:thick} follows whenever $\widehat{\mathcal{C}} \gtrsim \tau_\lambda$ is satisfied.  For $\tau_\lambda \gg 1$, the condition $\widehat{\mathcal{C}} \gtrsim \tau_\lambda$ is equivalent to the RSLA static diffusion criterion of Equation~\eqref{rsla:staticdiffreq} when the background flow velocity is negligible.  Also, note that this condition is sufficient to ensure that the relevant acoustic waves are unaltered by the RSLA, but it may be more restrictive than necessary.

For the IVP, the analysis is much more difficult, but the behavior of the acoustic wave mode in the various regimes previously discussed should mirror the behavior of this mode for the BVP.  To investigate the behavior of the IVP, we solve the corresponding dispersion relation numerically using the Newton-Raphson method.  We do this for several values of $\widehat{\mathcal{C}} \in [10, \mathcal{C}]$ in both the gas-energy- and radiation-energy-dominated cases, and examine the behavior of the phase speed and temporal damping rate as a function of optical depth per wavelength $\tau_\lambda$.  The dispersion relation to be solved is the fifth-order polynomial equation given by
\begin{eqnarray}
  (Z^2-1) Z \nonumber \\
  \times \left[ 1 - \frac{3}{\widehat{\mathcal{C}}^2} \left( Z + i \frac{\widehat{\mathcal{C}}\tau_\lambda}{2\pi} \right)^2 - i \frac{2\gamma \mathcal{P}_0 \tau_\lambda}{\pi \widehat{\mathcal{C}}} \left( Z + i \frac{\widehat{\mathcal{C}}\tau_\lambda}{2\pi} \right) \right] \nonumber \\
  + i \frac{2 (\gamma-1) \mathcal{P}_0 \mathcal{C} \tau_\lambda}{\pi} (\gamma Z^2-1) \nonumber \\
  \times \left[ 1 - \frac{3}{\widehat{\mathcal{C}}^2} \left(1 - \frac{\widehat{\mathcal{C}}}{3(\gamma-1)\mathcal{C}} \right) Z \left( Z + i \frac{\widehat{\mathcal{C}}\tau_\lambda}{2\pi} \right) \right] \nonumber \\
  + i \frac{2 (\gamma-1) \mathcal{P}_0 \tau_\lambda}{\pi \widehat{\mathcal{C}}} Z \nonumber \\
  \times \left[ i \frac{2 \gamma \mathcal{P}_0 \mathcal{C} \tau_\lambda}{3\pi} + \frac{1}{3(\gamma-1)} Z + i \frac{\mathcal{C} \tau_\lambda}{2\pi} \right] = 0, \nonumber \\ \label{radlinearwave:ivp:dispersion}
\end{eqnarray}
where $Z \equiv 1/z = \omega/(a_0 k)$, and $\tau_\lambda$, $\mathcal{P}_0$, $\mathcal{C}$, and $\widehat{\mathcal{C}}$ are defined as before.  Equation~\eqref{radlinearwave:ivp:dispersion} is a fifth-degree, complex polynomial whose roots, in general, must be found numerically.  There are 3 principal wave modes represented by the solutions to Equation~\eqref{radlinearwave:ivp:dispersion}:  an entropy mode that is always purely damped and non-propagating; an acoustic mode propagating in the $\pm x$-direction at either the adiabatic, isothermal, or radiation-modified sound speed; and a radiation mode propagating in the $\pm x$-direction at a phase speed of $\widehat{\mathcal{C}}/\sqrt{3}$ in the optically thin regime, but non-propagating in the so-called \emph{quiet regime} at larger optical depth \citep{Lowrie:1999}.  

\begin{figure}
  \centering
  \epsscale{1}
  \plotone{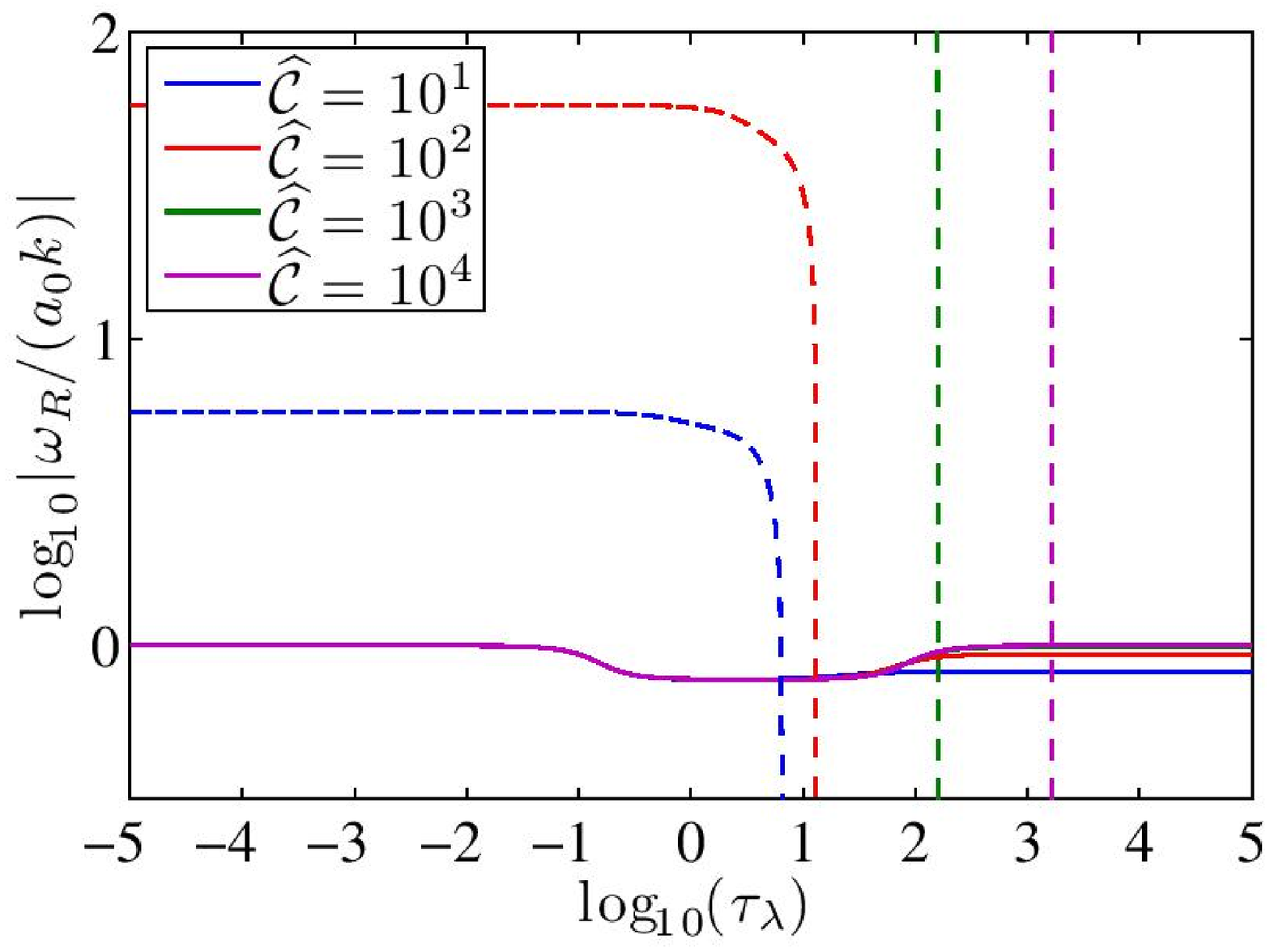}
  \plotone{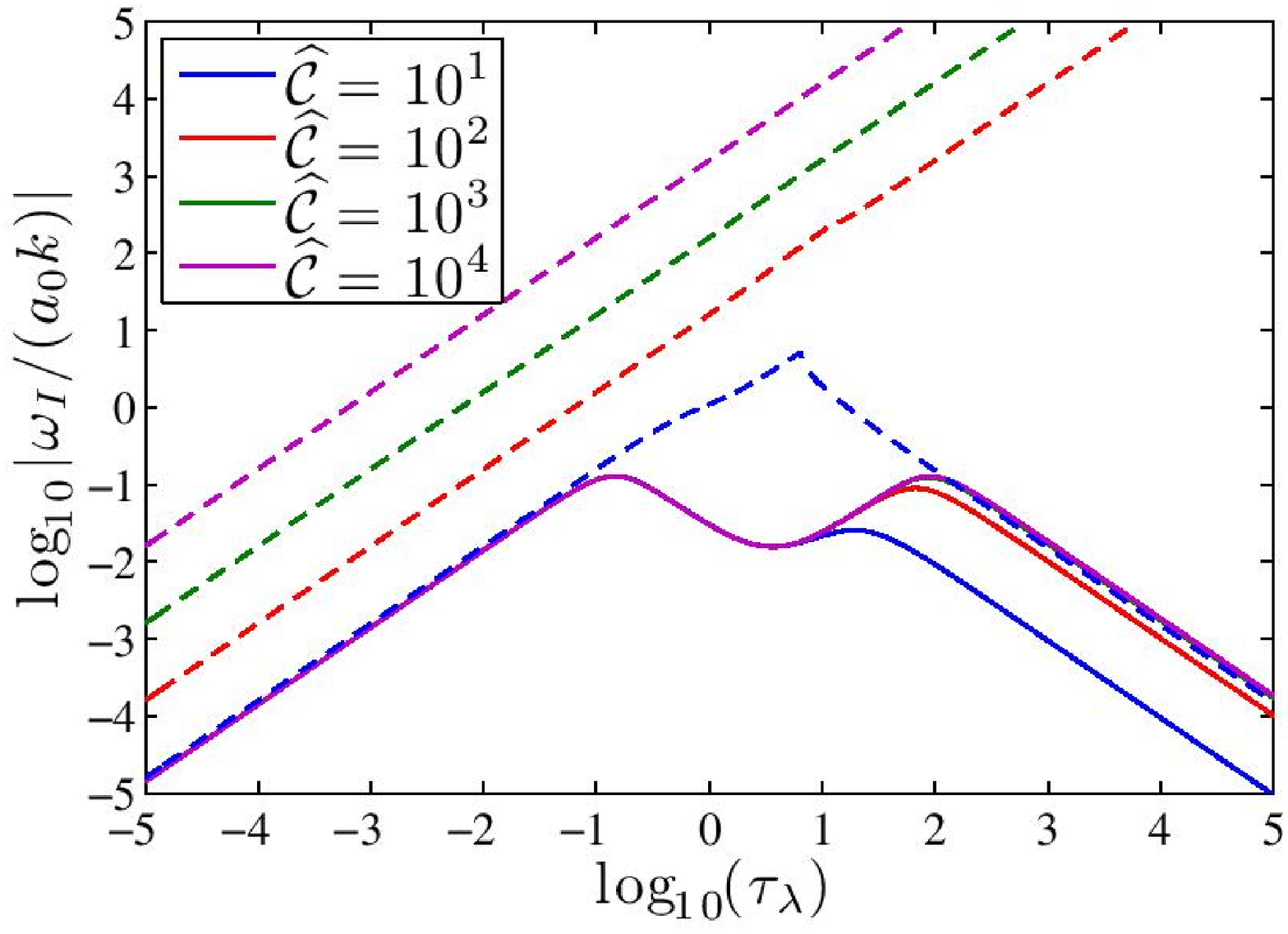}
  \caption{Phase speed (top) and damping rate (bottom) as a function of $\tau_\lambda$ for the acoustic (solid) and radiation (dashed) modes of the linear RHD wave dispersion relation (equation~\ref{radlinearwave:ivp:dispersion}) in the gas-energy-dominated case with $\mathcal{P}_0 = 10^{-3}$ and $\mathcal{C} = 10^4$.  In each plot, the value of $\widehat{\mathcal{C}}$ corresponding to each curve increases from bottom to top, from $\widehat{\mathcal{C}} = 10$ to $\widehat{\mathcal{C}} = \mathcal{C}$.  
  \label{radlinearwave:plot:smallp:dispersion}}
\end{figure}

\begin{figure}
  \centering
  \epsscale{1}
  \plotone{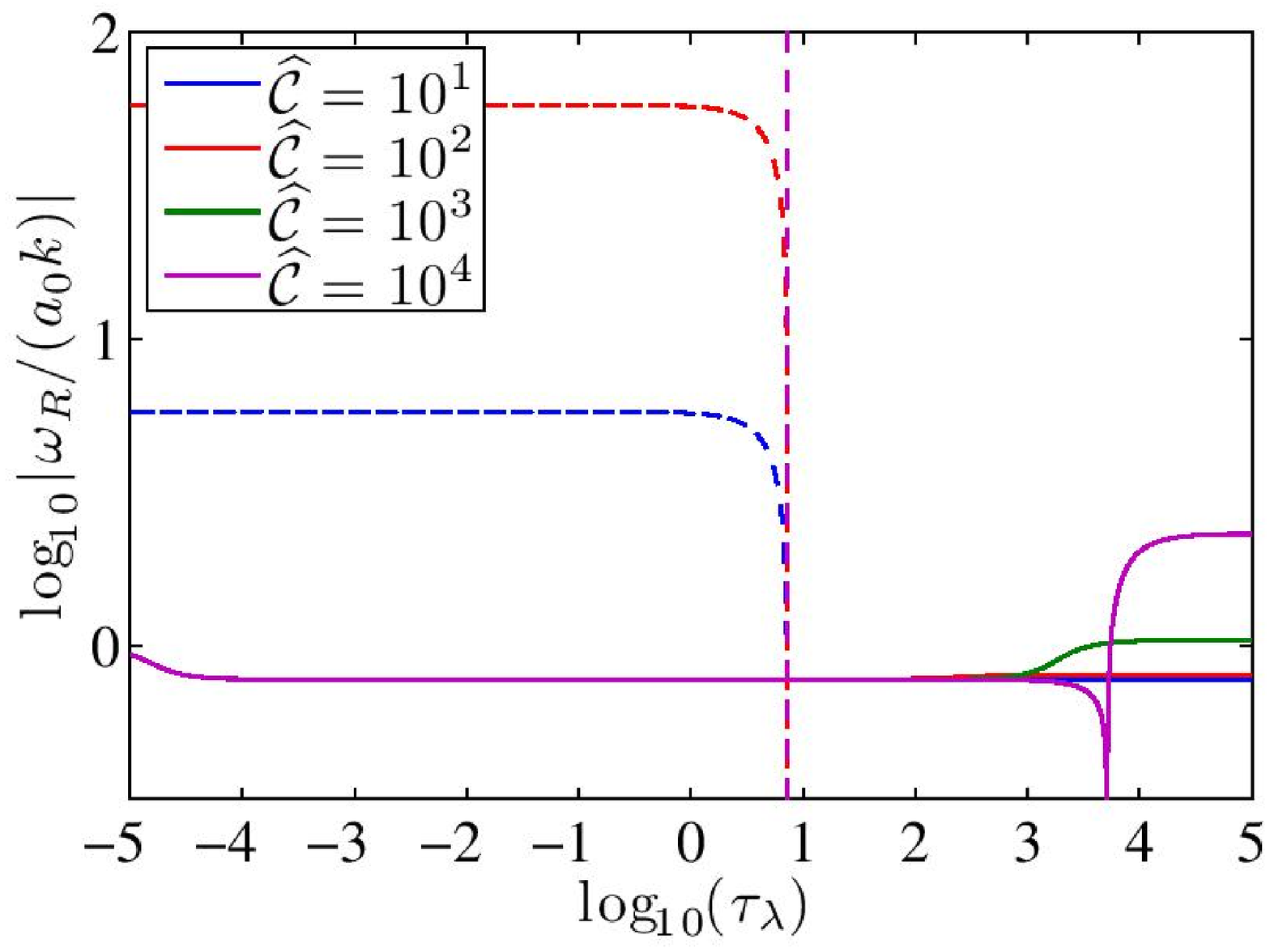}
  \plotone{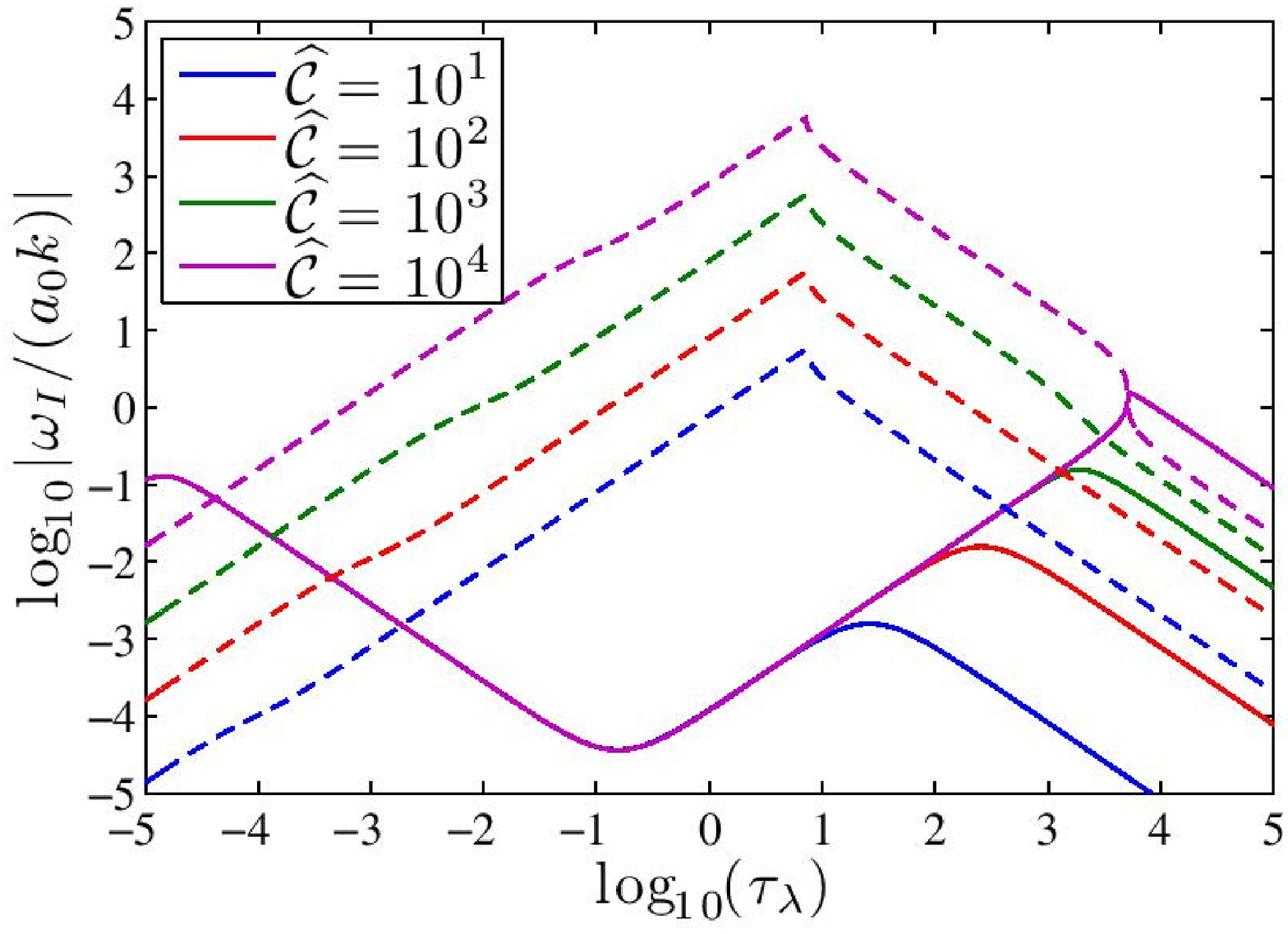}
  \caption{Same as Figure~\ref{radlinearwave:plot:smallp:dispersion} in the radiation-energy-dominated case with $\mathcal{P}_0 = 10^{1}$, $\mathcal{C} = 10^4$, and various values of $\widehat{\mathcal{C}}$ approaching $\mathcal{C}$.  In each plot, the value of $\widehat{\mathcal{C}}$ corresponding to each curve increases from bottom to top, from $\widehat{\mathcal{C}} = 10$ to $\widehat{\mathcal{C}} = 10^4 = \mathcal{C}$.  
  \label{radlinearwave:plot:bigp:dispersion}}
\end{figure}

Figure~\ref{radlinearwave:plot:smallp:dispersion} shows the phase speed $\omega_{\rm R}/(a_0 k) = {\rm Re}(Z)$ and damping rate $\omega_{\rm I}/(a_0 k) = {\rm Im}(Z)$ as a function of optical depth per wavelength $\tau_\lambda$ for the acoustic and radiation modes of the dispersion relation of Equation~\eqref{radlinearwave:ivp:dispersion} in the gas-energy-dominated case with $\mathcal{P}_0 = 10^{-3}$, $\mathcal{C} = 10^4$, $\gamma = \case{5}{3}$, and for values of the reduced speed of light ranging from $\widehat{\mathcal{C}}=10$ to $\widehat{\mathcal{C}} = \mathcal{C}$.  For small and large $\tau_\lambda$, the acoustic mode propagates at the adiabatic sound speed $a_0$ and is weakly damped, as predicted by Equations~\eqref{radlinearwave:acoustic:thin} and~\eqref{radlinearwave:acoustic:thick}, respectively, in the analysis of the BVP described above.  For $1/(\mathcal{P}_0 \mathcal{C}) \lesssim \tau_\lambda \lesssim \mathcal{P}_0 \mathcal{C}$, the acoustic mode propagates at the isothermal sound speed, $a_{\rm iso} \equiv a_0/\sqrt{\gamma}$, and is more strongly damped.  Furthermore, the phase speed and damping rate for the RSLA (i.e., $\widehat{\mathcal{C}} < \mathcal{C}$) solutions agree with the $\widehat{\mathcal{C}} = \mathcal{C}$ solution when $\widehat{\mathcal{C}}/\tau_\lambda$ is sufficiently large, as predicted by the analysis of the BVP.

Figure~\ref{radlinearwave:plot:bigp:dispersion} shows the same phase speed and damping rate plots as Figure~\ref{radlinearwave:plot:smallp:dispersion}, but in the radiation-energy-dominated case with $\mathcal{P}_0 = 10^1$.  Once more, for small $\tau_\lambda$, the acoustic mode propagates at the adiabatic sound speed $a_0$ and is weakly damped, as predicted by Equation~\eqref{radlinearwave:acoustic:thin}.  However, for large $\tau_\lambda$, the acoustic mode propagates at the phase speed predicted by Equation~\eqref{radlinearwave:bigp:radiationacousticmode}, which is approximately equal to the radiation-modified acoustic speed $a_0^*$ given in Equation~\eqref{radlinearwave:radiationacousticspeed} when $\widehat{\mathcal{C}} \approx \mathcal{C}$.  Once again, the solutions of the RSLA dispersion relation in Equation~\eqref{radlinearwave:ivp:dispersion} with $\widehat{\mathcal{C}} < \mathcal{C}$ agree with the $\widehat{\mathcal{C}} = \mathcal{C}$ solution wherever $\widehat{\mathcal{C}}/\tau_\lambda$ is sufficiently large.

To test the code, we examine the behavior of the propagating acoustic mode (either adiabatic, isothermal, or radiation-modified) of the IVP for a range of optical depths per wavelength from $\tau_\lambda = 10^{-2}$ to $\tau_\lambda = 10^2$.  We impose periodic boundary conditions on a one-dimensional domain spanning a single wavelength with grid resolution $N=512$.  We continue to use $\gamma = \case{5}{3}$ and $\mathcal{C} = 10^4$ as before, and for our first test, we use $\mathcal{P}_0 = 10^{-3}$ for the gas-energy-dominated case.  We set $\widehat{\mathcal{C}} = 10$ for $\tau_\lambda \le 10$, but for $\tau_\lambda = 100 \gg \mathcal{P}_0 \mathcal{C} = 10$, we use $\widehat{\mathcal{C}} = 100$.  The gas and radiation variables are initialized with a linear combination of right- and left-propagating eigenmodes in order to produce a standing linear wave of amplitude $10^{-6}$, and we evolve the solution for 10 wave periods.  We measure the phase speed (i.e., $\omega_{\rm R}$) and damping rate (i.e., $\omega_{\rm I}$) of the mode by regularly sampling the density solution several times per wave period at some particular anti-node (amplitude extremum) of the wave.  Once the amplitude maxima have been located in time, we measure and average the first 10 wave periods to determine the phase speed, then fit an exponential-decay envelope to the waveform to determine the damping rate.  

\begin{figure}
  \centering
  \epsscale{1}
  \plotone{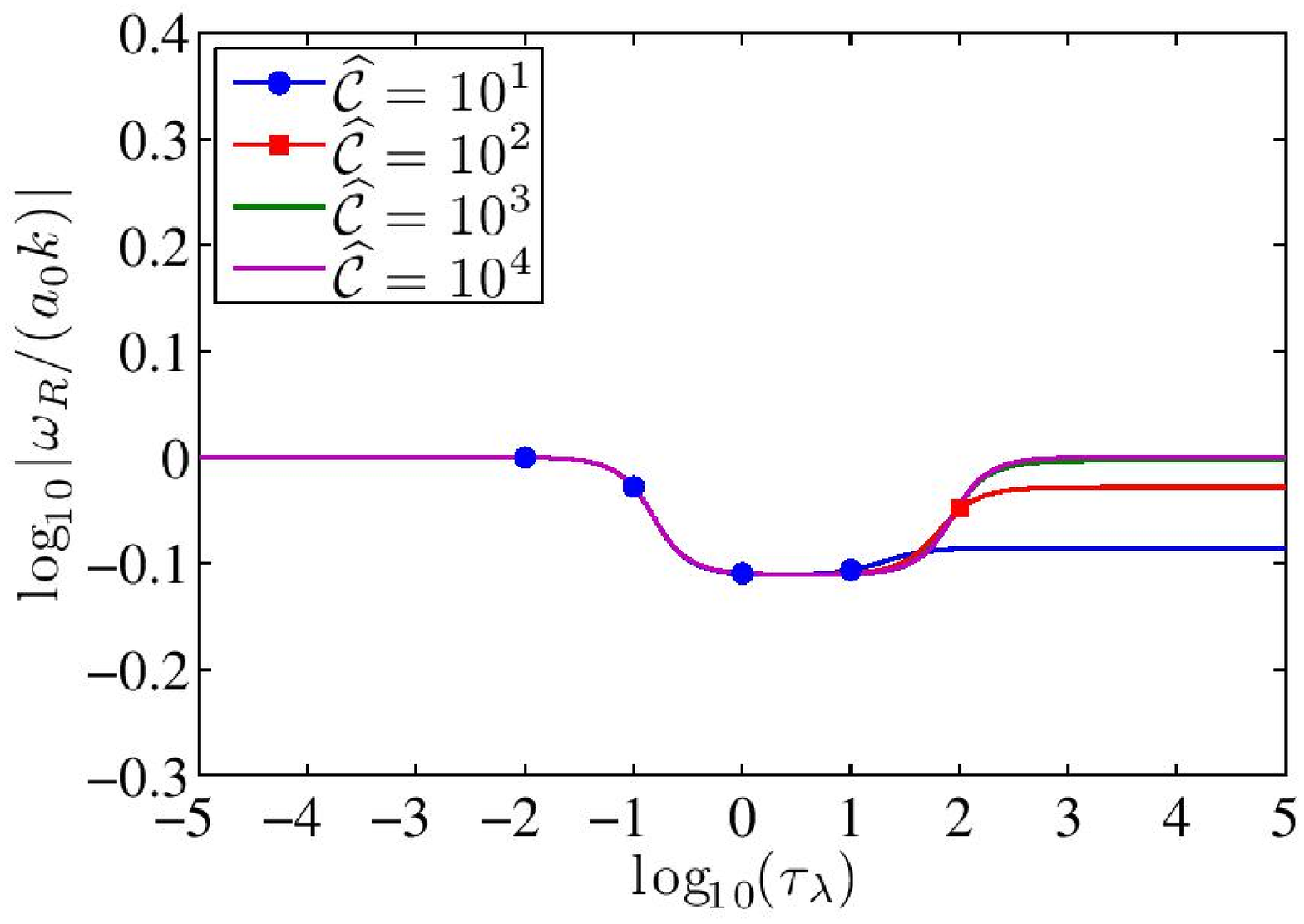}
  \plotone{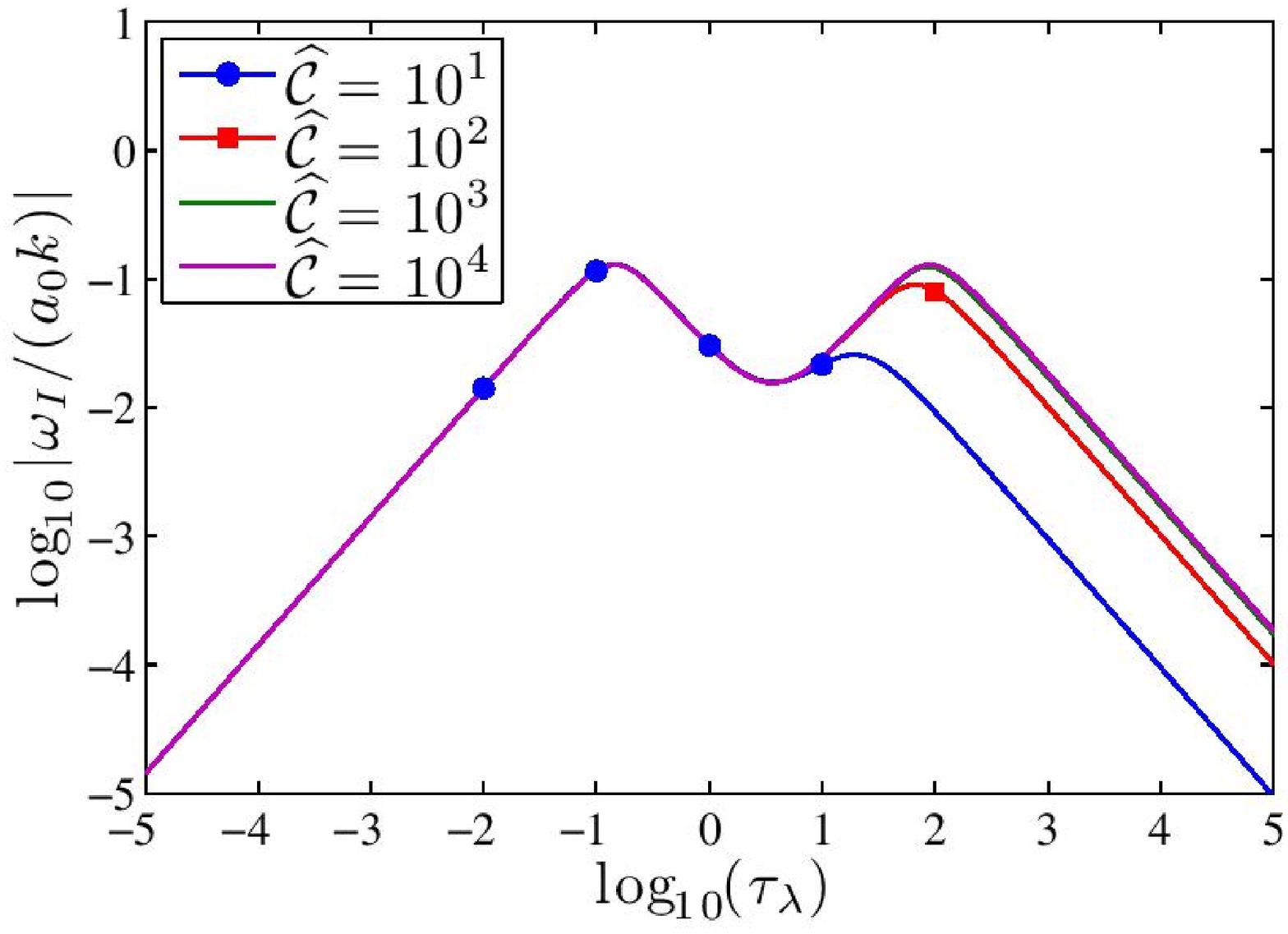}
  \caption{Computed phase speed (top) and damping rate (bottom) as a function of $\tau_\lambda$ for the acoustic mode in the gas-energy-dominated case with $\mathcal{P}_0 = 10^{-3}$, $\mathcal{C} = 10^4$, and $\gamma = \case{5}{3}$.  We use a grid resolution of $N=512$ and evolve a standing wave for 10 periods.  In each plot, we show the semi-analytic solution of Equation~\eqref{radlinearwave:ivp:dispersion} for various values of $\widehat{\mathcal{C}}$ along with the numerical results using $\widehat{\mathcal{C}} = 10$ for all $\tau_\lambda \le 10$, and $\widehat{\mathcal{C}} = 100$ for the $\tau_\lambda = 100$ case.
  \label{radlinearwave:plot:p1}}
\end{figure}

\begin{figure}
  \centering
  \epsscale{1}
  \plotone{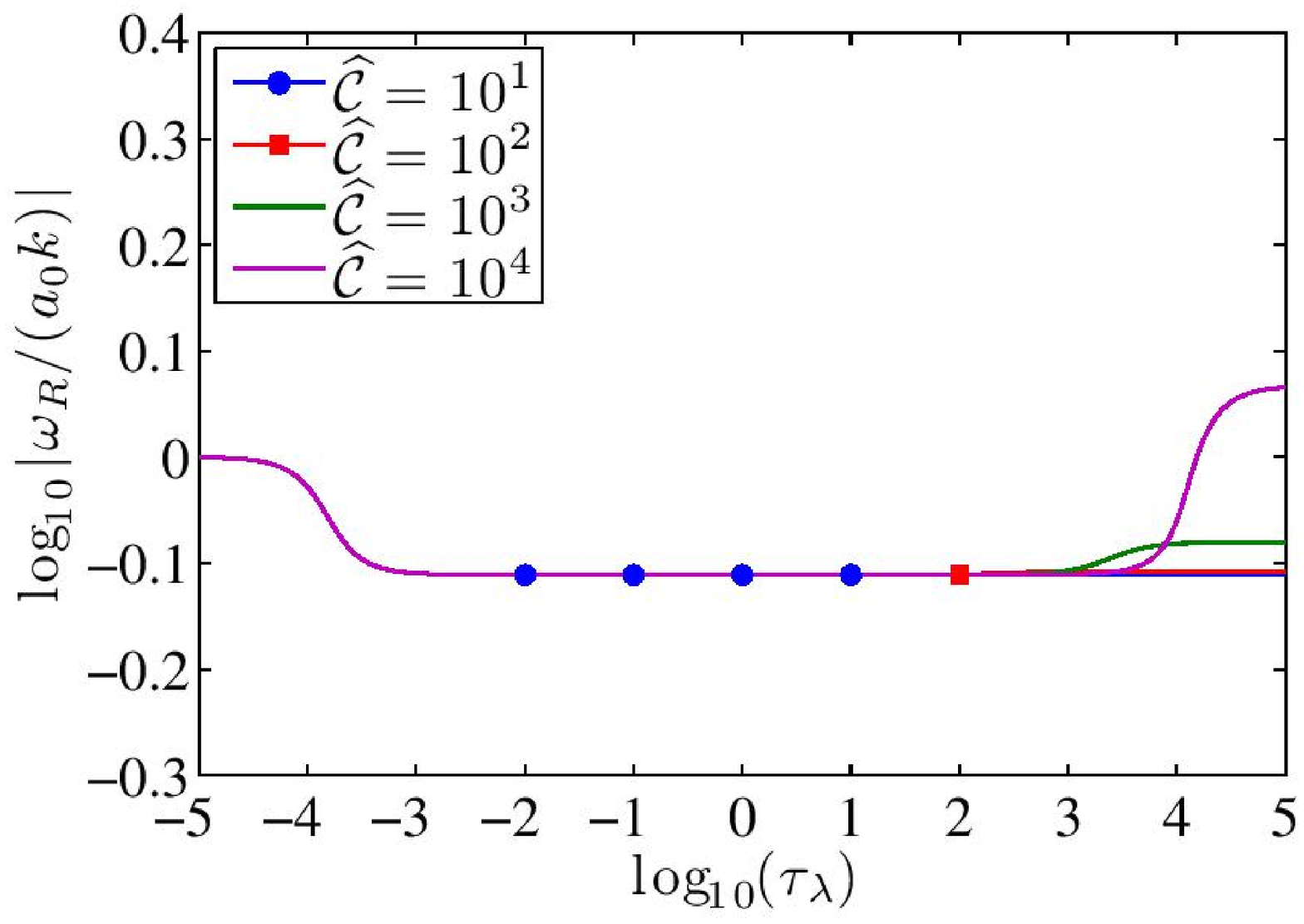}
  \plotone{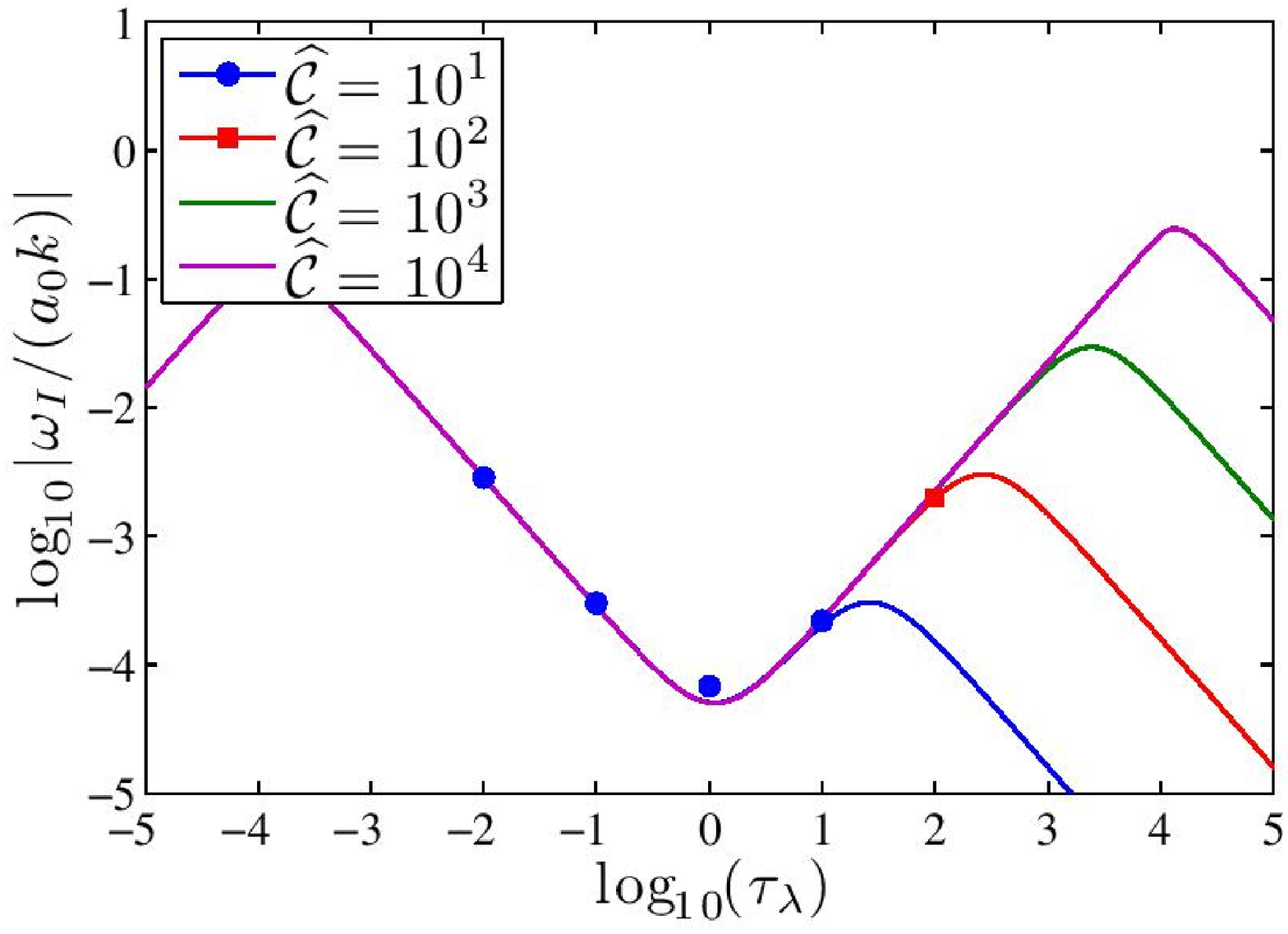}
  \caption{Same as Figure~\ref{radlinearwave:plot:p1} in the case of equal gas and radiation energies with $\mathcal{P}_0 = 1$, $\mathcal{C} = 10^4$, and $\gamma = \case{5}{3}$.  The numerical results are computed using $\widehat{\mathcal{C}} = 10$ for all $\tau_\lambda \le 10$, while for the $\tau_\lambda = 100$ case we use $\widehat{\mathcal{C}} = 100$.
  \label{radlinearwave:plot:p2}}
\end{figure}

\begin{figure}
  \centering
  \epsscale{1}
  \plotone{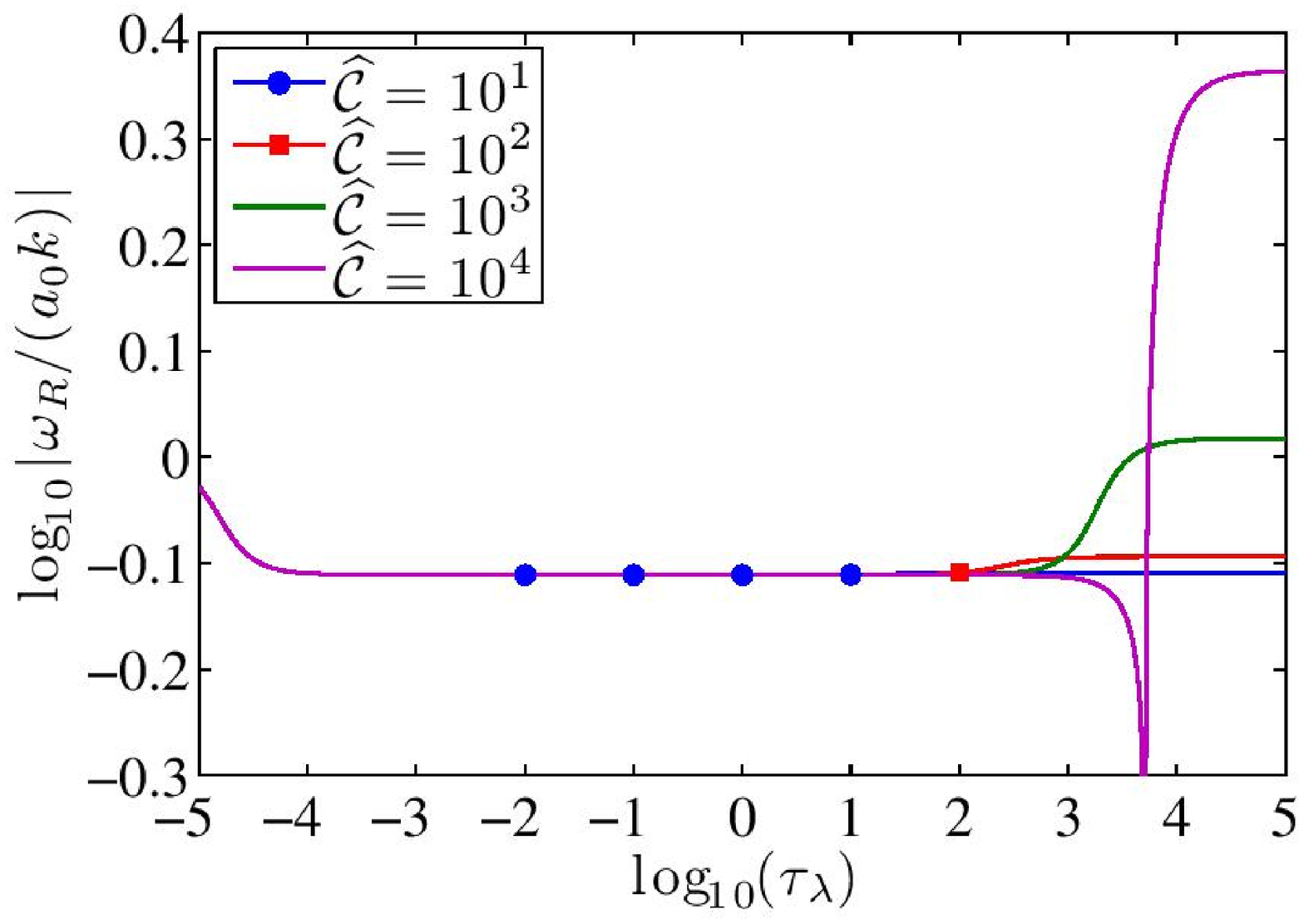}
  \plotone{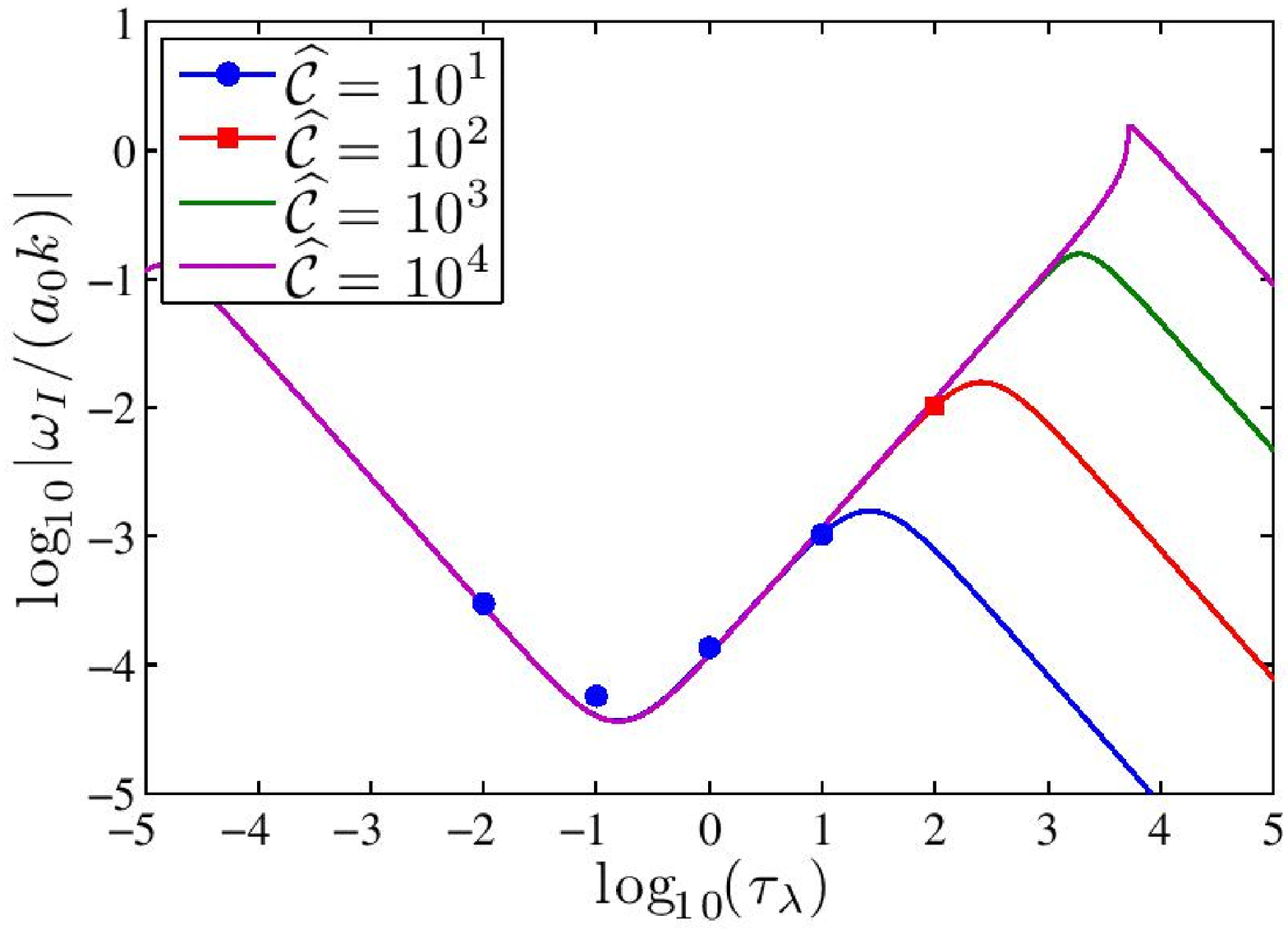}
  \caption{Same as Figure~\ref{radlinearwave:plot:p1} in the radiation-energy-dominated case with $\mathcal{P}_0 = 10^{1}$, $\mathcal{C} = 10^4$, and $\gamma = \case{5}{3}$.  The numerical results are computed using $\widehat{\mathcal{C}} = 10$ for all $\tau_\lambda \le 10$, and $\widehat{\mathcal{C}} = 100$ for the $\tau_\lambda = 100$ case.
  \label{radlinearwave:plot:p3}}
\end{figure}

\begin{figure}
  \centering
  \epsscale{1}
  \plotone{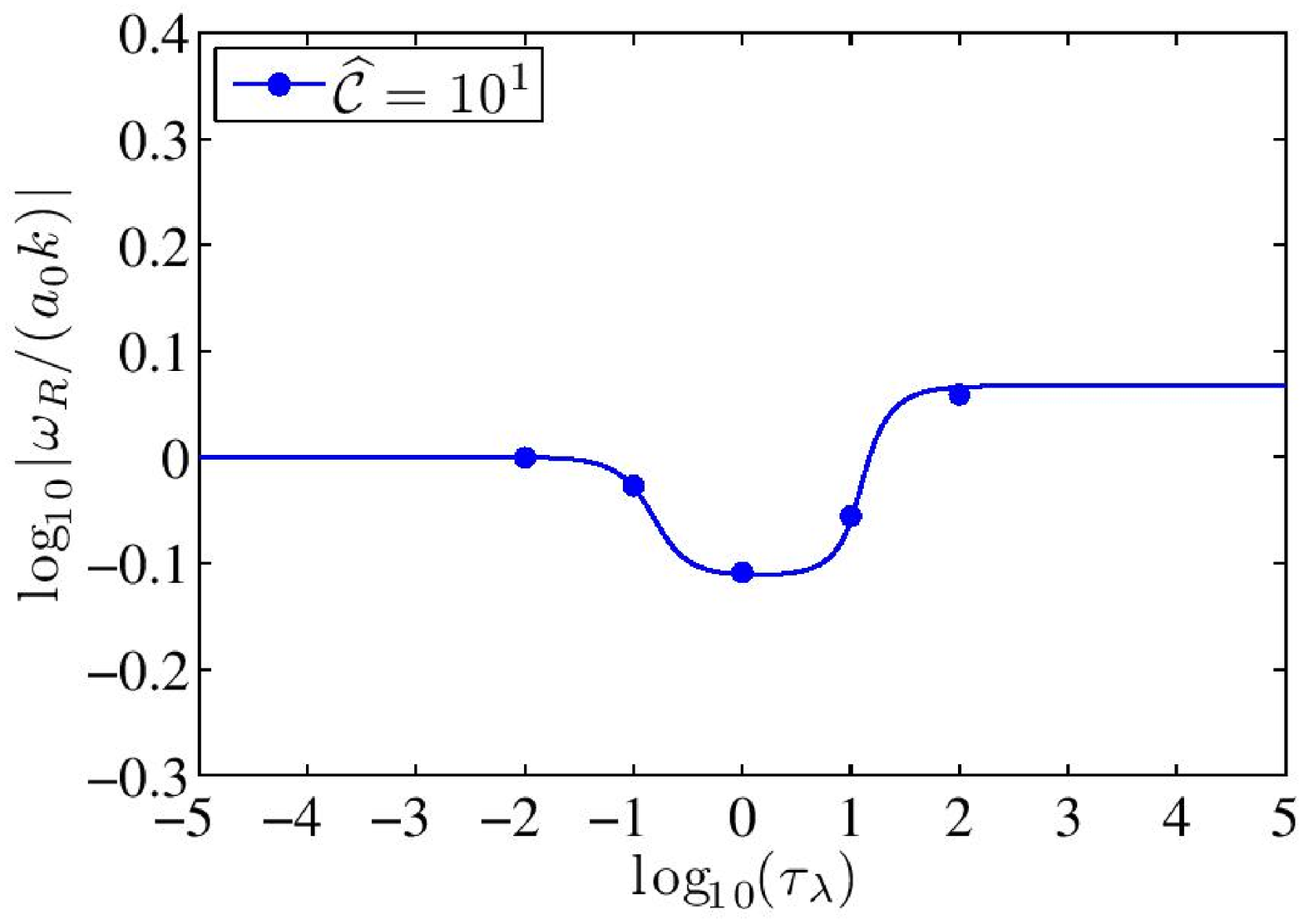}
  \plotone{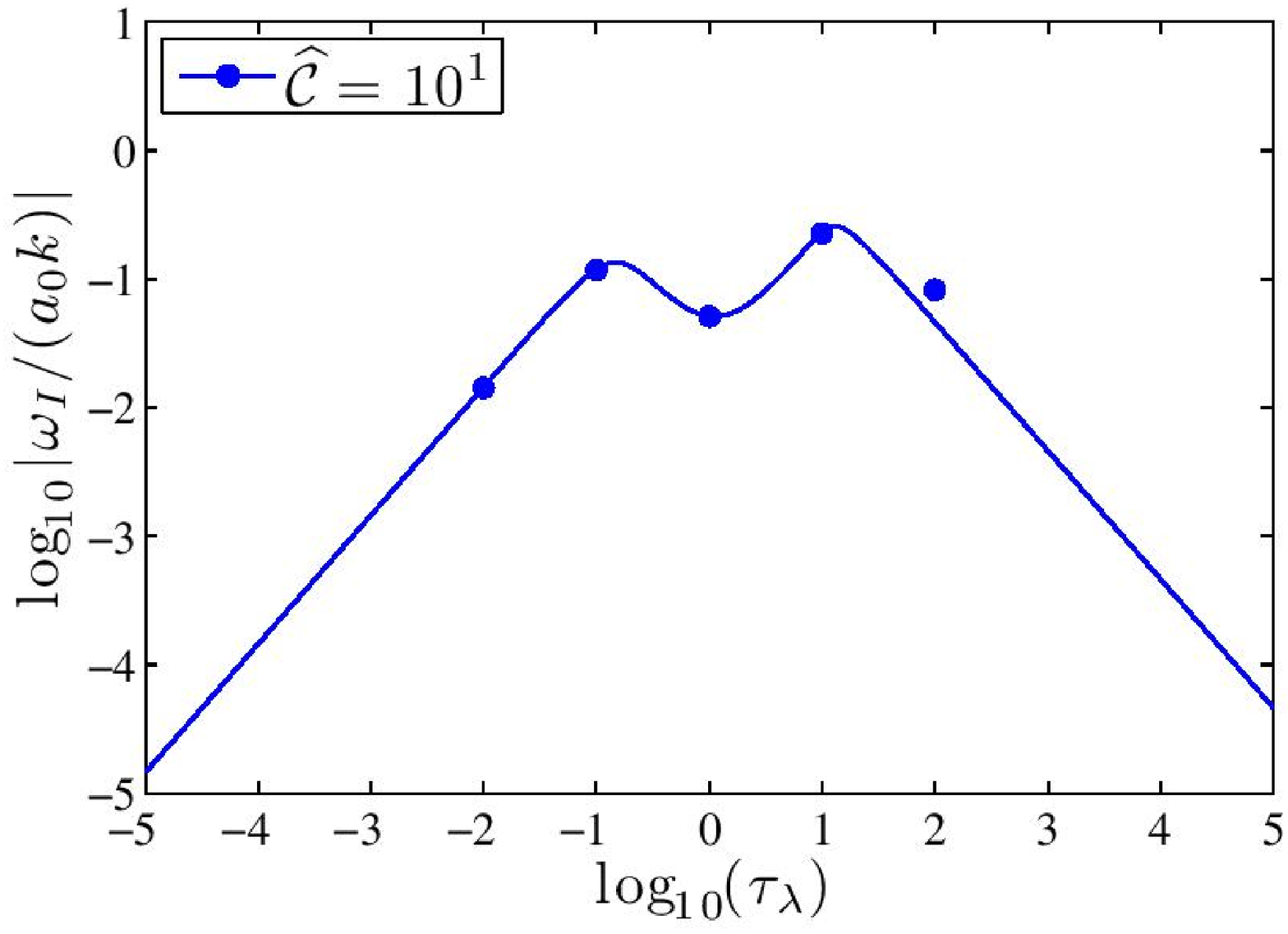}
  \caption{Same as Figure~\ref{radlinearwave:plot:p1} in the relativistic case with $\mathcal{P}_0 = 1$, $\mathcal{C} = 10$, and $\gamma = \case{5}{3}$.  The numerical results are computed using $\widehat{\mathcal{C}} = 10 = \mathcal{C}$ for all $\tau_\lambda$.  \label{radlinearwave:plot:p4}}
\end{figure}

Figure~\ref{radlinearwave:plot:p1} shows the computed phase speed and damping rate for each value of $\tau_\lambda$ considered.  For each $\tau_\lambda \le 10$, our computations adopt $\widehat{\mathcal{C}} = 10$.  The resulting values of $\omega_{\rm R}$ and $\omega_{\rm I}$ are in good agreement with the semi-analytic solution of Equation~\eqref{radlinearwave:ivp:dispersion} for $\widehat{\mathcal{C}} = 10$.  In addition, these computed values agree with the solution of Equation~\eqref{radlinearwave:ivp:dispersion} for $\widehat{\mathcal{C}} = 10^4 = \mathcal{C}$, i.e., the true solution.  For $\tau_\lambda \gtrsim 10$, the solution of the dispersion relation using $\widehat{\mathcal{C}} = 10$ departs significantly from the true ($\widehat{\mathcal{C}}=10^4$) solution, while our $\widehat{\mathcal{C}} = 100$ semi-analytic solution remains close for large $\tau_\lambda$.  By using $\widehat{\mathcal{C}} = 100$ for our numerical computation at $\tau_\lambda = 100$, we obtain a phase speed and damping rate consistent with the semi-analytic $\widehat{\mathcal{C}} = 100$ solution and close to the $\widehat{\mathcal{C}} = 10^4 = \mathcal{C}$ solution.  We can not expect to obtain better values for large $\tau_\lambda$ in this test, since $\widehat{\mathcal{C}} = 100$ is at the limit of what we can feasibly do with our algorithm.  

Next, we repeat this experiment for $\mathcal{P}_0 = 1$ and $\mathcal{P}_0 = 10$ in order to consider cases where the gas and radiation energies are similar and where the radiation energy is dominant.  The results for these cases are shown in Figures~\ref{radlinearwave:plot:p2} and~\ref{radlinearwave:plot:p3}, respectively.  We use all of the same parameters as the previous test, including the same values of $\tau_\lambda$ and the corresponding values of $\widehat{\mathcal{C}}$.  Again, we find that the numerical results agree with the solutions of Equation~\eqref{radlinearwave:ivp:dispersion} and that provided $\widehat{\mathcal{C}} \gtrsim \tau_\lambda$, the RSLA solution agrees with the $\widehat{\mathcal{C}} = \mathcal{C}$ solution.  This is fortunate, since $\mathcal{P}_0 \tau_\lambda \gg 1$ indicates that the gas and radiation momenta are strongly coupled, and the fact that our algorithm employs an operator splitting between the solutions of the gas and radiation subsystems means that there is no guarantee that the combined momentum will be conserved.  There is some discrepancy in both of these tests for the computed values of the smallest damping rates, which are on the order of $10^{-4}$.  This discrepancy may be due to numerical diffusion or to measurement error, since the waves are hardly damped at all in 10 periods.

From the above tests, we conclude that provided $\widehat{\mathcal{C}} \gg \max\{1,\tau_\lambda\}$, which is equivalent to Equation~\eqref{rsla:staticdiffreq} for $v_{\rm max} = a_0$, the RSLA does not affect the character of linear waves, and both propagation speeds and damping rates are recovered using our numerical code.  

In our last test, we consider a nearly relativistic gas with $\mathcal{C} = 10$, in which case we can use $\widehat{\mathcal{C}} = \mathcal{C}$ in order to investigate the behavior of the linear waves when the RSLA is not employed.  We use $\mathcal{P}_0 = 1$, and set all other parameters as before.  The results shown in Figure~\ref{radlinearwave:plot:p4} indicate that there is some error for the largest value of $\mathcal{P}_0 \tau_\lambda$, which must be caused by the splitting error.  Nonetheless, the algorithm does compute the radiation-modified acoustic speed reasonably well in this case.  

\vspace{1em}
{\subsubsection{Radiative Shocks}  \label{radshock}}

Another test of the fully coupled, non-equilibrium system is the shock of a cold, optically thick medium in the presence of radiation.  The classical analysis is described by \cite{Zeldovich:2002} and \cite{Mihalas:1999}.  More recently, \cite{Lowrie:2008} have described a semi-analytic method for obtaining the full family of non-equilibrium shock solutions parameterized by the shock Mach number.

In the diffusion limit, the non-equilibrium radiation energy equation in non-dimensional form reduces to \citep[see][Section~97]{Mihalas:1999}
\begin{eqnarray}
  \frac{c}{\hat{c}} \partial_t T_{\rm rad}^4 + \frac{4}{3} \partial_x (vT_{\rm rad}^4) - \partial_x \left( \frac{c}{3\rho\kappa_0} \partial_x T_{\rm rad}^4 \right) \nonumber \\
  = c\rho\kappa_0 (T_{\rm gas}^4 - T_{\rm rad}^4) + \frac{1}{3} v \partial_x T_{\rm rad}^4, \label{radshock:radenergy}
\end{eqnarray}
where $a_{\rm R} T_{\rm rad}^4 \equiv \mathcal{E}_0$ defines $T_{\rm rad}$, the comoving-frame radiation temperature.  Note that a steady-state solution satisfying Equation~\eqref{radshock:radenergy} will not depend on $\hat{c}$.  The corresponding hydrodynamic equations in non-dimensional form are given by
\begin{eqnarray}
  \partial_t \rho + \partial_x (\rho v) = 0, \\
  \partial_t (\rho v) + \partial_x \left( \rho v^2 + P + \frac{1}{3} \mathcal{P}_0 T_{\rm rad}^4 \right) = 0, \\
  \partial_t E + \partial_x [(E+P)v] \nonumber \\
  = -\mathcal{P}_0 \left[ c\rho\kappa_0 (T_{\rm gas}^4-T_{\rm rad}^4) + \frac{1}{3} v \partial_x T_{\rm rad}^4 \right], \label{radshock:gasenergy}
\end{eqnarray}
where $\mathcal{P}_0 \equiv a_{\rm R} T_0^4/(\rho_0 a_0^2)$ is the dimensionless pressure ratio, and $T_0$, $\rho_0$, and $a_0$ are the gas temperature, density, and adiabatic sound speed, respectively, in the upstream state.  For given values of $\gamma$, $\kappa_0$, and $\mathcal{P}_0$, the structure of the shock solution can be entirely characterized by the upstream Mach number, $\mathcal{M}_0$.

We use the non-dimensional parameters $\gamma = \case{5}{3}$, $\kappa_0 = 1$, $\mathcal{P}_0 = 1 \times 10^{-4}$, and $\mathcal{M}_0 = 3$ to set the upstream state, then calculate the downstream state according to the Rankine--Hugoniot jump conditions \citep[see][equation~8]{Lowrie:2008}.  We use Dirichlet boundary conditions for the gas and radiation variables on a one-dimensional grid whose size is determined by setting the fractional temperature change equal to $\Delta T/T_0 = 10^{-4}$ in both the upstream and downstream states.  We use a resolution of $N=1024$ zones and allow the solution to evolve for several sound-crossing times until the radiative shock solution has reached a steady state.  Although we compute the solution in the rest frame of the shock with the interface initially located at the origin, the accumulation of small numerical errors will cause the computed shock solution to migrate by a small amount, much less than the size of the computational domain.  It is therefore necessary to track the shock front, which can be done by minimizing the relative residuals of the hydrodynamic jump conditions in a manner similar to the method described by \cite{Lowrie:2008}.

\begin{figure}
  \centering
  \epsscale{1}
  \plotone{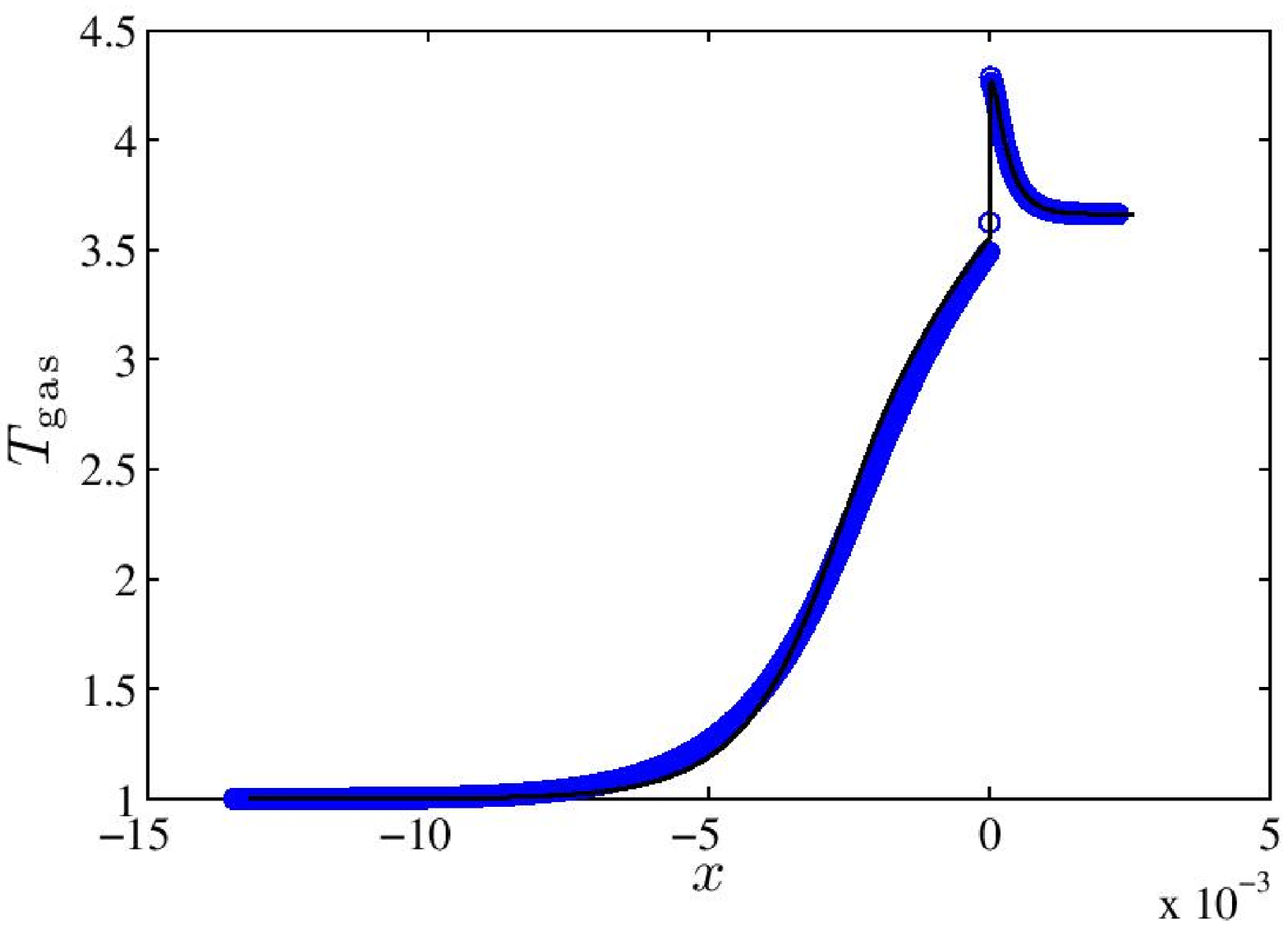}
  \plotone{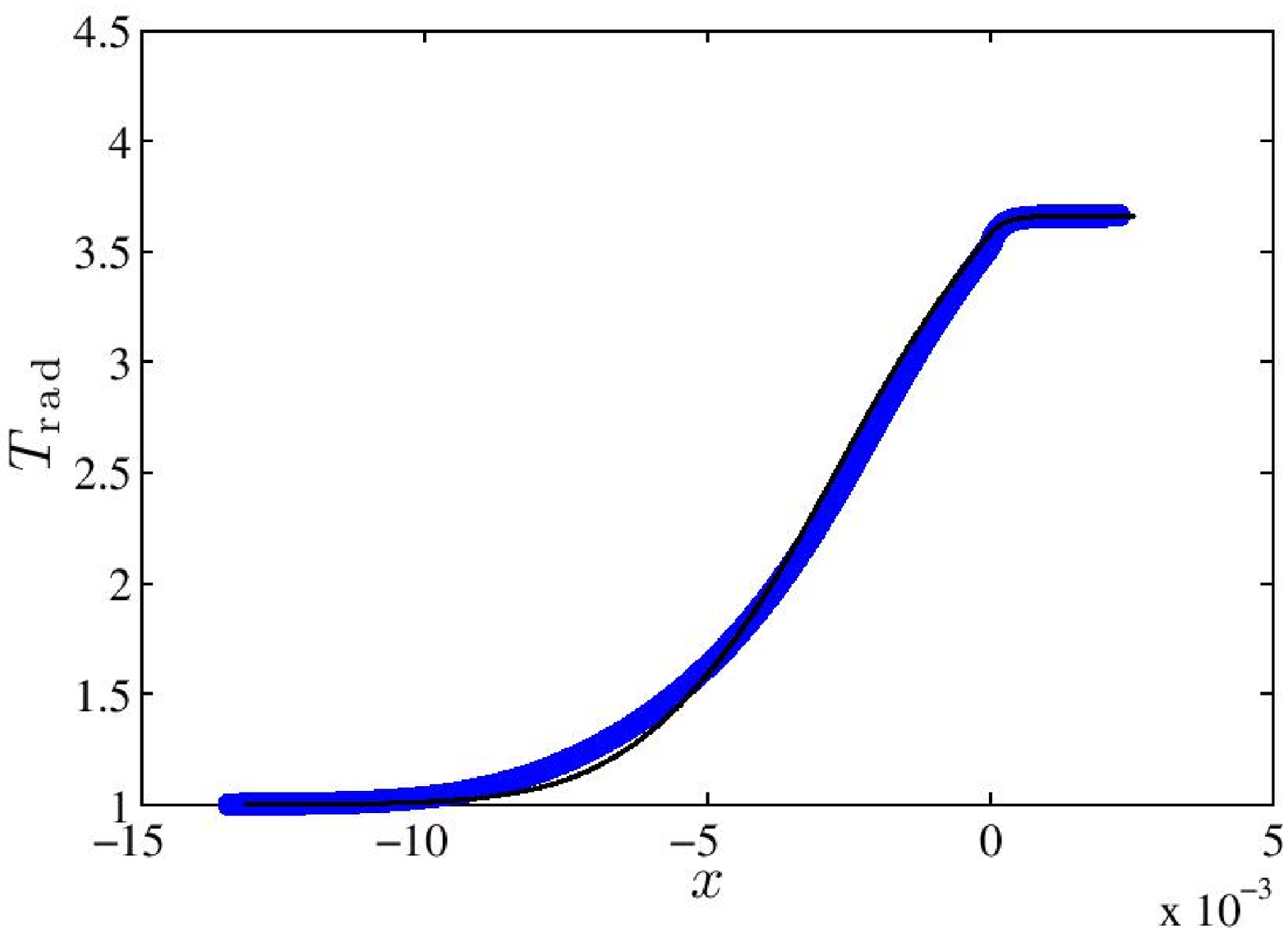}
  \caption{Semi-analytic (solid line) and computed (circles) solutions of the gas (top) and radiation (bottom) temperature profiles for a sub-critical, non-equilibrium radiative shock with $\mathcal{M}_0=3$.  \label{radshock:temperature}}
\end{figure}

\begin{figure}
  \centering
  \epsscale{1}
  \plotone{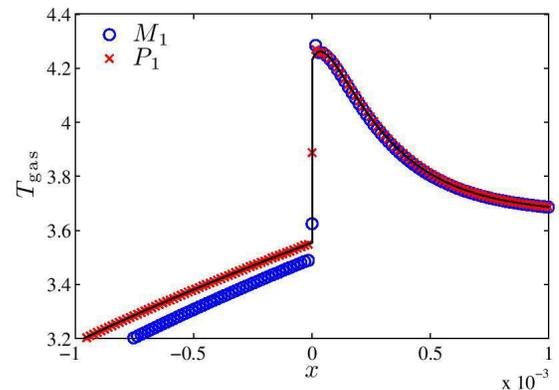}
  \caption{Detail of the Zel'dovich spike in the semi-analytic (solid line) and numerical (circles) solutions of the gas temperature profile shown in Figure~\ref{radshock:temperature} using the $M_1$ model.  For reference, we also include the gas temperature profile numerically obtained using the Eddington ($P_1$) approximation (exes) instead of computed eigenvalues.  \label{radshock:spike}}
\end{figure}

Figure~\ref{radshock:temperature} shows the gas and radiation temperature profiles for the non-equilibrium radiative shock with $\mathcal{M}_0=3$.  This value of the upstream Mach number yields a sub-critical shock, i.e., the gas temperature in the radiatively heated shock precursor is less than the downstream value, for this parameter set.  Figure~\ref{radshock:spike} shows the detail of the Zel'dovich spike in the gas temperature profile of the upstream solution near the shock front.  These figures show good agreement with the semi-analytic solution, although there is some resolution-independent discrepancy in the gas and radiation temperatures of the shock precursor for the solution using computed eigenvalues.  We ran the same test with the eigenvalues fixed at $\lambda_{1,3} = \mp \hat{c}/\sqrt{3}$ (see Section~\ref{m1}) according to the Eddington approximation and include these results in Figure~\ref{radshock:spike} for comparison with the computed-eigenvalue solution.

The relative errors of the computed-eigenvalue solution with respect to the semi-analytic model are $1.7\%$ for the density, $6.1\%$ for the gas temperature, and $7.8\%$ for the radiation temperature, except at the shock interface where the solution is discontinuous.  For the fixed-eigenvalue solution, the relative errors are $0.42\%$ for the density, $0.49\%$ for the gas temperature, and $0.42\%$ for the radiation temperature.  Tests with other Mach numbers also agree well with the semi-analytic solution.  This suggests that the error observed in Figure~\ref{radshock:temperature}, i.e., for the computed-eigenvalue solution, is primarily a result of the Eddington approximation in the semi-analytic model \citep{Lowrie:2008}, not of the RSLA, which is expected to have no effect on the spatial variation of steady-state solutions (see Section~\ref{rsla}).  

\vspace{1em}
{\subsubsection{Radiative Momentum-driven Expanding Shell} \label{radshell}}

As a final test, we consider the evolution of a spherical, dusty shell of gas with expansion driven by absorption of radiation momentum from a central source.  The problem set-up is a modified version of that described in Appendix A of \cite{Ostriker:2011}.  For simplicity, here we assume an isothermal equation of state under the condition of radiative equilibrium and neglect the gravitational potential.  The problem considers an idealized GMC of mass $M_{\rm GMC}$ that forms stars of total mass $M_*$ with efficiency $\varepsilon_{\rm GMC} = M_*/M_{\rm GMC}$ over its lifetime.  The remaining gas of mass $M_{\rm sh} \equiv (1-\varepsilon_{\rm GMC})M_{\rm GMC}$ is ejected as an expanding, spherical shell of (variable) radius $r$ due to the radiation force from the stellar component, which we model here as a centrally located cluster with (fixed) radius $r_*$ and luminosity per unit mass $\Psi \equiv L_*/M_*$ typical of young, luminous clusters.  Here, as in \cite{Ostriker:2011}, we consider just the effects of reprocessed IR continuum radiation, as the corresponding radiation force exceeds that of the primary UV/optical streaming photons by a factor $\sim \tau_{\rm IR}$ when the dusty shell is optically thick.  The source function for the central luminous cluster is given by
\begin{equation}
	j_*(r) = \frac{L_*}{(2\pi \sigma_*^2)^{3/2}} \exp \left(-\frac{r^2}{2\sigma_*^2}\right).  \label{radshell:sourceprofile}
\end{equation}

Assuming the ejected shell is thin, i.e., $H \ll r$ for a shell of thickness $H$ at radius $r$, the volume of the shell is approximately $V(r) \approx 4\pi r^2 H$, the density of the shell is approximately $\rho_{\rm sh}(r) \approx M_{\rm sh}/(4 \pi r^2 H)$, and the optical depth across the shell is approximately $\tau_{\rm sh}(r) \approx M_{\rm sh} \kappa_0/ (4 \pi r^2)$, where $\kappa_0$ is the absorption opacity of the dust, which is hydrodynamically coupled to the gas, to infrared radiation.  For $\tau_{\rm sh} \gtrsim 1$, the diffuse radiation reprocessed by the dust applies a force $F_{\rm rad} \approx L_* \tau_{\rm sh}/c$ on the shell.  Neglecting gravitational and internal pressure forces, the outward acceleration of the shell is
\begin{equation}
	\ddot{r} = \frac{L_* \kappa_0}{4\pi r^2 c},  \label{radshell:odeaccel}
\end{equation}
which is independent of the shell's thickness.\footnote{Note that the inward gravitational acceleration of the shell, arising from both the gravitational force of the central cluster as well as the self-gravitational force of the shell itself, has the same $r^{-2}$ dependence as the acceleration given in Equation~\eqref{radshell:odeaccel}; hence, the net acceleration may be reduced correspondingly depending on the relative strengths of the gravitational and radiation forces.  \citet[][see their Equation A1]{Ostriker:2011} point out that in this case the shell can become unbound only if $\epsilon_{\rm GMC} > \epsilon_{\rm min} \equiv [\Psi\kappa_0/(2\pi c G) - 1]^{-1}$; \citet[][see their Equation 17]{Murray:2010} reached a similar conclusion. \label{radshell:epsmin}}  With $\ddot{r}$ given by Equation~\eqref{radshell:odeaccel}, the characteristic dynamical time for acceleration is given by
\begin{equation}
t_{\rm dyn} \equiv \left(\frac{\ddot{r}}{r}\right)^{-1/2} = \left(\frac{L_* \kappa_0}{4\pi r^3 c}\right)^{-1/2},
\end{equation}
and the characteristic velocity produced is given by
\begin{equation}
v_{\rm dyn} \equiv \frac{r}{t_{\rm dyn}} = \left(\frac{L_* \kappa_0}{4\pi r c}\right)^{1/2}.  \label{radshell:vdyn}
\end{equation}

To non-dimensionalize the problem, we set the length unit to $r_0$, set the density unit to $\rho_0 \equiv M_{\rm sh}/(\case{4}{3} \pi r_0^3)$, the density of a uniform spherical cloud of gas with radius $r_0$ and mass $M_{\rm sh}$, and the speed unit to $a_0$, the isothermal sound speed.  The corresponding time unit is $t_0 \equiv r_0/a_0$.

Choosing dimensional parameters $\Psi = 2000 \text{ erg}\text{ s}^{-1}\text{ g}^{-1}$, $M_{\rm GMC} = 10^6 M_\odot$, $\kappa_0 = 20 \text{ cm}^2\text{ g}^{-1}$, together with $r_0 = 5\text{ pc}$, $a_0 = 2 \times 10^5 \text{ cm}\text{ s}^{-1}$, and with an efficiency of $\varepsilon_{\rm GMC}=0.5$,\footnote{Although we neglect gravitational forces here, this value of $\epsilon_{\rm GMC}$ corresponds to the minimum efficiency that would be required for the shell to become unbound if gravitational forces were included in the net acceleration using the given parameter set.  See Footnote~\ref{radshell:epsmin}.} we obtain the reference dynamical Mach number
\begin{eqnarray}
	\mathcal{M}_0 &\equiv& \frac{v_{\rm dyn}(r_0)}{a_0} = \left( \frac{L_* \kappa_0}{4\pi r_0 c a_0^2} \right)^{1/2} \nonumber \\ 
	&\approx& 13 \Bigg[ \frac{(\varepsilon_{\rm GMC}/0.5) (\Psi/2000 \text{ erg}\text{ s}^{-1}\text{ g}^{-1})}{(r_0/5\text{ pc})(a_0/2 \times 10^5 \text{ cm}\text{ s}^{-1})^2} \nonumber \\
	&& \times (M_{\rm GMC}/10^6 M_\odot)(\kappa_0/ 20 \text{ cm}^2\text{ g}^{-1})\Bigg]^{1/2},  \label{radshell:mach0}
\end{eqnarray}
and the reference optical depth across a thin shell of mass $M_{\rm sh}$ at radius $r_0$ by
\begin{eqnarray}
	\tau_0 &\equiv& \tau_{\rm sh}(r_0) = \frac{M_{\rm sh} \kappa_0}{4\pi r_0^2} \nonumber \\
	&\approx& 6.6 \Bigg[ \frac{([1 - \varepsilon_{\rm GMC}]/0.5) (M_{\rm GMC}/10^6 M_\odot)}{(r_0/5\text{ pc})^2}  \nonumber \\
	&& \times (\kappa_0/ 20 \text{ cm}^2\text{ g}^{-1}) \Bigg].  \label{radshell:tau0}
\end{eqnarray}

We begin with a spherical cloud of uniform density $\rho_{\rm cl} = M_{\rm sh}/(\case{4}{3} \pi r_0^3) = 3\rho_0(H/r_0)$, and at time $t=0$, we turn on a source with emission profile given by Equation~\eqref{radshell:sourceprofile} in order to investigate the premise that the ejected gas forms a thin shell.  In a time $t_{\rm dyn}/t_0 = \mathcal{M}_0^{-1} \approx 0.076$, a thin shell should form near the radius $r_0$.  Since the problem is spherically symmetric, we can simplify the full three-dimensional problem by restricting our computation to the octant $(x,y,z) \ge 0$ and by imposing reflection boundary conditions on the inner boundaries, effectively doubling the resolution.  

According to the RSLA static diffusion criterion of Equation~\eqref{rsla:staticdiffreq}, we should choose a value of $\hat{c}/a_0$ such that $v_{\rm dyn} \sim a_0\mathcal{M}_0 \ll \hat{v}_{\rm diff} \sim \hat{c}/\tau_{\rm cl}$ for this test, where $\tau_{\rm cl} = \rho_{\rm cl} \kappa_0 r_0 = 3 \tau_0$ is the optical depth from the center of the cloud, i.e., such that $\hat{c}/a_0 \gg 3 \mathcal{M}_0 \tau_0$.  Yet for $\mathcal{M}_0 = 13$ and $\tau_0 = 6.6$, this would require $\hat{c}/a_0 \gg 260$, which is impractically large.  Instead, we choose $\hat{c}/a_0 = 260$, the consequence of which is that $v_{\rm diff}$ may not be large enough compared to $v_{\rm dyn}$ for the radiation to properly diffuse through the medium, hence the radiation force acting on the shell may initially be too large.  However, we relax this requirement here since we are primarily interested in the qualitative behavior of shell formation in this test; we shall investigate the quantitative behavior more carefully in our next test.

\begin{figure}
  \centering
  \epsscale{1}
	\plotone{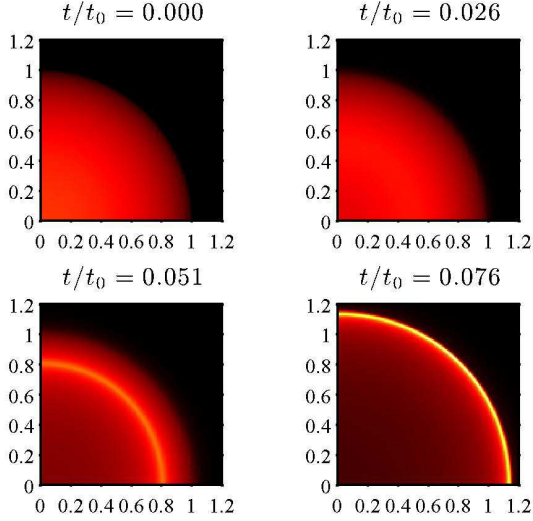}
  \caption{Pseudocolor images of the gas surface density at regular time intervals for the thin shell formation of gas ejected by a central radiation source.  The linear scale ranges from $\Sigma = 0$ (black) to $\Sigma = 1.34$ (white).  The source radiation model is described by Equation~\eqref{radshell:sourceprofile} and gas is initially distributed as a uniformly dense sphere of radius $r_0$.  By time $t = 0.076t_0$, a thin shell has formed near $r=r_0$.  \label{radshell:thinshell}}
\end{figure}

We use a uniform, three-dimensional grid with a resolution of $N^3 = 128^3$ zones in the domain $(x,y,z) \in [0,1.2r_0]^3$ and impose outflow boundary conditions on the outer boundaries.  Snapshots of the surface density $\Sigma \equiv \int \rho\,dz$ at regular time intervals are shown in Figure~\ref{radshell:thinshell}, demonstrating that a thin shell has indeed formed around $r=r_0$ by $t = 0.076 t_0$.

For our second test, we assume that at time $t=0$ a shell of thickness $H$ is located at initial radius $r_0 = 5 \text{ pc}$ with zero initial velocity.  We then evolve the shell and compare with the analytic solution.  This reference solution is obtained by first rewriting Equation~\eqref{radshell:odeaccel} in 
terms of the variables $\tilde{r} \equiv r/r_0$ and $\tilde{t} \equiv t/t_0$ to obtain the simplified ODE
\begin{equation}
	\frac{d^2 \tilde{r}}{d \tilde{t}^2} = \frac{\mathcal{M}_0^2}{\tilde{r}^2}.  \label{radshell:odeaccelreduced} 
\end{equation}
Equation~\eqref{radshell:odeaccelreduced} can then be integrated to obtain the shell velocity
\begin{equation}
	\frac{d \tilde{r}}{d \tilde{t}} = \mathcal{M}_0 \sqrt{2} \left( 1 - \frac{1}{\tilde{r}} \right)^{1/2}, \label{radshell:odevel}
\end{equation}
for $\tilde{r} \ge 1$.  Equation~\eqref{radshell:odevel} can then be integrated once more to obtain
\begin{equation}
	\tilde{t} = \frac{1}{\mathcal{M}_0\sqrt{2}} \left[\sqrt{\tilde{r}}\sqrt{\tilde{r}-1} + \ln\left(\sqrt{\tilde{r}} + \sqrt{\tilde{r}-1}\right)\right], \label{radshell:odepos}
\end{equation}
for $\tilde{r} \ge 1$.

In order to preserve the original ordering of time scales under the RSLA, we must choose $\hat{c}$ such that $v_{\rm max} \ll \hat{v}_{\rm diff}$ at all times during this test, which is equivalent to the RSLA static diffusion criterion in Equation~\eqref{rsla:staticdiffreq}, requiring that $\hat{c} \gg \tau_{\rm max} v_{\rm max}$.  The optical depth across the shell is given by $\tau_{\rm sh} \sim \tau_0(r/r_0)^{-2} \le \tau_0 = 6.6$.  The maximum signal speed in the gas is given by $v_{\rm max} \equiv v_{\rm flow} + c_{\rm eff}$, where $v_{\rm flow} \equiv a_0 (d\tilde{r}/d\tilde{t})$ is the typical flow speed with $d\tilde{r}/d\tilde{t} \sim \mathcal{M}_0 = 13 \gg 1$ given by Equation~\eqref{radshell:odevel}, and $c_{\rm eff} \sim a_0$ is the effective sound speed given by Equation~\eqref{algorithm:ceff}, hence $v_{\rm max} \sim a_0 \mathcal{M}_0 \gg a_0$.  The RSLA static diffusion criterion requires that $v_{\rm diff} \sim \hat{c}/\tau_{\rm max} \gg v_{\rm max}$ in the flow.  Since $\tau_{\rm max} \le \tau_0$ and $v_{\rm max} \lesssim \mathcal{M}_0 a_0$ for the range $r_0 \le r \le 2 r_0$ of our simulation, if we adopt $\hat{c}/a_0 = 10 \tau_0 \mathcal{M}_0 = 860$, then the we have $v_{\rm diff} > 10 v_{\rm max}$ and the RSLA static diffusion criterion will be satisfied.  Although this value of $\hat{c}$ might seem prohibitively large, recall that the typical gas time step is set by $v_{\rm max} \gg a_0$.  Our choice of $\hat{c}$ results in a gas-to-radiation time step ratio $R \equiv \Delta t_{\rm gas}/\Delta t_{\rm rad} = \hat{c}/v_{\rm max} \sim 10\tau_0 = 66$ and in practice $R$ is in the range 53-73 with an average of $\sim 55$, which is computationally feasible.

Additionally, the internal gas pressure force in the shell is given by $F_{\rm press} \sim 4\pi r^2 H \,\nabla P \sim 4\pi r^2 \rho_0 a_0^2$, which implies that $F_{\rm rad}/F_{\rm press} \sim \mathcal{M}_0^2 H r_0^3 / r^4$.  If we assume $H/r \gtrsim 0.1$, the gas pressure forces will remain small until $r \sim 2.6 r_0 = 13 \text{ pc}$.  We evolve the shell from radius $r = 5 \text{ pc}$ to $r = 10 \text{ pc}$ in our test.  Thus, neglecting the internal pressure forces in the model described by Equation~\eqref{radshell:odeaccel} is justified.

Since the radiation field is quasi-static with respect to the gas and the condition of radiative equilibrium has been imposed, the initial radiation flux profile $F_*(r)$ can be approximated by solving the ODE $\nabla \cdot \mathbf{F}_* = j_*(r)$ to obtain
\begin{equation}
	F_*(r) = \frac{L_*}{4\pi r^2} \left[ {\rm erf}\left(\frac{r}{\sqrt{2}\sigma_*}\right) - \frac{2r}{\sqrt{2\pi\sigma_*^2}} \exp\left(-\frac{r^2}{2\sigma_*^2}\right)\right],  \label{radshell:fluxprofile}
\end{equation}
where the bracketed expression in Equation~\eqref{radshell:fluxprofile} rapidly approaches 1 as $r/\sigma_* \to \infty$.  To approximate the initial profile for the radiation energy density, we then solve the ODE
\begin{equation}
	\nabla \cdot \mathbb{P} = -\rho \kappa_0 \frac{\mathbf{F}}{c},  \label{radshell:odepressure}
\end{equation}
using the $M_1$ closure relation to relate $\mathbb{P}$ to $\mathcal{E}$ and $\mathbf{F}$.  

For a spherically symmetric system, Equation~\eqref{radshell:odepressure} reduces to
\begin{equation}
	\frac{1}{r^2} \partial_r (r^2 P_{rr}) - \frac{1}{r} (P_{\phi\phi} + P_{\theta\theta}) = -\rho\kappa_0 \frac{F}{c},
\end{equation}
where we have expanded the radial component of the divergence of the radiation pressure tensor.  Recalling that ${\rm tr} \mathbb{P}=\mathcal{E}$, expressing $P_{rr}=\mathcal{E}\chi$ in terms of the Eddington factor, $\chi$, and using $\mathcal{E}=F/(cf)$ and $\chi=\case{1}{3}(5-2\sqrt{4-3f^2})$ to eliminate $\mathcal{E}$ and $\chi$ in favor of the known flux function $F(r)$ and the unknown reduced flux $f(r)$, we obtain the ODE
\begin{eqnarray}
	\partial_r f &=& \frac{3f\sqrt{4-3f^2}}{5\sqrt{4-3f^2}-8} \Bigg[ \frac{\partial_r \ln F}{3} (5-2\sqrt{4-3f^2}) \nonumber \\
	&&+ \frac{2}{r} (2-\sqrt{4-3f^2}) + \rho\kappa_0 f \Bigg].  \label{radshell:odefluxm1}
\end{eqnarray}
It is evident that the nonlinear ODE in Equation~\eqref{radshell:odefluxm1} is singular at the critical value $f_{\rm crit}=2\sqrt{3}/5$.  A necessary condition for this to be a \emph{regular} singularity of the ODE is that the bracketed expression in Equation~\eqref{radshell:odefluxm1} also vanish at the point $r_{\rm crit}$ for which $f(r_{\rm crit})=f_{\rm crit}$.  In general, this point will depend on the specific profiles $\rho(r)$ and $F(r)$, and its location can be determined by solving the nonlinear equation
\begin{equation}
	3 \,\partial_r \ln F + \frac{4}{r} + 2\sqrt{3} \rho\kappa_0 = 0  \label{radshell:rcrit}
\end{equation}
numerically for $r$, taking $F(r)$ from Equation~\eqref{radshell:fluxprofile}.  Any regular solution of Equation~\eqref{radshell:odefluxm1} must pass through the critical point $(r_{\rm crit}, f_{\rm crit})$, hence this point can be used as an internal boundary condition from which Equation~\eqref{radshell:odefluxm1} can be integrated outward to either larger or smaller $r$ to obtain the semi-analytic solution $f(r)$.  Once this solution has been obtained, we have $\mathcal{E}(r) = F_*(r)/[c f(r)]$.

\begin{figure}
  \centering
  \epsscale{1}
  \plotone{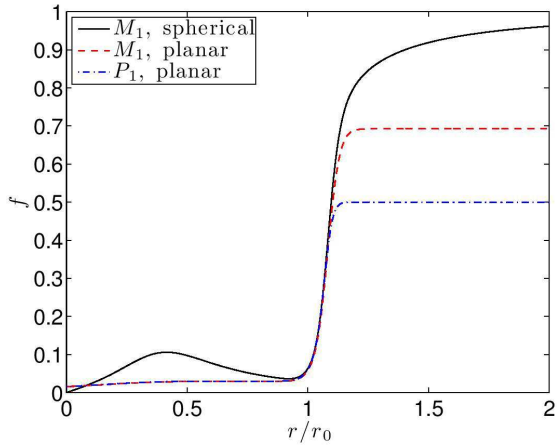}
  \caption{Initial equilibrium profile of the reduced flux, $f$, for the spherically symmetric thin shell problem with the $M_1$ closure relation (solid line).  For reference, the approximate planar model solutions with the $M_1$ (dashed line) and $P_1$ (i.e., Eddington; dash-dotted line) closure relations are also shown.  \label{radshell:fr_initial}}
\end{figure}

We model the initial density of the shell using the Gaussian profile given by
\begin{equation}
	\rho_{\rm sh}(r) = \frac{M_{\rm sh}}{4\pi r^2\,\sqrt{2\pi \sigma_{\rm sh}^2}} \exp \left( -\frac{(r-r_0)^2}{2\sigma_{\rm sh}^2} \right),  \label{radshell:rhoprofile}
\end{equation}
where $\sigma_{\rm sh} \equiv H/(2\sqrt{2\ln 2})$ is the half-width of the shell.  The profile in Equation~\eqref{radshell:rhoprofile} is a smooth function of $r$ whose volume integral rapidly approaches the shell mass, $M_{\rm sh}$, as $|r-r_0| \to \infty$.  For our choice of $\tau_0 = M_{\rm sh} \kappa_0/(4\pi r_0^2) \approx 6.6$, we solve Equation~\eqref{radshell:rcrit} iteratively via Newton's method to obtain $r_{\rm crit} \approx 1.128 \,r_0$.  Integrating Equation~\eqref{radshell:odefluxm1} outward and inward from $r_{\rm crit}$ via a fourth-order Runge-Kutta scheme, we obtain the semi-analytic solution $f(r)$ shown in Figure~\ref{radshell:fr_initial}.  For reference, we also show the solutions for the $P_1$ (i.e., Eddington) and $M_1$ closures in planar geometry.  Note in Figure~\ref{radshell:fr_initial} that all three solutions are approximately equal in the vicinity of the shell, but far from the shell where geometric effects become important, the planar solutions do not adequately describe the true, spherically symmetric radiation field.

We initialize our test using Equation~\eqref{radshell:rhoprofile} for the radial density profile, $v_0=0$ for the initial velocity, Equation~\eqref{radshell:fluxprofile} for the radial flux profile, and the semi-analytic solution of Equation~\eqref{radshell:odefluxm1} for the radial energy density profile.  To prevent the gas time steps from becoming prohibitively small, we enforce a density floor of $\rho_{\rm min} \equiv 10^{-8}\rho_0$ initially as well as after each gas integration.  We use a uniform, three-dimensional grid of resolution $N^3=128^3$ on the domain $(x,y,z) \in [0,2\,r_0]^3$, with reflection boundary conditions on the inner boundaries and outflow boundary conditions on the outer boundaries, to enhance efficiency.  When gas evolution is turned off, the code holds the predicted spherical $M_1$ solution shown in Figure~\ref{radshell:fr_initial} very well; it also relaxes to this solution when started from other initial conditions.  When gas evolution is turned on, from Equation~\eqref{radshell:odepos} we expect the shell to reach radius $r=2\,r_0$ at time $t_{\rm final} \approx 0.12t_0$.

\begin{figure}
  \centering
  \epsscale{1}
  \plotone{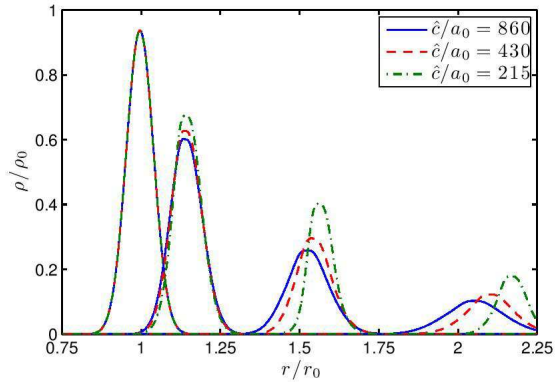}
  \caption{Density averaged over radial shells at regular time intervals $t/t_0 = \{0,0.042,0.083,0.124\}$ and for several values of $\hat{c}/a_0$ in the radiatively driven expanding shell problem. \label{radshell:data_rho_crad_all}}
\end{figure}

\begin{figure}
  \centering
  \epsscale{1}
	\plotone{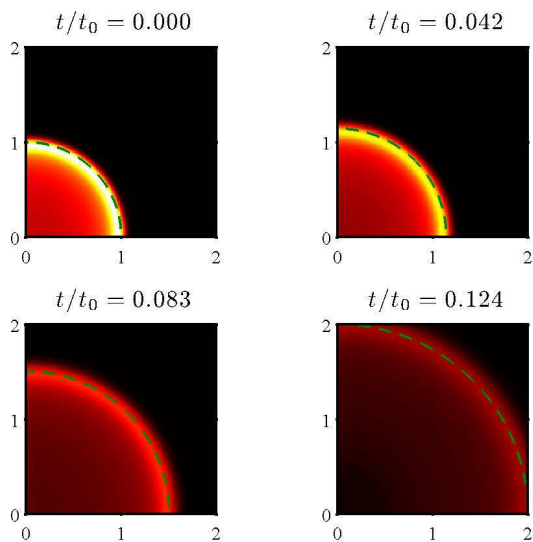}
  \caption{Pseudocolor images of the gas surface density $\Sigma \equiv \int \rho\,dz$ at regular time intervals for the radiatively driven expanding shell problem with $\hat{c}/a_0 = 860$.  The linear scale ranges from $\Sigma = 0$ (black) to $\Sigma = 0.706$ (white).  For reference, we plot a curve (dash) indicating the shell radius given by the semi-analytic solution of Equation~\eqref{radshell:odepos} at the indicated time.  \label{radshell:data_d}}
\end{figure}

In Figure~\ref{radshell:data_rho_crad_all}, we plot the gas density $\rho$, averaged over spherical shells, at the same time intervals for a range of values of $\hat{c}/a_0$, including the sufficiently large value $\hat{c}/a_0 = 860$ as well as one-half and one-quarter of this value.  It is clear from this figure that for too small a value of $\hat{c}/a_0$ the radiation force is too strong, hence the shell remains thinner and expands more rapidly than expected.  Figure~\ref{radshell:data_d} shows a series of snapshots of $\Sigma$ at regular time intervals along with a reference curve indicating the shell radius given by the semi-analytic solution of Equation~\eqref{radshell:odepos}, where it can be seen that the shell remains uniformly spherical and thin as it expands.  

\begin{figure}
  \centering
  \epsscale{1}
  \plotone{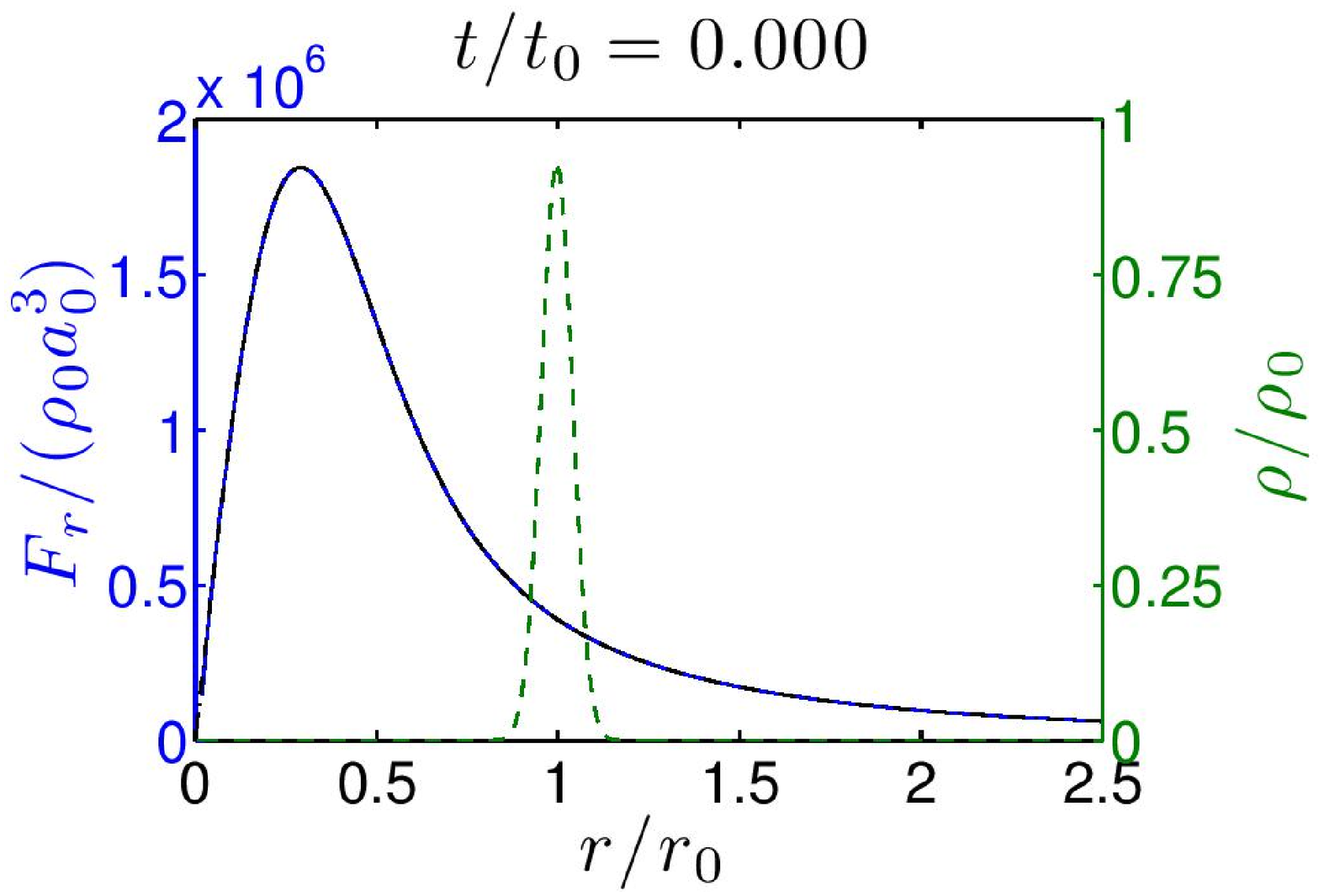}
  \plotone{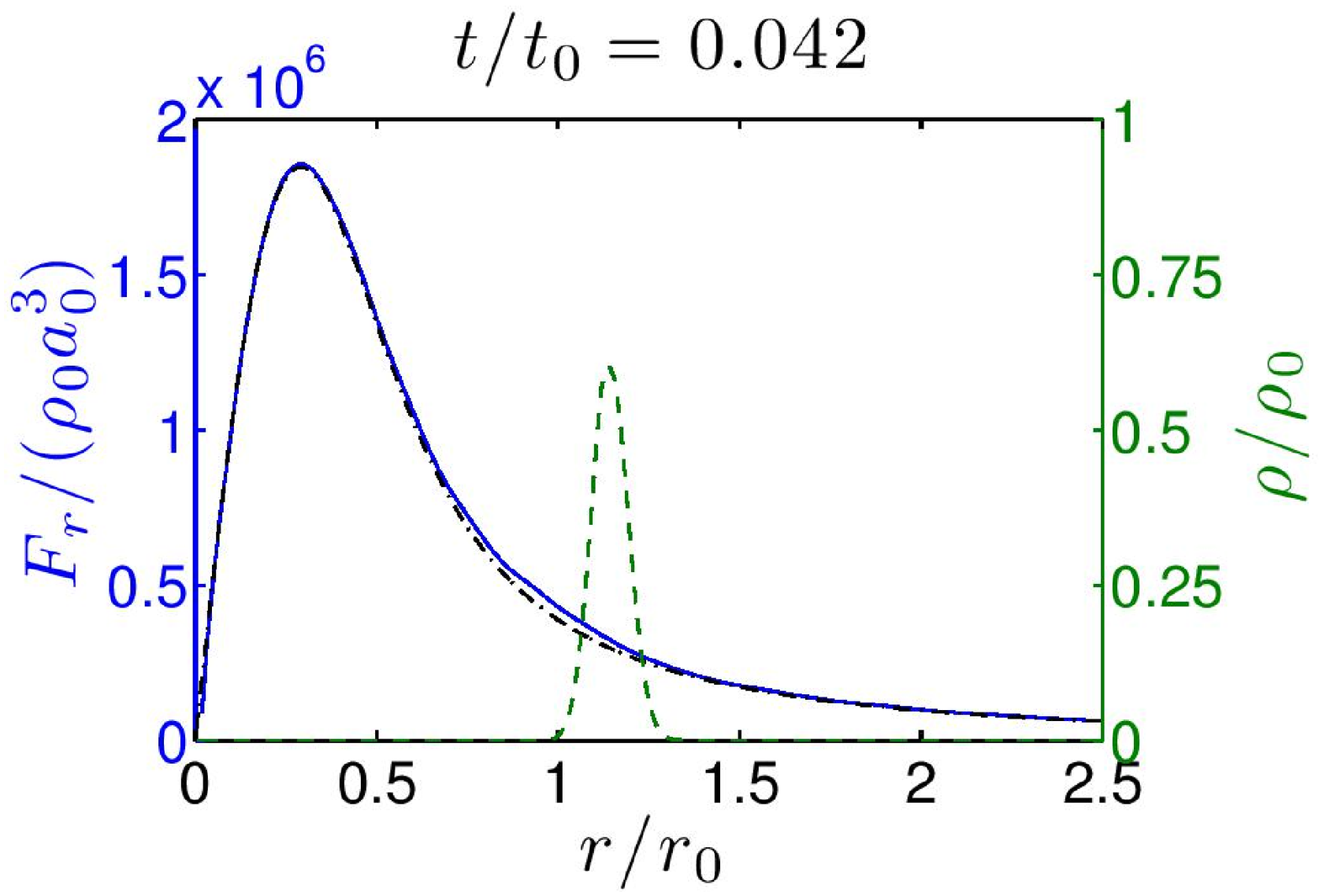}
  \plotone{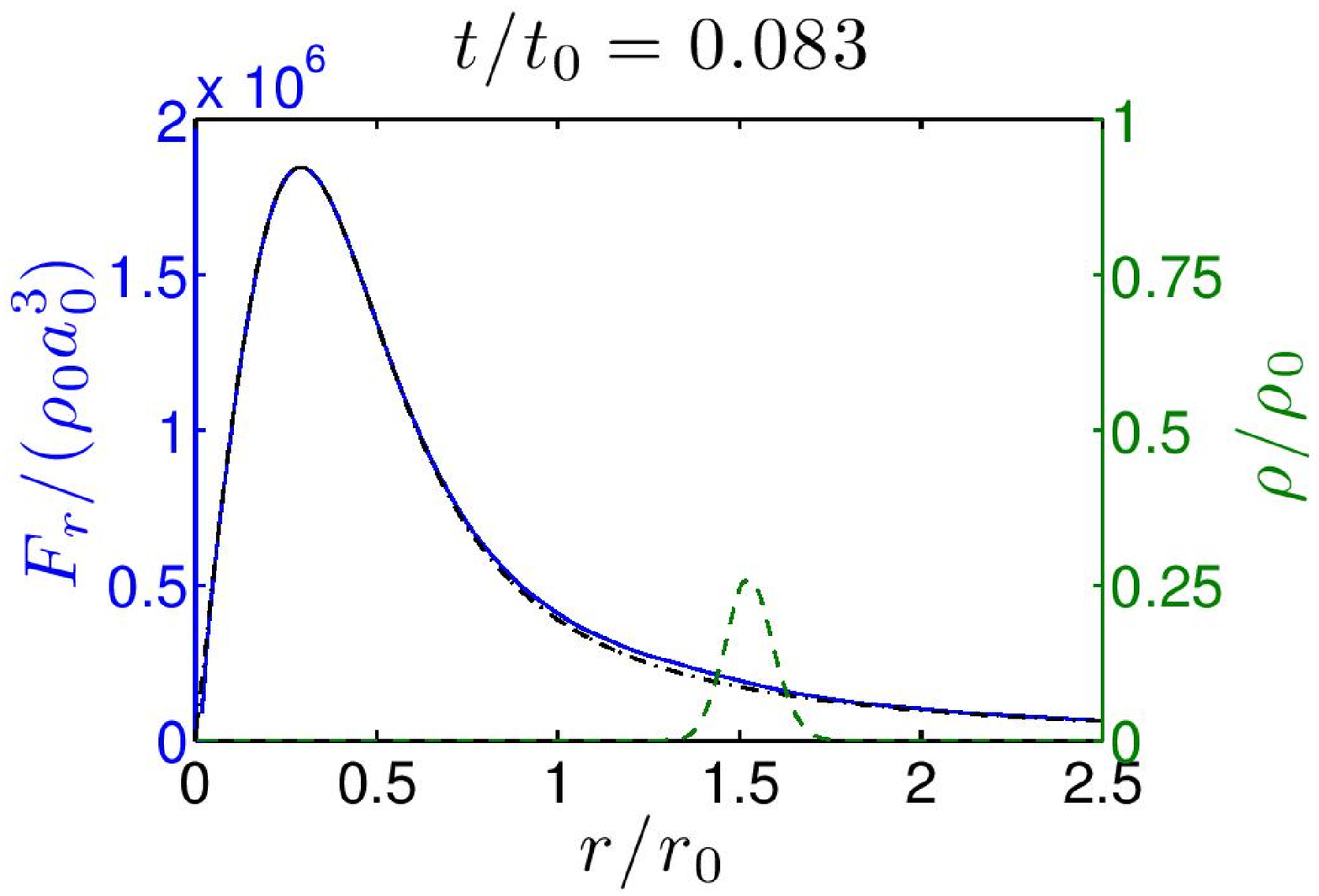}
  \plotone{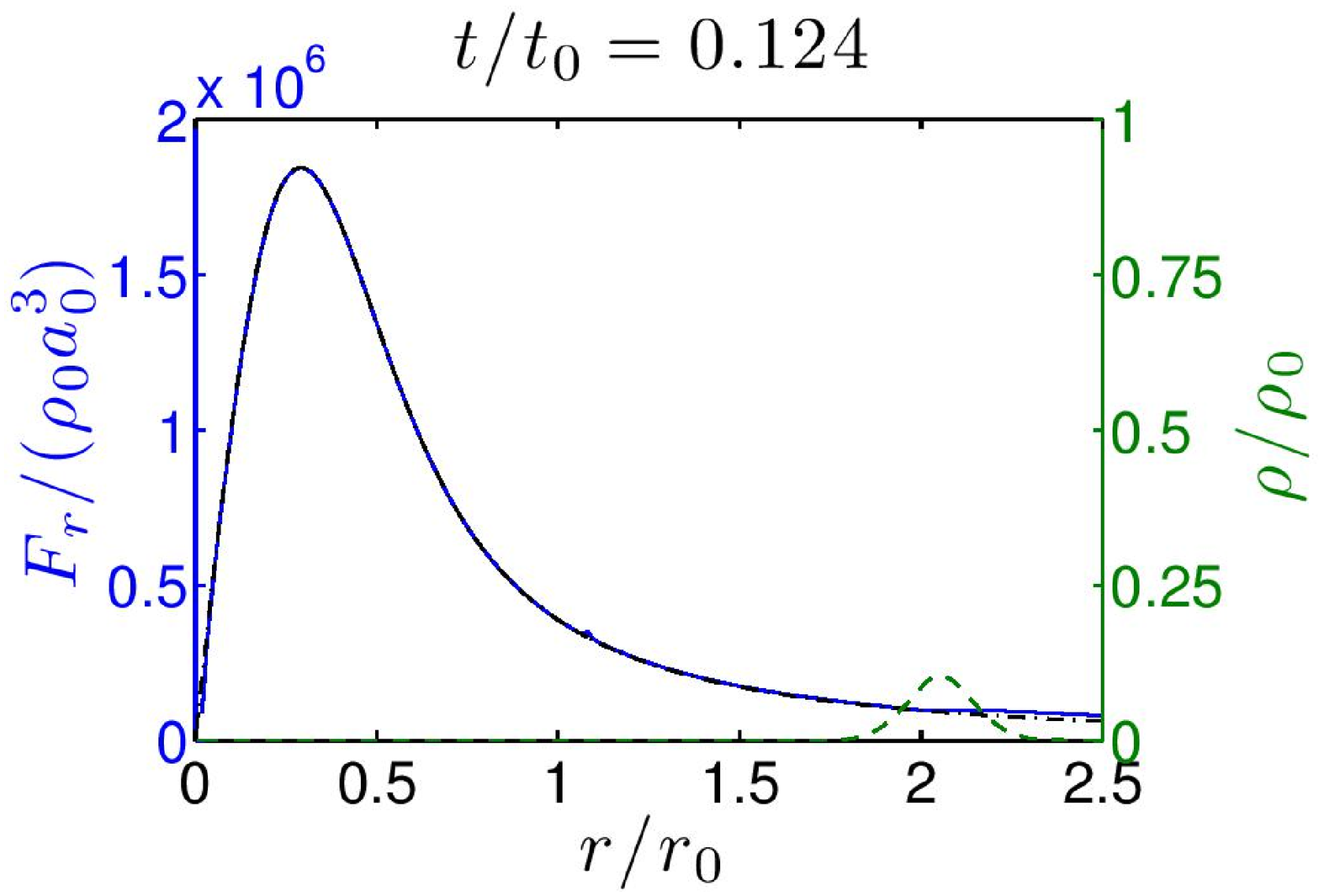}
  \caption{Radiative flux (solid) averaged over radial shells at regular time intervals for the radiatively driven expanding shell problem with $\hat{c}/a_0 = 860$.  For reference, we also plot the equilibrium solution (dash-dot) given by Equation~\eqref{radshell:fluxprofile}, and the density profile (dash).  \label{radshell:data_F}}
\end{figure}

\begin{figure}
  \centering
  \epsscale{1}
  \plotone{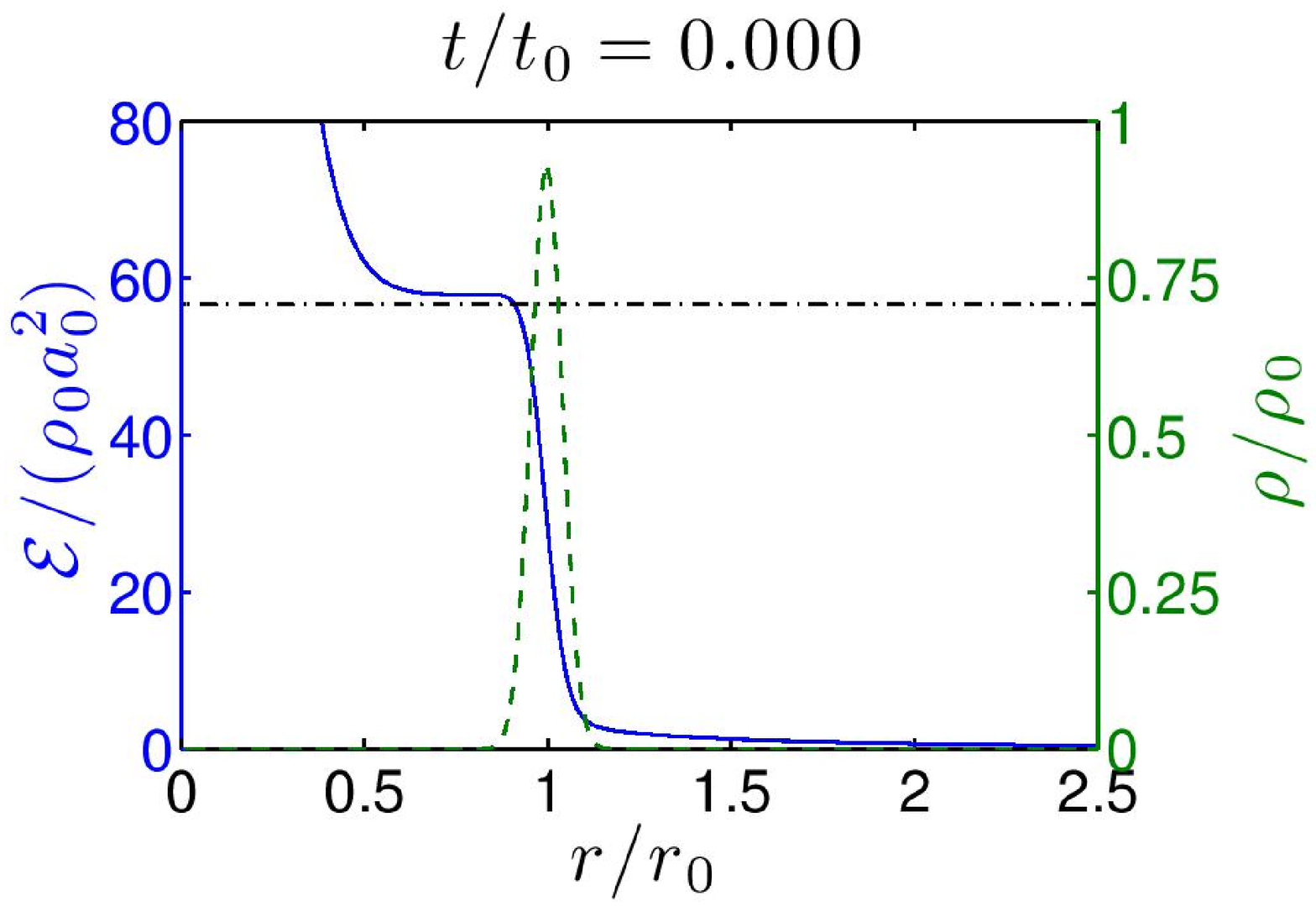}
  \plotone{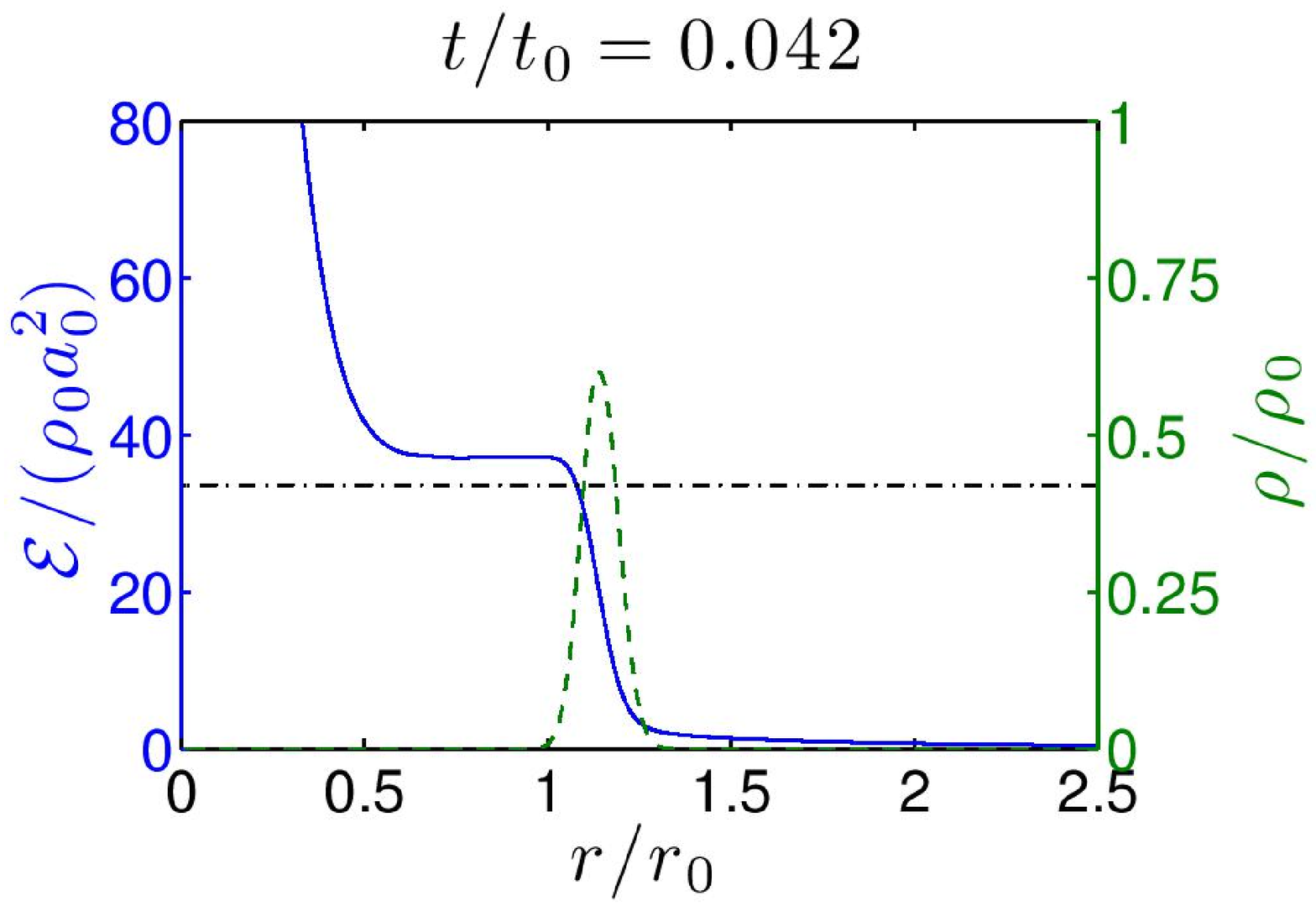}
  \plotone{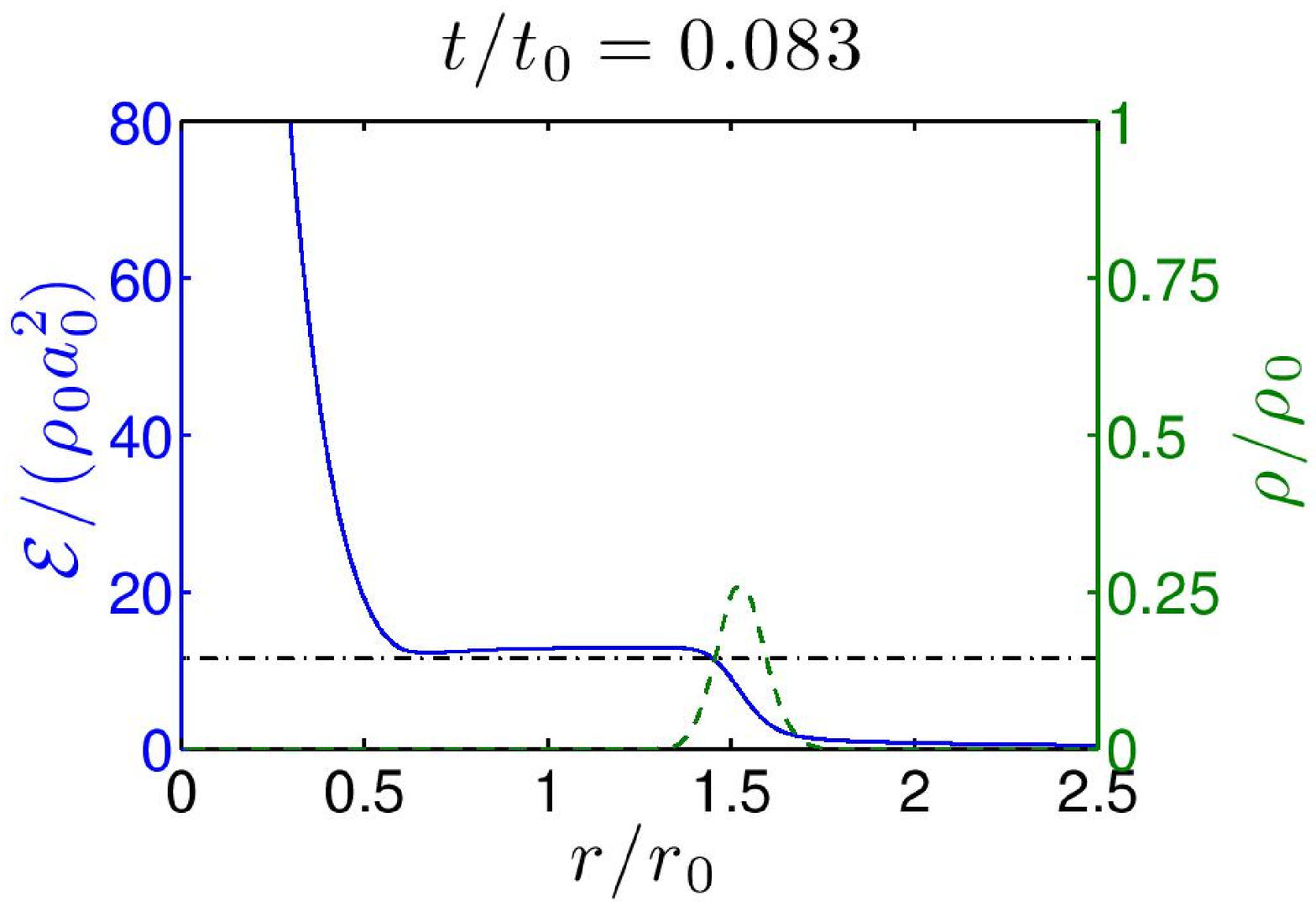}
  \plotone{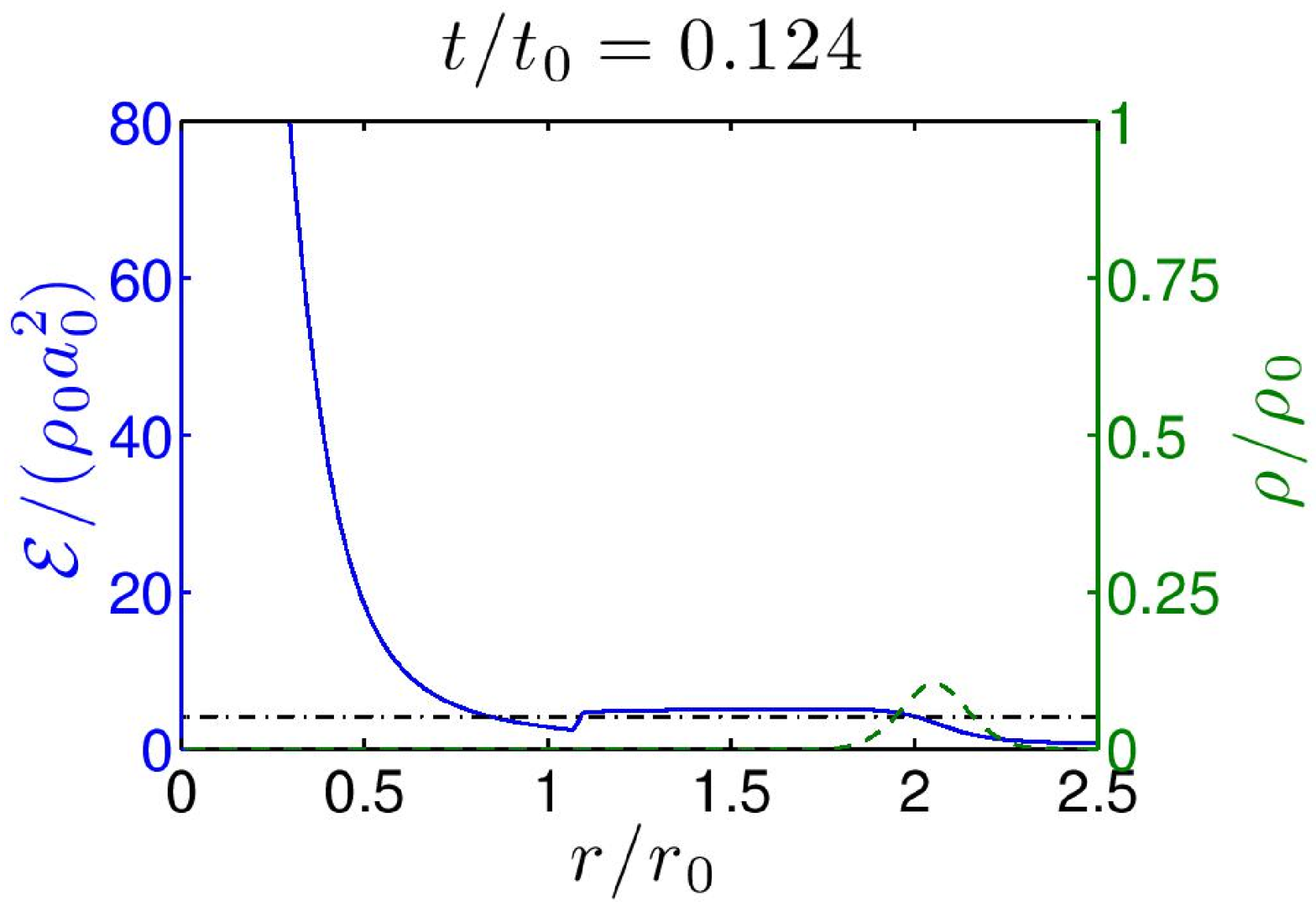}
  \caption{The same as Figure~\ref{radshell:data_F} for the radiative energy density (solid).  For reference, we also plot the density profile (dash) and the model solution $\mathcal{E}_{\rm Edd}(r)$ defined by Equation~\eqref{radshell:Ermodel} (dash-dot), where we use $r=\langle r \rangle$, the average radial coordinate defined by Equation~\eqref{radshell:ravgdef}.  \label{radshell:data_Er}}
\end{figure}

Line plots of the radiative flux measured along the positive $x$-axis as well as the flux profile modeled in Equation~\eqref{radshell:fluxprofile} are shown in Figure~\ref{radshell:data_F} for the same time intervals with $\hat{c}/a_0 = 860$.  The numerical flux is close to the total flux corresponding to the underlying source function described in Equation~\eqref{radshell:sourceprofile}.  This total flux is the sum of the directed flux from the source and the diffusive flux from radiation reprocessed by the gas in the shell, which further indicates that the radiation has sufficient time to diffuse through the shell in a gas time step.  Figure~\ref{radshell:data_Er} shows line plots of the radiative energy density, averaged over spherical shells, at the same time intervals with $\hat{c}/a_0 = 860$.  For reference, we also show the radiative energy density in the interior of the shell predicted by the plane-parallel model using the Eddington closure relation.  This model is given by
\begin{equation}
	\mathcal{E}_{\rm Edd}(r) \equiv \frac{3 F_*(r)}{c} \left(\tau_{\rm sh}(r) + \frac{2}{3}\right),  \label{radshell:Ermodel}
\end{equation}
where $F_*(r)$ is flux given in Equation~\eqref{radshell:fluxprofile} for a shell of radius $r$, and $\tau_{\rm sh}(r) \equiv M_{\rm sh} \kappa_0/(4\pi r^2)$ is the total optical depth across a uniformly dense shell of mass $M_{\rm sh}$ and  radius $r$.  At a given time $t$, we estimate the shell radius using $r = \langle r \rangle$, where
\begin{equation}
	\langle r \rangle \equiv \frac{\int \rho r \,dV}{\int \rho \,dV},  \label{radshell:ravgdef}
\end{equation}
is the density-weighted radial coordinate of the shell computed by the code.

\begin{figure}
  \centering
  \epsscale{1}
  \plotone{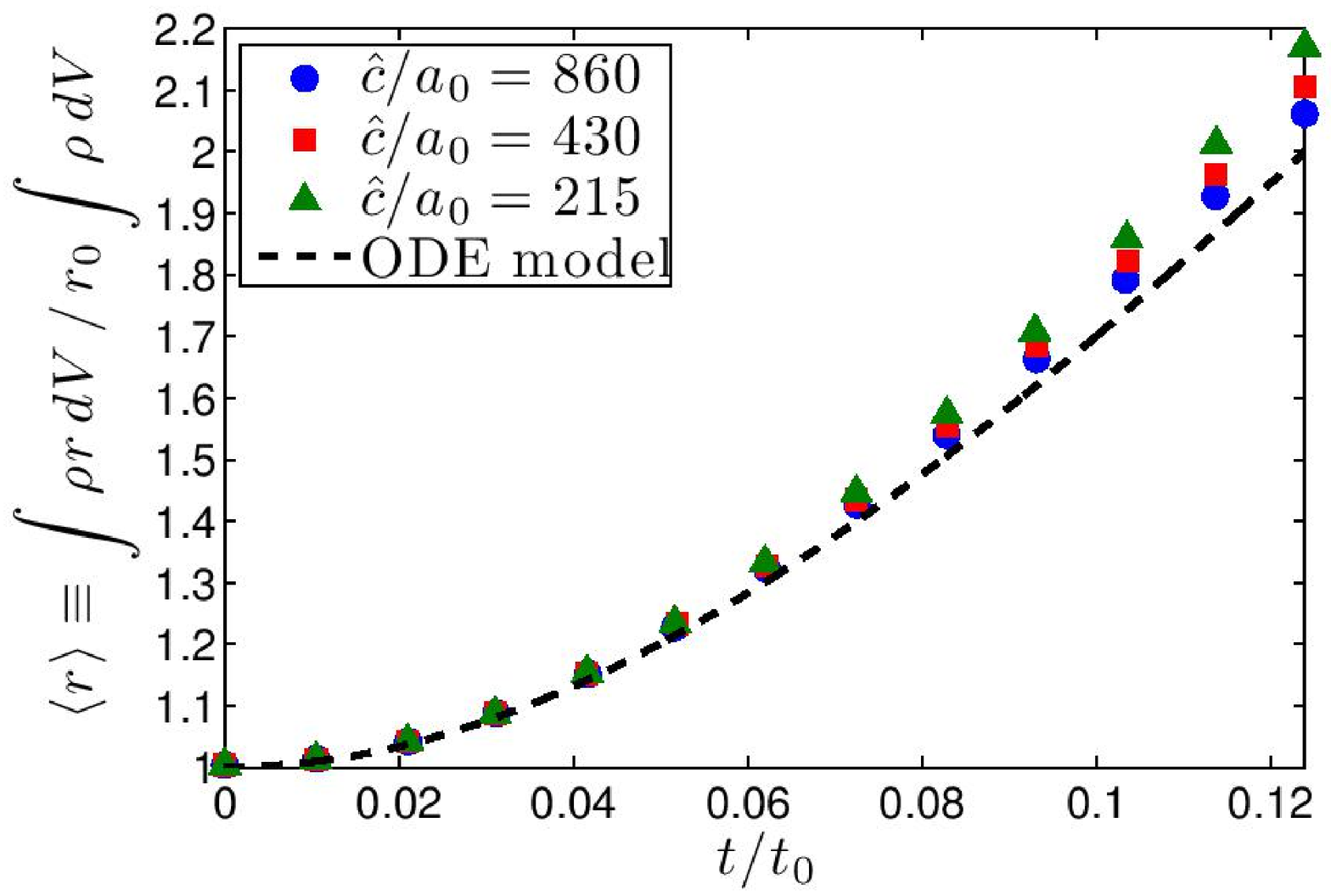}
  \plotone{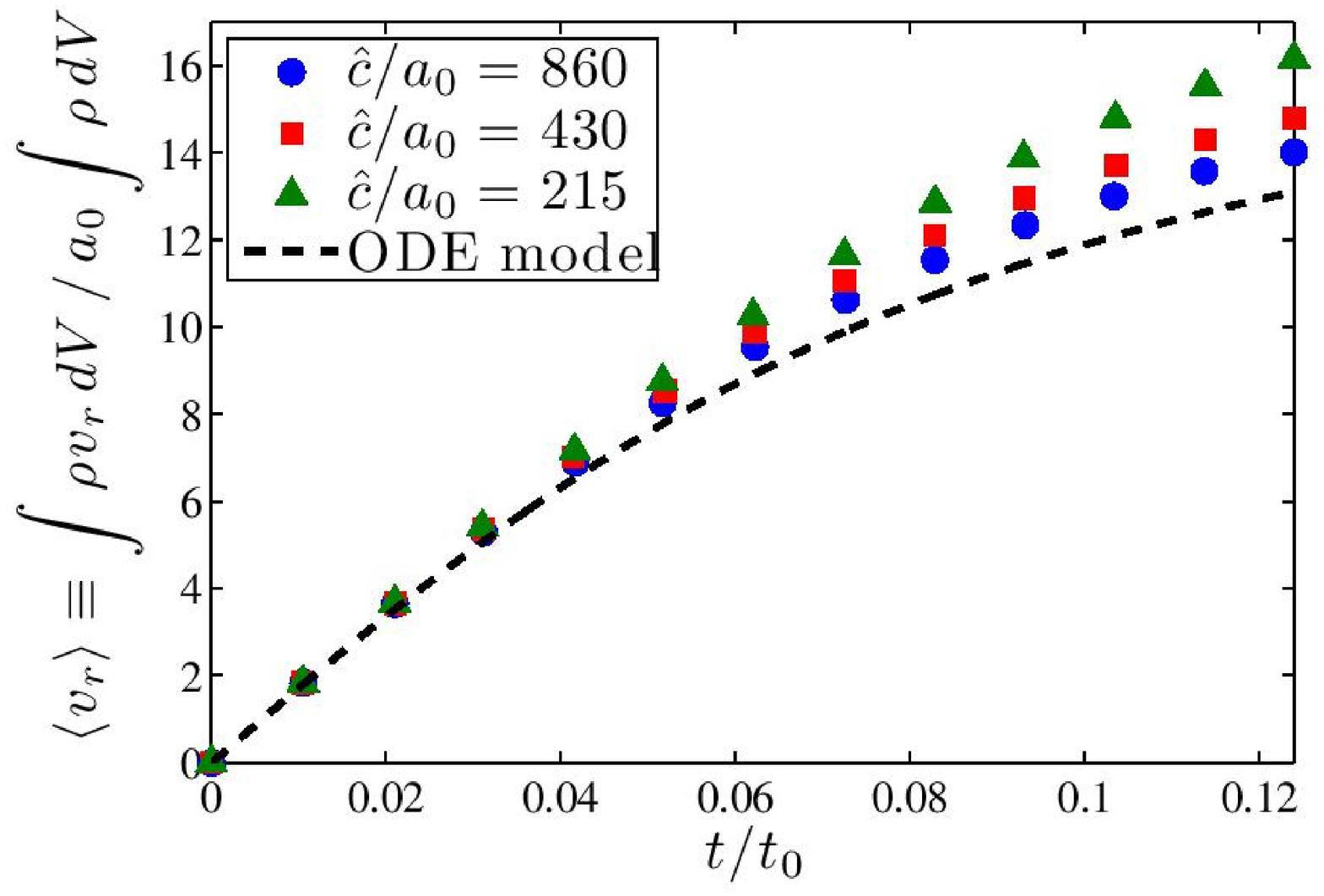}
  \caption{Density-averaged radial position, $\langle r \rangle$ (top), and velocity, $\langle v_r \rangle$ (bottom), of the radiation-driven shell at regular time intervals and for several values of $\hat{c}$ in the radiatively driven expanding shell problem.  For $\hat{c}/a_0 = 860$, the maximum relative errors for the computed solution (circles) are $3.1\%$ and $7.0\%$ for $\langle r \rangle$ and $\langle v_r \rangle$, respectively. \label{radshell:data_rv}}
\end{figure}

To compare our computed solution with the predicted ODE solutions given by Equations~\eqref{radshell:odevel} and~\eqref{radshell:odepos}, we compute $\langle r \rangle$, defined in Equation~\eqref{radshell:ravgdef}, as well as the density-weighted radial velocity defined by
\begin{equation}
	\langle v_r \rangle \equiv \frac{\int \rho (\mathbf{v} \cdot \hat{\mathbf{r}}) \,dV}{\int \rho \,dV},  \label{radshell:vravgdef}
\end{equation}
with the volume integrals in Equations~\eqref{radshell:ravgdef} and~\eqref{radshell:vravgdef} performed over the entire grid.  Figure~\ref{radshell:data_rv} shows the data for the quantities $\langle r \rangle$ and $\langle v_r \rangle$ along with the models given by the semi-analytic solutions of Equations~\eqref{radshell:odepos} and~\eqref{radshell:odevel}, respectively, as the shell expands.  For $\hat{c}/a_0 = 860$, the maximum relative errors for $\langle r \rangle$ and $\langle v_r \rangle$ are approximately $3.1\%$ and $7.0\%$, respectively, showing good agreement with the model.  For $\hat{c}/a_0 = 430$ the maximum relative errors are $5.3\%$ and $13\%$, and for $\hat{c}/a_0 = 215$ they are $8.5\%$ and $23\%$, respectively.  The sizes of the relative errors when $\hat{c}/a_0 < 860$ underscores the importance of taking $\hat{c}/a_0$ sufficiently large in order to preserve the proper ordering of time scales under the RSLA.

\vspace{1em}
{\section{Conclusion}  \label{conclusion}}

We have described a module for the {\it Athena} code that solves the gray equations of RHD using the $M_1$ closure of the radiation moment equations and an explicit update for the radiation transport terms, which we have designed primarily to study star formation in galactic disks and in GMCs.  We employ the RSLA with subcycling of the radiation variables in order to reduce computational costs, and have identified the regime of applicability of this approximation.  Our implementation, {\it Hyperion}, is based on the {\it Athena} astrophysical MHD code \citep{Stone:2008} and has been tested over a wide range of optical depths and energy ranges.  Like {\it Athena}, {\it Hyperion} is dimensionally unsplit in one, two, and three dimensions.  

We have verified our algorithm using a wide variety of novel and established quantitative tests, including propagation of linear RHD waves, strong advection of radiation, non-equilibrium radiative shocks, force balance in optically thin and optically thick systems, and radiative momentum-driven expanding shells.  We have also carried out a basic timing benchmark comparing the performance of {\it Hyperion} with that of the FLD module in the {\it Enzo} code \citep{Reynolds:2009}, which suggests a clear advantage of our method over others that require expensive matrix inversions, in cases where the RSLA is practical (generally optical depths $\lesssim 10$).

There are some limitations to our algorithm that might be improved in future versions.  For example, our algorithm does not perform as efficiently in the dynamic diffusion regime as other fully implicit methods, since the dynamical properties of this regime do not permit us to reduce $\hat{c}$ to computationally feasible levels.  Also, operator splitting of transport and source terms in our algorithm introduces an error that is formally first-order in the relevant time step, although in practice, the coefficient is typically small \citep{Leveque:2002}.  The $M_1$ approximation is known to have limited accuracy in situations where the optical depth is low and there are multiple radiation sources.  For treating these applications, our RSLA semi-explicit algorithm can be extended by substituting for the $M_1$ closure relation a directly computed estimate of the Eddington tensor (e.g., as already implemented in {\it Athena} by \cite{Davis:2012} using the short-characteristics solution of the transfer equation).

Further comparison of our method with other more exact RHD solvers will be important for defining where our approximate---but relatively inexpensive---method is most advantageous to use in astrophysical applications.

\acknowledgments
The authors are grateful to the referee, Nick Gnedin, for careful reading of our manuscript and preparation of a thorough and insightful report.  The authors would also like to thank Yan-Fei Jiang, Richard Klein, and Jim Stone for helpful suggestions during the course of this project, and Dan Reynolds for supplying the problem set-up used in the comparison with {\it Enzo}'s FLD module.  This work was supported by grant AST 0908185 from the National Science Foundation.

\bibliography{apj-jour,references}

\begin{thebibliography}{49}
\expandafter\ifx\csname natexlab\endcsname\relax\def\natexlab#1{#1}\fi

\bibitem[{{Aubert} \& {Teyssier}(2008)}]{Aubert:2008}
{Aubert}, D., \& {Teyssier}, R. 2008, \mnras, 387, 295

\bibitem[{{Audit} {et~al.}(2002){Audit}, {Charrier}, {Chi{\`e}ze}, \&
  {Dubroca}}]{Audit:2002}
{Audit}, E., {Charrier}, P., {Chi{\`e}ze}, J.~P., \& {Dubroca}, B. 2002,
  arXiv:astro-ph/0206281

\bibitem[{{Barkana} \& {Loeb}(2001)}]{Barkana:2001}
{Barkana}, R., \& {Loeb}, A. 2001, \physrep, 349, 125

\bibitem[{{Ciotti} \& {Ostriker}(2007)}]{Ciotti:2007}
{Ciotti}, L., \& {Ostriker}, J.~P. 2007, \apj, 665, 1038

\bibitem[{{Commer{\c c}on} {et~al.}(2011){Commer{\c c}on}, {Teyssier}, {Audit},
  {Hennebelle}, \& {Chabrier}}]{Commercon:2011}
{Commer{\c c}on}, B., {Teyssier}, R., {Audit}, E., {Hennebelle}, P., \&
  {Chabrier}, G. 2011, \aap, 529, A35

\bibitem[{{Davis} {et~al.}(2012){Davis}, {Stone}, \& {Jiang}}]{Davis:2012}
{Davis}, S.~W., {Stone}, J.~M., \& {Jiang}, Y.-F. 2012, \apjs, 199, 9

\bibitem[{{Falle}(1991)}]{Falle:1991}
{Falle}, S.~A.~E.~G. 1991, \mnras, 250, 581

\bibitem[{{Frank} {et~al.}(2012){Frank}, {Hauck}, \& {Olbrant}}]{frank:2012}
{Frank}, M., {Hauck}, C.~D., \& {Olbrant}, E. 2012, arXiv:1208.0772

\bibitem[{{Fryxell} {et~al.}(2000){Fryxell}, {Olson}, {Ricker}, {Timmes},
  {Zingale}, {Lamb}, {MacNeice}, {Rosner}, {Truran}, \& {Tufo}}]{Fryxell:2000}
{Fryxell}, B., {et~al.} 2000, \apjs, 131, 273

\bibitem[{{Gardiner} \& {Stone}(2005)}]{Gardiner:2005}
{Gardiner}, T.~A., \& {Stone}, J.~M. 2005, J. Comp. Phys., 205, 509

\bibitem[{{Gardiner} \& {Stone}(2008)}]{Gardiner:2008}
---. 2008, J. Comp. Phys., 227, 4123

\bibitem[{{Gittings} {et~al.}(2008){Gittings}, {Weaver}, {Clover}, {Betlach},
  {Byrne}, {Coker}, {Dendy}, {Hueckstaedt}, {New}, {Oakes}, {Ranta}, \&
  {Stefan}}]{Gittings:2008}
{Gittings}, M., {et~al.} 2008, Computational Science and Discovery, 1, 015005

\bibitem[{{Gnedin} \& {Abel}(2001)}]{Gnedin:2001}
{Gnedin}, N.~Y., \& {Abel}, T. 2001, \na, 6, 437

\bibitem[{{Gong} \& {Ostriker}(2013)}]{Gong:2013}
{Gong}, H., \& {Ostriker}, E.~C. 2013, \apjs, 204, 8

\bibitem[{{Gonz{\'a}lez} {et~al.}(2007){Gonz{\'a}lez}, {Audit}, \&
  {Huynh}}]{Gonzalez:2007}
{Gonz{\'a}lez}, M., {Audit}, E., \& {Huynh}, P. 2007, \aap, 464, 429

\bibitem[{{Hayes} \& {Norman}(2003)}]{Hayes:2003}
{Hayes}, J.~C., \& {Norman}, M.~L. 2003, \apjs, 147, 197

\bibitem[{{Hirose} {et~al.}(2009){Hirose}, {Krolik}, \& {Blaes}}]{Hirose:2009}
{Hirose}, S., {Krolik}, J.~H., \& {Blaes}, O. 2009, \apj, 691, 16

\bibitem[{{Jiang} {et~al.}(2012){Jiang}, {Stone}, \& {Davis}}]{Jiang:2012}
{Jiang}, Y.-F., {Stone}, J.~M., \& {Davis}, S.~W. 2012, \apjs, 199, 14

\bibitem[{{Johnson} \& {Klein}(2010)}]{Johnson:2010}
{Johnson}, B.~M., \& {Klein}, R.~I. 2010, \jqsrt, 111, 723

\bibitem[{{Krumholz} {et~al.}(2007){Krumholz}, {Klein}, {McKee}, \&
  {Bolstad}}]{Krumholz:2007}
{Krumholz}, M.~R., {Klein}, R.~I., {McKee}, C.~F., \& {Bolstad}, J. 2007, \apj,
  667, 626

\bibitem[{{Krumholz} \& {Thompson}(2012)}]{Krumholz:2012}
{Krumholz}, M.~R., \& {Thompson}, T.~A. 2012, \apj, 760, 155

\bibitem[{LeVeque(2002)}]{Leveque:2002}
LeVeque, R.~J. 2002, Finite volume methods for hyperbolic problems, Cambridge
  texts in applied mathematics (Cambridge: Cambridge University Press)

\bibitem[{{Levermore}(1984)}]{Levermore:1984}
{Levermore}, C.~D. 1984, \jqsrt, 31, 149

\bibitem[{{Levermore} \& {Pomraning}(1981)}]{Levermore:1981}
{Levermore}, C.~D., \& {Pomraning}, G.~C. 1981, \apj, 248, 321

\bibitem[{{Lowrie} \& {Edwards}(2008)}]{Lowrie:2008}
{Lowrie}, R.~B., \& {Edwards}, J.~D. 2008, Shock Waves, 18, 129

\bibitem[{{Lowrie} \& {Morel}(2001)}]{Lowrie:2001}
{Lowrie}, R.~B., \& {Morel}, J.~E. 2001, \jqsrt, 69, 475

\bibitem[{{Lowrie} {et~al.}(1999){Lowrie}, {Morel}, \&
  {Hittinger}}]{Lowrie:1999}
{Lowrie}, R.~B., {Morel}, J.~E., \& {Hittinger}, J.~A. 1999, \apj, 521, 432

\bibitem[{{Marshak}(1958)}]{Marshak:1958}
{Marshak}, R.~E. 1958, Physics of Fluids, 1, 24

\bibitem[{{Mihalas} \& {Auer}(2001)}]{Mihalas:2001}
{Mihalas}, D., \& {Auer}, L.~H. 2001, \jqsrt, 71, 61

\bibitem[{{Mihalas} \& {Klein}(1982)}]{Mihalas:1982}
{Mihalas}, D., \& {Klein}, R.~I. 1982, J. Comp. Phys., 46, 97

\bibitem[{{Mihalas} \& {Mihalas}(1983)}]{Mihalas:1983}
{Mihalas}, D., \& {Mihalas}, B.~W. 1983, \apj, 273, 355

\bibitem[{Mihalas \& Weibel-Mihalas(1999)}]{Mihalas:1999}
Mihalas, D., \& Weibel-Mihalas, B. 1999, Foundations of Radiation
  Hydrodynamics, Dover Books on Physics (Mineola: Dover Publications)

\bibitem[{{Murray} {et~al.}(2010){Murray}, {Quataert}, \&
  {Thompson}}]{Murray:2010}
{Murray}, N., {Quataert}, E., \& {Thompson}, T.~A. 2010, \apj, 709, 191

\bibitem[{{Ostriker} \& {Shetty}(2011)}]{Ostriker:2011}
{Ostriker}, E.~C., \& {Shetty}, R. 2011, \apj, 731, 41

\bibitem[{{Petkova} \& {Springel}(2011)}]{Petkova:2011}
{Petkova}, M., \& {Springel}, V. 2011, \mnras, 415, 3731

\bibitem[{{Pomraning}(1979)}]{Pomraning:1979}
{Pomraning}, G.~C. 1979, \jqsrt, 21, 249

\bibitem[{{Reynolds} {et~al.}(2009){Reynolds}, {Hayes}, {Paschos}, \&
  {Norman}}]{Reynolds:2009}
{Reynolds}, D.~R., {Hayes}, J.~C., {Paschos}, P., \& {Norman}, M.~L. 2009, J.
  Comp. Phys., 228, 6833

\bibitem[{{Saad} \& {Schultz}(1986)}]{Saad:1986}
{Saad}, Y., \& {Schultz}, M.~H. 1986, SIAM J. Sci. and Stat. Comput., 7, 856

\bibitem[{{Sincell} {et~al.}(1999){Sincell}, {Gehmeyr}, \&
  {Mihalas}}]{Sincell:1999}
{Sincell}, M.~W., {Gehmeyr}, M., \& {Mihalas}, D. 1999, Shock Waves, 9, 391

\bibitem[{{Stone} \& {Gardiner}(2009)}]{Stone:2009}
{Stone}, J.~M., \& {Gardiner}, T. 2009, \na, 14, 139

\bibitem[{{Stone} {et~al.}(2008){Stone}, {Gardiner}, {Teuben}, {Hawley}, \&
  {Simon}}]{Stone:2008}
{Stone}, J.~M., {Gardiner}, T.~A., {Teuben}, P., {Hawley}, J.~F., \& {Simon},
  J.~B. 2008, \apjs, 178, 137

\bibitem[{{Su} \& {Olson}(1996)}]{Su:1996}
{Su}, B., \& {Olson}, G.~L. 1996, \jqsrt, 56, 337

\bibitem[{{Swesty} \& {Myra}(2009)}]{Swesty:2009}
{Swesty}, F.~D., \& {Myra}, E.~S. 2009, \apjs, 181, 1

\bibitem[{{Thompson} {et~al.}(2005){Thompson}, {Quataert}, \&
  {Murray}}]{Thompson:2005}
{Thompson}, T.~A., {Quataert}, E., \& {Murray}, N. 2005, \apj, 630, 167

\bibitem[{{Turner} \& {Stone}(2001)}]{Turner:2001}
{Turner}, N.~J., \& {Stone}, J.~M. 2001, \apjs, 135, 95

\bibitem[{{van der Holst} {et~al.}(2011){van der Holst}, {T{\'o}th}, {Sokolov},
  {Powell}, {Holloway}, {Myra}, {Stout}, {Adams}, {Morel}, {Karni}, {Fryxell},
  \& {Drake}}]{van-der-Holst:2011}
{van der Holst}, B., {et~al.} 2011, \apjs, 194, 23

\bibitem[{{Vaytet} {et~al.}(2011){Vaytet}, {Audit}, {Dubroca}, \&
  {Delahaye}}]{Vaytet:2011}
{Vaytet}, N.~M.~H., {Audit}, E., {Dubroca}, B., \& {Delahaye}, F. 2011, \jqsrt,
  112, 1323

\bibitem[{{Zel'dovich} \& {Raizer}(2002)}]{Zeldovich:2002}
{Zel'dovich}, Y.~B., \& {Raizer}, Y.~P. 2002, Physics of shock waves and
  high-temperature hydrodynamic phenomena, Dover books on physics (Mineola:
  Dover Publications)

\bibitem[{{Zhang} {et~al.}(2011){Zhang}, {Howell}, {Almgren}, {Burrows}, \&
  {Bell}}]{Zhang:2011}
{Zhang}, W., {Howell}, L., {Almgren}, A., {Burrows}, A., \& {Bell}, J. 2011,
  \apjs, 196, 20

\end{thebibliography}
\clearpage

\end{document}